\documentclass[english,12pt]{article}
\usepackage[T1]{fontenc}
\usepackage[utf8]{inputenc}
\usepackage[pdftex]{graphicx}
\usepackage{verbatim}
\usepackage[width=\textwidth]{caption}
\usepackage{amsmath,amssymb,bm}
\usepackage{a4wide}
\usepackage[english]{babel}
\usepackage[]{float}
\usepackage[]{placeins}
\usepackage{flafter}
\usepackage[]{longtable}
\usepackage{verbatim}
\usepackage{amsopn}
\usepackage{dsfont}
\usepackage{color}
\usepackage{lscape}
\usepackage{array}
\usepackage{natbib}
\usepackage{pdfpages}
\usepackage{xr}
\makeatletter

\newtheorem{thm}{Theorem}
\newtheorem{cor}{Corollary}
\newtheorem{prop}{Proposition}
\newtheorem{lem}{Lemma}
\newtheorem{hyp}{Assumption}
\newtheorem{mydef}{Regression}

\newcommand{\eps}{\varepsilon}

\newcommand{\st}[1]{\texttt{#1}}
\newcommand{\pl}{\text{pl}}
\newcommand{\DID}{\text{DID}}
\newcommand{\DIDM}{\DID_{\text{M}}}

\renewcommand{\section}{\@startsection{section}{2}{0mm}{-1.5\baselineskip}{1\baselineskip}{\normalfont\large\bfseries}}
\renewcommand{\subsection}{\@startsection{subsection}{2}{0mm}{-1.2\baselineskip}{1\baselineskip}{\normalfont\normalsize\bfseries}}
\renewcommand{\subsubsection}{\@startsection{subsubsection}{3}{0mm}{-0.8\baselineskip}{0.4\baselineskip}{\normalfont\normalsize\itshape}}
\date{March 5, 2020}
\begin{document}

\title{Two-way fixed effects estimators with heterogeneous treatment effects\thanks{We are very grateful to Olivier Desch\^esnes, Guido Imbens, Peter Kuhn, Kyle Meng, Jesse Shapiro, Dick Startz, Doug Steigerwald, Cl\'emence Tricaud, Gonzalo Vazquez-Bare, members of the UCSB econometrics research group, and seminar participants at Bergen, CIREQ Econometrics conference, CREST, Goteborg, Gothenburg, Groningen, ITAM, Pompeu Fabra, Stanford, SMU, Tinbergen Institute, UCL, UCLA, UC Davis, UCSB, USC, and Warwick for their helpful comments. Xavier D'Haultf\oe uille gratefully acknowledges financial
support from the research grants Otelo (ANR-17-CE26-0015-041) and the Labex Ecodec: Investissements d’Avenir (ANR-11-IDEX-0003/Labex Ecodec/ANR-11-LABX-0047).}}

\author{Cl\'{e}ment de Chaisemartin\thanks{University of California at Santa Barbara, clementdechaisemartin@ucsb.edu%
} \and Xavier D'Haultf\oe{}uille%
\thanks{CREST-ENSAE, xavier.dhaultfoeuille@ensae.fr}}

\maketitle ~\vspace{-1cm}

\begin{abstract}
Linear regressions with period and group fixed effects are widely used to estimate treatment effects. We show that they estimate weighted sums of the average treatment effects (ATE) in each group and period, with weights that may be negative. Due to the negative weights, the linear regression coefficient may for instance be negative while all the ATEs are positive. We propose another estimator that solves this issue. In the two applications we revisit, it is significantly different from the linear regression estimator.
\end{abstract}
 \textbf{Keywords:} linear regressions, fixed effects, heterogeneous treatment effects, difference-in-differences.

\medskip
\textbf{JEL Codes:} C21, C23

%Conventions d'ecritures:
%indicator variable et pas dummy variable
%period plutot que time periods
%period and/or group fixed effects et pas period and/or group dummies
%dans la lit review: "corresponds to Regression XXXX" signifie que la regression correspond exactement a l'une de nos regressions. "is similar to Regression XXXX" signifie que s'en rapproche, mais pas exactement.
%Dans une somme de poids*parametres, on met d'abord le poids, puis le parametre.
%On utilise crochets que quand parentheses a l'interieur.
%Quand on discute différentes hypothèses, on essaye d'écrire "Condition XXXX is rarely/often met" pour les hyp fully testables, et "Assumption XXXX is plausible/implausible" pour celles non fully testables.

\newpage

\section{Introduction}

A popular method to estimate the effect of a treatment on an outcome is to compare over time groups experiencing different evolutions of their exposure to treatment. In practice, this idea is implemented by estimating regressions that control for group and time fixed effects. Hereafter, we refer to those as two-way fixed effects (FE) regressions. We conducted a survey, and found that 20\% of all empirical articles published by the American Economic Review (AER) between 2010 and 2012 have used a two-way FE regression to estimate the effect of a treatment on an outcome. When the treatment effect is constant across groups and over time, such regressions estimate that effect under the standard ``common trends'' assumption. However, it is often implausible that the treatment effect is constant. For instance, the minimum wage's effect on employment may vary across US counties, and may change over time. This paper examines the properties of two-way FE regressions when the constant effect assumption is violated.

\medskip
We start by assuming that all observations in the same $(g,t)$ cell have the same treatment and that the treatment is binary, as is for instance the case when the treatment is a county-level law. We consider the regression of $Y_{i,g,t}$, the outcome of unit $i$ in group $g$ at period $t$ on group fixed effects, period fixed effects, and $D_{g,t}$, the treatment in group $g$ at period $t$. Let $\widehat{\beta}_{fe}$ denote the coefficient of $D_{g,t}$, and let $\beta_{fe}$ denote its expectation. Under the common trends assumption, we show that $\beta_{fe}$ is equal to a weighted sum of the treatment effect in each treated $(g,t)$ cell:	
\begin{align}\label{eq:beta_fe_intro}
\beta_{fe}& =E\left(\sum_{(g,t):D_{g,t}=1}W_{g,t}\Delta_{g,t}\right).
\end{align}
$\Delta_{g,t}$ is the average treatment effect (ATE) in group $g$ and period $t$ and the weights $W_{g,t}$s sum to one but may be negative. Negative weights arise because $\widehat{\beta}_{fe}$ is a weighted sum of several difference-in-differences (DID), which compare the evolution of the outcome between consecutive time periods across pairs of groups. However, the “control group” in some of those comparisons may be treated at both periods. Then, its treatment effect at the second period gets differenced out by the DID, hence the negative weights.

\medskip
The negative weights are an issue when the ATEs are heterogeneous across groups or periods. Then, one could have that $\beta_{fe}$ is negative while all the ATEs are positive. For instance, $1.5\times 1-0.5\times 4$, a weighted sum of $1$ and $4$, is strictly negative. Using the data set of \cite{gentzkow2011}, we find that 40\% of the weights attached to $\beta_{fe}$ are negative, so $\beta_{fe}$ is not robust to heterogeneous effects.\footnote{\label{foot:gentzkowbetafd} \cite{gentzkow2011} do not estimate $\beta_{fe}$, but $\beta_{fd}$, the treatment coefficient in the first-difference regression defined below. 46\% of the weights attached to $\beta_{fd}$ are strictly negative.}

\medskip
Researchers may want to know how serious that issue is in the application they consider. We show that conditional on all treatments, the absolute value of the expectation of $\widehat{\beta}_{fe}$ divided by the standard deviation of the weights is equal to the minimal value of the standard deviation of the ATEs across the treated $(g,t)$ cells under which the average treatment on the treated (ATT) may actually have the opposite sign than that coefficient. One can estimate that ratio to assess the robustness of the two-way FE coefficient. If that ratio is close to 0, that coefficient and the ATT can be of opposite signs even under a small and plausible amount of treatment effect heterogeneity. In that case, treatment effect heterogeneity would be a serious concern for the validity of that coefficient. On the contrary, if that ratio is very large, that coefficient and the ATT can only be of opposite signs under a very large and implausible amount of treatment effect heterogeneity.

\medskip
Finally, we propose a new estimator, $\DIDM$, that is valid even if the treatment effect is heterogeneous over time or across groups. It estimates the average treatment effect across all the $(g,t)$ cells whose treatment changes from $t-1$ to $t$. It relies on common trends assumptions on both potential outcomes. Those conditions are partly testable, and we propose a test that amounts to looking at pre-trends. This test differs from the standard event study pre-trends test \citep[see][]{autor2003}, which has been shown to be invalid when treatment effects are heterogeneous \citep[see][]{abraham2018}. We show that our estimator is asymptotically normal. We compute it in the data sets of \cite{gentzkow2011} and \cite{vella1998whose}, and in both cases we find that it is significantly different from $\widehat{\beta}_{fe}$.\footnote{In both cases, our estimator is also significantly different from $\widehat{\beta}_{fd}$.} Our estimator can be used in applications where, for each pair of consecutive dates, there are groups whose treatment does not change. We estimate that this condition is satisfied for around 80\% of the papers using two-way fixed effects regressions found in our survey of the AER.

\medskip
Overall, our paper has implications for applied researchers estimating two-way fixed effects regressions. First, we recommend that they compute the weights attached to their regression and the ratio of $|\widehat{\beta}_{fe}|$ divided by the standard deviation of the weights. To do so, they can use the \st{twowayfeweights} Stata package that is available from the SSC repository. If many weights are negative, and if the ratio is not very large, we recommend that they compute our new estimator, using the \st{fuzzydid} and \st{did\_multiplegt} Stata packages, also available from the SSC repository \citep[see][for explanations on how to use the former package]{deChaisemartin18}.

\medskip
We extend our results in several important directions. First, another commonly-used regression is the first-difference regression of $Y_{g,t}-Y_{g,t-1}$, the change in the mean outcome in group $g$, on period fixed effects and on $D_{g,t}-D_{g,t-1}$, the change in the treatment. We let $\beta_{fd}$ denote the expectation of the coefficient of $D_{g,t}-D_{g,t-1}$. We show that under common trends, $\beta_{fd}$ also identifies a weighted sum of treatment effects, with potentially some negative weights. Second, in our Web Appendix we show that our results extend to fuzzy designs, where the treatment varies within $(g,t)$ cells, and to two-way fixed effects regressions with a non-binary treatment and with covariates.

\medskip
Our paper is related to the DID literature. Our main result generalizes Theorem 1 in \cite{deChaisemartin15b}. When the data has two groups and two periods, the Wald-DID estimand considered therein is equal to $\beta_{fe}$ and $\beta_{fd}$. Our results on $\beta_{fe}$ and $\beta_{fd}$ are thus extensions of that theorem to the case with multiple periods and groups.\footnote{In fact, a preliminary version of our main result appeared in a working paper version of \cite{deChaisemartin15b} \citep[see Theorems S1 and S2 in][]{deChaisemartin15c}.} Moreover, our $\DIDM$ estimator is related to the Wald-TC estimator with many groups and periods proposed in \cite{deChaisemartin15b}, and to the multi-period DID estimator proposed by \cite{imai2018}. In Section \ref{sec:alternative_estimand}, we explain the differences between those three estimators.

\medskip
More recently, \cite{borusyak2016}, \cite{abraham2018}, \cite{athey2018}, \cite{callaway2018}, and \cite{goodman2018} study the special case of staggered adoption designs, where the treatment of a group is weakly increasing over time. Those papers derive some important results specific to that design that we do not consider here. Still, some of the results in those papers are related to ours, and we describe precisely those connections later in the paper.
The most important dimension on which our paper differs from those is that our results apply to any two-way fixed effects regressions, not only to those with staggered adoption. In our survey of the AER papers estimating two-way fixed effects regressions, less than 10\% have a staggered adoption design. This suggests that while staggered adoptions are an important research design, they may account for a relatively small minority of the applications where two-way fixed effects regressions have been used.

\medskip
The paper is organized as follows. Section \ref{sec:regressions} introduces the set-up. Section \ref{sec:sharpcase} presents our decomposition results. Section \ref{sec:alternative_estimand} introduces our alternative estimator. Section \ref{sec:extensions} briefly describes some of the extensions covered in our Web Appendix. Section \ref{sec:appli} presents our survey of the articles published in the AER, and our two empirical applications.

\section{Set up}\label{sec:regressions}

One considers observations that can be divided into $G$ groups and $T$ periods. For every $(g,t)\in \{1,...,G\}\times \{1,...,T\}$, let $N_{g,t}$ denote the number of observations in group $g$ at period $t$, and let $N=\sum_{g,t}N_{g,t}$ be the total number of observations. The data may be an individual-level panel or repeated cross-section data set where groups are, say, individuals' county of birth. The data could also be a cross-section where cohort of birth plays the role of time. For instance, \cite{Duflo01} compares the schooling of different cohorts in Indonesia, some of which were exposed to a school construction program. It is also possible that for all $(g,t)$, $N_{g,t}=1$, e.g. a group is one individual or firm. All of the above are special cases of the data structure we consider.

\medskip
One is interested in measuring the effect of a treatment on some outcome. Throughout the paper we assume that treatment is binary, but our results apply to any ordered treatment, as we show in Section \ref{sub:non_binary_ordered_treatment} of the Web Appendix. Then,
for every $(i,g,t)\in \{1,...,N_{g,t}\}\times \{1,...,G\}\times \{1,...,T\}$, let $D_{i,g,t}$ and $(Y_{i,g,t}(0),Y_{i,g,t}(1))$ respectively denote the treatment status and the potential outcomes without and with treatment of observation $i$ in group $g$ at period $t$.

\medskip
The outcome of observation $i$ in group $g$ and period $t$ is $Y_{i,g,t}=Y_{i,g,t}(D_{i,g,t})$.
For all $(g,t)$, let
\begin{equation*}
D_{g,t}=\frac{1}{N_{g,t}}\sum_{i=1}^{N_{g,t}}D_{i,g,t},~Y_{g,t}(0)=\frac{1}{N_{g,t}}\sum_{i=1}^{N_{g,t}}Y_{i,g,t}(0),~Y_{g,t}(1)=\frac{1}{N_{g,t}}\sum_{i=1}^{N_{g,t}}Y_{i,g,t}(1),\text{ and }Y_{g,t}=\frac{1}{N_{g,t}}\sum_{i=1}^{N_{g,t}}Y_{i,g,t}.
\end{equation*}
$D_{g,t}$ denotes the average treatment in group $g$ at period $t$, while $Y_{g,t}(0)$, $Y_{g,t}(1)$, and $Y_{g,t}$ respectively denote the average potential outcomes without and with treatment and the average observed outcome in group $g$ at period $t$.

\medskip
Throughout the paper, we maintain the following assumptions.
\begin{hyp}\label{hyp:supp_gt}
	(Balanced panel of groups) For all $(g,t)\in \{1,...,G\}\times  \{1,...,T\}$, $N_{g,t}>0$.
\end{hyp}
Assumption \ref{hyp:supp_gt} requires that no group appears or disappears over time. This assumption is often satisfied. Without it, our results still hold but the notation becomes more complicated as the denominators of some of the fractions below may then be equal to zero.
\begin{hyp}\label{hyp:sharp}
	(Sharp design) For all $(g,t)\in \{1,...,G\}\times  \{1,...,T\}$ and $i\in\{1,...,N_{g,t}\}$, $D_{i,g,t}=D_{g,t}$.
\end{hyp}
Assumption \ref{hyp:sharp} requires that units' treatments do not vary within each $(g,t)$ cell, a situation we refer to as a sharp design. This is for instance satisfied when the treatment is a group-level variable, for instance a county- or a state-law. This is also mechanically satisfied when $N_{g,t}=1$. In our survey in Section \ref{sub:litreview}, we find that almost 80\% of the papers using two-way fixed effects regressions and published in the AER between 2010 and 2012 consider sharp designs. We focus on sharp designs because of their prevalence, but in Section \ref{sec:fuzzycase} of the Web Appendix, we show that all the results in Sections \ref{sec:sharpcase}-\ref{sec:alternative_estimand} below can be extended to fuzzy designs.
\begin{hyp}\label{hyp:independent_groups}
	(Independent groups) The vectors $(Y_{g,t}(0),Y_{g,t}(1),D_{g,t})_{1\leq t\leq T}$ are mutually independent.
\end{hyp}
We consider $D_{g,t}$, $Y_{g,t}(0)$, $Y_{g,t}(1)$ as random variables. For instance, aggregate random shocks may affect the average potential outcomes of group $g$ at period $t$. The treatment status of group $g$ at period $t$ may also be random. The expectations below are taken with respect to the distribution of those random variables. Assumption \ref{hyp:independent_groups} allows for the possibility that the treatments and potential outcomes of a group may be correlated over time, but it requires that the potential outcomes and treatments of different groups be independent.
\begin{hyp}\label{hyp:strong_exogeneity}
	(Strong exogeneity) For all $(g,t)\in \{1,...,G\}\times\{2,...,T\}$, \\ $E(Y_{g,t}(0)-Y_{g,t-1}(0)|D_{g,1},...,D_{g,T})=E(Y_{g,t}(0)-Y_{g,t-1}(0))$.
\end{hyp}
Assumption \ref{hyp:strong_exogeneity} requires that the shocks affecting a group's $Y_{g,t}(0)$ be mean independent of that group's treatment sequence. This rules out the possibility that a group gets treated because it experiences negative shocks, the so-called Ashenfelter's dip \citep[see][]{ashenfelter1978estimating}. Assumption \ref{hyp:strong_exogeneity} is related to the strong exogeneity condition in panel data models, which, as is well-known, is necessary to obtain the consistency of the fixed effects estimator \citep[see, e.g.,][]{wooldridge2002}.

\medskip
We now define the FE regression described in the introduction.\footnote{\label{foot:well_def} Throughout the paper, we assume that $D_{g,t}$ in Regression \ref{reg1} and $D_{g,t}-D_{g,t-1}$ in Regression \ref{reg2} below are not collinear with the other independent variables in those regressions, so $\widehat{\beta}_{fe}$ and $\widehat{\beta}_{fd}$ are well-defined.}
\begin{mydef}\label{reg1}
(Fixed-effects regression)

\medskip
Let $\widehat{\beta}_{fe}$ denote the coefficient of $D_{g,t}$ in an OLS regression of $Y_{i,g,t}$ on group fixed effects, period fixed effects, and $D_{g,t}$. Let $\beta_{fe}=E\left[\widehat{\beta}_{fe}\right]$.\footnote{As the independent variables in Regression \ref{reg1} are constant within each $(g,t)$ cell, Regression \ref{reg1} is equivalent to a $(g,t)$-level regression of $Y_{g,t}$ on group and period fixed effects and $D_{g,t}$, weighted by $N_{g,t}$.}
\end{mydef}

\medskip
For all $g$ and $t$, let $N_{g,.}=\sum_{t=1}^{T}N_{g,t}$ and $N_{.,t}=\sum_{g=1}^{G}N_{g,t}$ respectively denote the total number of observations in group $g$ and in period $t$. For any variable $X_{g,t}$ defined in each $(g,t)$ cell, let $X_{g,.}=\sum_{t=1}^{T}(N_{g,t}/N_{g,.})X_{g,t}$ denote the average value of $X_{g,t}$ in group $g$, let $X_{.,t}=\sum_{g=1}^{G}(N_{g,t}/N_{.,t})X_{g,t}$ denote the average value of $X_{g,t}$ in period $t$, and let $X_{.,.}=\sum_{g,t}(N_{g,t}/N)X_{g,t}$ denote the average value of $X_{g,t}$. For instance, $D_{3,.}$ and $D_{.,2}$ respectively denote the average treatment in group 3 across time and in period 2 across groups, whereas $Y_{.,.}$ denotes the average value of the outcome across groups and time. Finally, for any variable $X_{g,t}$, we let $\boldsymbol{X}$ denote the vector $(X_{g,t})_{(g,t)\in \{1,...,G\}\times \{1,...,T\}}$ collecting the values of that variable in each $(g,t)$ cell. For instance, $\boldsymbol{D}$ is the vector $(D_{g,t})_{(g,t)\in \{1,...,G\}\times \{1,...,T\}}$ collecting the treatments of all the $(g,t)$ cells.

\section{Two-way fixed effects regressions}
\label{sec:sharpcase}

\subsection{A decomposition result} % (fold)
\label{sub:FE_as_weighted_average_of_average_treatment_effects_sharp}

We study the FE regression under the following common trends assumption.
\begin{hyp}\label{hyp:common_trends}
(Common trends)
\medskip
For $t\geq 2$, $E(Y_{g,t}(0) - Y_{g,t-1}(0))$ does not vary across $g$.
\end{hyp}
Assumption \ref{hyp:common_trends} requires that the expectation of the outcome without treatment follow the same evolution over time in every group. When $t$ represents birth cohorts, Assumption \ref{hyp:common_trends} requires that the outcome difference between consecutive cohorts be the same across groups.

\medskip
Let $N_1=\sum_{i,g,t}D_{i,g,t}$ denote the number of treated units, let
$$\Delta^{TR} =  \frac{1}{N_1} \sum_{(i, g,t):D_{g,t}=1}\left[Y_{i,g,t}(1)-Y_{i,g,t}(0)\right]$$
denote the average treatment effect across all treated units, and let $\delta^{TR}=E\left[\Delta^{TR}\right]$
denote the expectation of that parameter, hereafter referred to as the ATT.
For any $(g,t)\in \{1,...,G\}\times \{1,...,T\}$, let $$\Delta_{g,t}=\frac{1}{N_{g,t}}\sum_{i=1}^{N_{g,t}} \left[Y_{i,g,t}(1)-Y_{i,g,t}(0)\right]$$
denote the ATE in cell $(g,t)$.
$\delta^{TR}$ is equal to the expectation of a weighted average of the treated cells' $\Delta_{g,t}$s:
\begin{equation}
\delta^{TR} =  E\left[\sum_{g,t:D_{g,t}=1} \frac{N_{g,t}}{N_1} \Delta_{g,t}\right].
\label{eq:decomp_Delta_TR}	
\end{equation}
Under the common trends assumption, we show that $\beta_{fe}$ is also equal to the expectation of a weighted sum of the $\Delta_{g,t}$s, with potentially some negative weights.

\medskip
Let $\eps_{g,t}$ denote the residual of observations in cell $(g,t)$ in the regression of $D_{g,t}$ on group and period fixed effects:\footnote{\label{foot:eps_does_not_vary_at_i_level} $\eps_{g,t}$ arises from a unit-level regression, where the dependent and independent variables only vary at the $(g,t)$ level. Therefore, all the units in the same $(g,t)$ cell have the same value of $\eps_{g,t}$.}
$$D_{g,t}=\alpha+\gamma_g+\lambda_t+\eps_{g,t}.$$
One can show that if the regressors in Regression \ref{reg1} are not collinear, the average value of $\eps_{g,t}$ across all treated $(g,t)$ cells differs from 0: $\sum_{(g,t):D_{g,t}=1}(N_{g,t}/N_1)\eps_{g,t}\neq 0$. Then we let $w_{g,t}$ denote $\eps_{g,t}$ divided by that average:
$$w_{g,t}=\frac{\eps_{g,t}}{\sum_{(g,t):D_{g,t}=1}\frac{N_{g,t}}{N_1}\eps_{g,t}}.$$

\begin{thm}
Suppose that Assumptions \ref{hyp:supp_gt}-\ref{hyp:common_trends} hold.  Then,\footnote{\label{foot:conditional_version_of_Theorem_1} In the proof, we show the following, stronger result: $$E\left[\widehat{\beta}_{fe}\middle|\mathbf{D}\right] =  \sum_{(g,t):D_{g,t}=1}\frac{N_{g,t}}{N_1}w_{g,t} E\left[\Delta_{g,t}|\mathbf{D}\right].$$}
\begin{align*}
\beta_{fe}  & =E\left[\sum_{(g,t):D_{g,t}=1}\frac{N_{g,t}}{N_1}w_{g,t}\Delta_{g,t}\right].
\end{align*}
\label{thm:main_sharp}
\end{thm}

This  result implies that in general, $\beta_{fe}\neq \delta^{TR}$, so $\widehat{\beta}_{fe}$ is a biased estimator of the ATT. To illustrate this, we consider a simple example of a staggered adoption design with two groups and three periods, and where the treatments are non-stochastic: group 1 is untreated at periods 1 and 2 and treated at period 3, while group 2 is untreated at period 1 and treated both at periods 2 and 3.\footnote{\label{foot:B&J_example} A similar example appears in \cite{borusyak2016}.} We also assume that $N_{g,t}/N_{g,t-1}$ does not vary across $g$: all groups experience the same growth of their number of observations from $t-1$ to $t$, a requirement that is for instance satisfied when the data is a balanced panel. Then, one can show that
\begin{equation*}\label{eq:closed_form_espfe}
\eps_{g,t}=D_{g,t}-D_{g,.}-D_{.,t}+D_{.,.},
\end{equation*}
thus implying that
\begin{align*}
\eps_{1,3}&=1-1/3-1+1/2=1/6, \\
\eps_{2,2}&=1-2/3-1/2+1/2=1/3,\\
\eps_{2,3}&=1-2/3-1+1/2=-1/6.
\end{align*}
The residual is negative in group 2 and period 3, because the regression predicts a treatment probability larger than one in that cell, a classic extrapolation problem with linear regressions. Then, under the common trends assumption, it follows from Theorem \ref{thm:main_sharp} and the fact that the treatments are non-stochastic that
\begin{align*}
\beta_{fe}&=1/2 E\left[\Delta_{1,3}\right]+E\left[\Delta_{2,2}\right]-1/2E\left[\Delta_{2,3}\right].
\end{align*}
$\beta_{fe}$ is equal to a weighted sum of the ATEs in group 1 at period 3, group 2 at period 2, and group 2 at period 3, the three treated $(g,t)$ cells. However, the weight assigned to each ATE differs from $1/3$, the proportion that each cell accounts for in the population of treated observations. Therefore, $\beta_{fe}$ is not equal to $\delta^{TR}$. Perhaps more worryingly, not all the weights are positive: the weight assigned to the ATE in group 2 period 3 is strictly negative. Consequently, $\beta_{fe}$  may be a very misleading measure of the treatment effect. Assume for instance that $E\left[\Delta_{1,3}\right]=E\left[\Delta_{2,2}\right]=1$ and $E\left[\Delta_{2,3}\right]=4$. At the period when they start receiving the treatment, both groups experience a modest positive ATE. But this effect builds over time and in period 3, one period after it has started receiving the treatment, group 2 now experiences a large ATE. Then,
\begin{align*}
\beta_{fe}&=1/2\times 1 +1-1/2 \times 4=-1/2.
\end{align*}
$\beta_{fe}$ is strictly negative, while $E\left[\Delta_{1,3}\right]$, $E\left[\Delta_{2,2}\right]$, and $E\left[\Delta_{2,3}\right]$ are all positive. More generally, the negative weights are an issue if the $E\left[\Delta_{g,t}\right]$s are heterogeneous, across groups or over time.\footnote{\label{foot:heterog} On the other hand, $\beta_{fe}$ does not rule out heterogeneous treatment effects within $(g,t)$ cells, as it is identified by variations across $(g,t)$ cells, and does not leverage any within-cell variation.} If $E\left[\Delta_{1,3}\right]=E\left[\Delta_{2,2}\right]=E\left[\Delta_{2,3}\right]=1$, then $\beta_{fe}=1=\delta^{TR}$.

\medskip
Here is some intuition as to why one weight is negative in this example. It follows from Equation \eqref{eq:GoodmanBacon} in the proof of Theorem \ref{thm:main_sharp} \citep[see also Theorem 1 in][]{goodman2018} that in this simple example, $\beta_{fe}=(\DID_1 +\DID_2)/2$, with
\begin{align*}
\DID_1& = E(Y_{2,2})-E(Y_{2,1})-\left(E(Y_{1,2})-E(Y_{1,1})\right),\\
\DID_2& = E(Y_{1,3})-E(Y_{1,2})-\left(E(Y_{2,3})-E(Y_{2,2})\right).
\end{align*}
The first DID compares the evolution of the mean outcome from period 1 to 2 in group 2 and in group 1. The second one compares the evolution of the mean outcome from period 2 to 3 in group 1 and in group 2. The control group in the second DID, group 2, is treated both in the pre and in the post period. Therefore, under the common trends assumption, it follows from Lemma \ref{lem:IV-DID} in Appendix \ref{sub:additional_notation_and_two_useful_lemmas} (a similar result appears in Lemma 1 of \cite{Chaisemartin2011fuzzy} and in Equation (13) of \cite{goodman2018}) that $\DID_1=E\left[\Delta_{2,2}\right]$, but
$$\DID_2 = E\left[\Delta_{1,3}\right] - (E\left[\Delta_{2,3}\right] - E\left[\Delta_{2,2}\right]).$$
$\DID_2$ is equal to the ATE in group 1 period 3, minus the change in group 2's ATE between periods 2 and 3. Intuitively, the mean outcome of groups 1 and 2 may follow different trends from period 2 to 3 either because group 1 becomes treated, or because group 2's ATE changes. The intuition that negative weights arise because $\widehat{\beta}_{fe}$ uses treated observations as controls also appears in \cite{borusyak2016}.

\medskip
We now generalize the previous illustration by characterizing the $(g,t)$ cells whose ATEs are weighted negatively by $\beta_{fe}$.

\begin{prop}\label{prop:staggered_CT_fe}
Suppose that Assumption \ref{hyp:supp_gt} holds and for all $t\geq 2$ $N_{g,t}/N_{g,t-1}$ does not vary across $g$. Then, for all $(g, t,t')$ such that $D_{g,t}=D_{g,t'}=1$, $D_{.,t}>D_{.,t'}$ implies $w_{g,t}<w_{g,t'}$. Similarly, for all $(g, g',t)$ such that $D_{g,t}=D_{g',t}=1$, $D_{g,.}>D_{g',.}$ implies $w_{g,t}<w_{g',t}$.
\end{prop}

Proposition \ref{prop:staggered_CT_fe} shows that $\beta_{fe}$ is more likely to assign a negative weight to periods where a large fraction of groups are treated, and to groups treated for many periods. Then, negative weights are a concern when treatment effects differ between periods with many versus few treated groups, or between groups treated for many versus few periods.

\medskip
Proposition \ref{prop:staggered_CT_fe} has interesting implications in staggered adoption designs, a special case of sharp designs defined as follows.

\begin{hyp}\label{hyp:staggered_adoption_design}
	(Staggered adoption designs) For all $g$, $D_{g,t}\geq D_{g,t-1}$ for all $t\geq 2$.
\end{hyp}
Assumption \ref{hyp:staggered_adoption_design} is satisfied in applications where groups adopt a treatment at heterogeneous dates \citep[see e.g.][]{athey2002}. In that design, \cite{borusyak2016} show that $\beta_{fe}$ is more likely to assign a negative weight to treatment effects at the last periods of the panel. This result is a special case of Proposition \ref{prop:staggered_CT_fe}: in staggered adoption designs, $D_{.,t}$ is increasing in $t$, so Proposition \ref{prop:staggered_CT_fe} implies that $w_{g,t}$ is decreasing in $t$.\footnote{\cite{borusyak2016} assume that the treatment effect of cell $(g,t)$ only depends on the number of periods since group $g$ has started receiving the treatment, whereas Proposition \ref{prop:staggered_CT_fe} does not rely on that assumption.} Proposition \ref{prop:staggered_CT_fe} also implies that in that design, groups that adopt the treatment earlier are more likely to receive some negative weights.

\medskip
Finally, in staggered adoption designs, \cite{athey2018} derive a decomposition of $\beta_{fe}$ that resembles to, but differs from, that in Theorem \ref{thm:main_sharp}. They derive their decomposition under the assumption that the dates at which each group starts receiving the treatment are randomly assigned, while we derive ours under a common trends assumption.

\subsection{Robustness to heterogeneous treatment effects}
\label{sub:robust}

Theorem \ref{thm:main_sharp} shows that in sharp designs with many groups and periods, $\widehat{\beta}_{fe}$ may be a misleading measure of the treatment effect under the standard common trends assumption, if the treatment effect is heterogeneous across groups and time periods. In the corollary below, we propose two robustness measures that can be used to assess how serious that concern is.

\medskip
Those robustness measures are defined conditional on $\bm{D}$, the vector stacking together the treatments of all the $(g,t)$ cells. Specifically, for all $(g,t)\in\{1,...,G\}\times\{1,...,T\}$, let $\widetilde{\Delta}_{g,t}=E\left(\Delta_{g,t}\middle|\bm{D}\right)$ denote the ATE in cell $(g,t)$ conditional on $\bm{D}$,\footnote{\label{foot:cond_versus_non_cond_ATE} $\widetilde{\Delta}_{g,t}$ may differ from $E(\Delta_{g,t})$. To see this, let us consider a simple example where $T=2$. Then, under Assumption \ref{hyp:independent_groups}, one has $\widetilde{\Delta}_{g,t}=E\left(\Delta_{g,t}\middle|D_{g,1},D_{g,2}\right)$. One may for instance have $E\left(\Delta_{g,1}\middle|D_{g,1}=0,D_{g,2}=0\right)< E\left(\Delta_{g,1}\middle|D_{g,1}=1,D_{g,2}=1\right)$, if a group is more likely to be treated if her treatment effect is initially high.} let $\widetilde{\Delta}^{TR}=E\left(\Delta^{TR}\middle|\bm{D}\right)$ denote the ATT conditional on $\bm{D}$, and let $\widetilde{\beta}_{fe}=E\left(\widehat{\beta}_{fe}\middle|\bm{D}\right)$. The first measure we consider is the minimal value of the standard deviation of the $\widetilde{\Delta}_{g,t}$s under which one could have that $\widetilde{\beta}_{fe}$ is of a different sign than $\widetilde{\Delta}^{TR}$. Therefore, this summary measure applies to $\widetilde{\beta}_{fe}$ and $\widetilde{\Delta}^{TR}$, rather than $\beta_{fe}$ and $\delta^{TR}$, the unconditional expectations of $\widehat{\beta}_{fe}$ and $\Delta^{TR}$ on which we have focused so far. However, one can show that when $G$, the number of groups,  goes to infinity, $\widetilde{\beta}_{fe}-\beta_{fe}$ and $\widetilde{\Delta}^{TR}-\delta^{TR}$ both converge to 0. So if the number of groups is large, $\widetilde{\beta}_{fe}$ and $\widetilde{\Delta}^{TR}$ should not differ much from $\beta_{fe}$ and $\delta^{TR}$, and our robustness measure ``almost'' applies to $\beta_{fe}$ and $\delta^{TR}$.

\medskip
Let
\begin{align*}
\sigma(\boldsymbol{\widetilde{\Delta}}) =&\left(\sum_{(g,t):D_{g,t}=1}\frac{N_{g,t}}{N_1}\left(\widetilde{\Delta}_{g,t} -\widetilde{\Delta}^{TR}\right)^2\right)^{1/2},\\
\sigma(\boldsymbol{w}) =&\left(\sum_{(g,t):D_{g,t}=1}\frac{N_{g,t}}{N_1}\left(w_{g,t}-1\right)^2\right)^{1/2}.
\end{align*}
$\sigma(\boldsymbol{\widetilde{\Delta}})$ is the standard deviation of the conditional ATEs, and $\sigma(\boldsymbol{w})$ is the standard deviation of the $\boldsymbol{w}$-weights,\footnote{\label{foot:weights1} One can show that $\sum_{(g,t):D_{g,t}=1}(N_{g,t}/N_1)w_{g,t}=1$.} across the treated $(g,t)$ cells. Let $n=\#\{(g,t):D_{g,t}=1\}$ denote the number of treated cells. For every $i\in\{1,...,n\}$, let $w_{(i)}$ denote the $i$th largest of the weights of the treated cells: $w_{(1)}\geq  w_{(2)} \geq ... \geq w_{(n)}$, and let $N_{(i)}$ and $\widetilde{\Delta}_{(i)}$ be the number of observations and the conditional ATE of the corresponding cell. Then, for any $k\in\{1,...,n\}$, let $P_k=\sum_{i\geq k}N_{(i)}/N_1$, $S_k=\sum_{i\geq k} (N_{(i)}/N_1) w_{(i)}$ and $T_k=\sum_{i\geq k}(N_{(i)}/N_1) w^2_{(i)}$.

\begin{cor}
Suppose that Assumptions \ref{hyp:supp_gt}-\ref{hyp:common_trends} hold.
\begin{enumerate}
\item If $\sigma(\boldsymbol{w})>0$, the minimal value of $\sigma(\boldsymbol{\widetilde{\Delta}})$ compatible with $\widetilde{\beta}_{fe}$ and $\widetilde{\Delta}^{TR}=0$ is
$$\underline{\sigma}_{fe}= \frac{|\widetilde{\beta}_{fe}|}{\sigma(\boldsymbol{w})}.$$
\item If $\widetilde{\beta}_{fe}\ne 0$ and at least one of the $w_{g,t}$ weights is strictly negative, the minimal value of $\sigma(\boldsymbol{\widetilde{\Delta}})$ compatible with $\widetilde{\beta}_{fe}$ and with $\widetilde{\Delta}_{g,t}$ of a different sign than $\widetilde{\beta}_{fe}$ for all $(g,t)$ is
$$\underline{\underline{\sigma}}_{fe}= \frac{|\widetilde{\beta}_{fe}|}{\left[T_s + S^2_s/(1-P_s)\right]^{1/2}},$$
where $s=\min\{i\in\{1,...,n\}:w_{(i)}<-S_{(i)}/(1-P_{(i)})\}.$
\end{enumerate}
\label{cor:sensib}
\end{cor}
$\underline{\sigma}_{fe}$ and $\underline{\underline{\sigma}}_{fe}$ can be estimated simply by replacing $\widetilde{\beta}_{fe}$ by $\widehat{\beta}_{fe}$. An estimator of $\underline{\sigma}_{fe}$ can be used to assess the robustness of $\widehat{\beta}_{fe}$ to treatment effect heterogeneity across groups and periods. If $\underline{\sigma}_{fe}$ is close to 0, $\widetilde{\beta}_{fe}$ and $\widetilde{\Delta}^{TR}$ can be of opposite signs even under a small and plausible amount of treatment effect heterogeneity. In that case, treatment effect heterogeneity would be a serious concern for the validity of $\widehat{\beta}_{fe}$. On the contrary, if $\underline{\sigma}_{fe}$ is very large, $\widetilde{\beta}_{fe}$ and $\widetilde{\Delta}^{TR}$ can only be of opposite signs under a very large and implausible amount of treatment effect heterogeneity. Then, treatment effect heterogeneity is less of a concern.

\medskip
Similarly, if $\underline{\underline{\sigma}}_{fe}$ is close to 0, one may have, say, $\widetilde{\beta}_{fe}>0$, while $\widetilde{\Delta}_{g,t}\leq 0$ for all $(g,t)$, even if the dispersion of the $\widetilde{\Delta}_{g,t}$s across $(g,t)$ cells is relatively small. Notice that $\underline{\underline{\sigma}}_{fe}$ is only defined if at least one of the weights is strictly negative: if all the weights are positive, then one cannot have that $\widetilde{\beta}_{fe}$ is of a different sign than all the $\widetilde{\Delta}_{g,t}$s.

\medskip
When some of the weights $w_{g,t}$ are negative, $\widehat{\beta}_{fe}$ may still be robust to heterogeneous treatment effects across groups and periods, provided the assumption below is satisfied.
\begin{hyp}\label{hyp:HTE_fe}
($\boldsymbol{w}$ uncorrelated with $\boldsymbol{\widetilde{\Delta}}$) $E\left[\sum_{(g,t):D_{g,t}=1}\frac{N_{g,t}}{N_1}(w_{g,t}-1)(\widetilde{\Delta}_{g,t}-\widetilde{\Delta}^{TR})\right]=0$.
\end{hyp}

\begin{cor}
If Assumptions \ref{hyp:supp_gt}-\ref{hyp:common_trends} and \ref{hyp:HTE_fe} hold, then $\beta_{fe}=\delta^{TR}$.
\label{cor:RH}
\end{cor}
Assumption \ref{hyp:HTE_fe} requires that the weights attached to the fixed effects estimator be uncorrelated with the conditional ATEs in the treated $(g,t)$ cells. This is often implausible. For instance, groups treated the most are also those with the lowest value of $w_{g,t}$, as shown in Proposition \ref{prop:staggered_CT_fe}. But those groups could also be those with the largest treatment effect. This would then induce a negative correlation between $\boldsymbol{w}$ and $\boldsymbol{\widetilde{\Delta}}$.
The plausibility of Assumption \ref{hyp:HTE_fe} can be assessed, by looking at whether $\boldsymbol{w}$ is correlated with a predictor of the treatment effect in each $(g,t)$ cell. In the two applications we revisit in Section \ref{sec:appli}, this test is rejected.

\subsection{Extension to the first-difference regression} % (fold)
\label{sub:FD}

Instead of Regression \ref{reg1}, many articles have estimated the first-difference regression defined below:
\begin{mydef}\label{reg2}
(First-difference regression)

\medskip
Let $\widehat{\beta}_{fd}$ denote the coefficient of $D_{g,t}-D_{g,t-1}$ in an OLS regression of $Y_{g,t}-Y_{g,t-1}$ on period fixed effects and $D_{g,t}-D_{g,t-1}$, among observations for which $t\geq 2$. Let $\beta_{fd}=E\left[\widehat{\beta}_{fd}\right]$.
\end{mydef}

When $T=2$ and $N_{g,2}/N_{g,1}$ does not vary across $g$, meaning that all groups experience the same growth of their number of units from period 1 to 2, one can show that $\widehat{\beta}_{fe}=\widehat{\beta}_{fd}$. $\widehat{\beta}_{fe}$ differs from $\widehat{\beta}_{fd}$ if $T>2$ or $N_{g,2}/N_{g,1}$ varies across $g$.

\medskip
We start by showing that a result similar to Theorem \ref{thm:main_sharp} also applies to $\widehat{\beta}_{fd}$. For any $(g,t)\in \{1,...,G\}\times \{2,...,T\}$, let $\eps_{fd,g,t}$ denote the residual of observations in group $g$ and at period $t$ in the regression of $D_{g,t}-D_{g,t-1}$ on period fixed effects, among observations for which $t\geq 2$. For any $g\in \{1,...,G\}$, let $\eps_{fd,g,1}=\eps_{fd,g,T+1}=0$. One can show that if the regressors in Regression \ref{reg2} are not perfectly collinear, $$\sum_{(g,t):D_{g,t}=1}\frac{N_{g,t}}{N_1}\left(\eps_{fd,g,t}- \frac{N_{g,t+1}}{N_{g,t}}\eps_{fd,g,t+1}\right)\neq 0.$$
Then we define
$$w_{fd,g,t}=\frac{\eps_{fd,g,t}- \frac{N_{g,t+1}}{N_{g,t}}\eps_{fd,g,t+1}}{\sum_{(g,t):D_{g,t}=1}\frac{N_{g,t}}{N_1}\left(\eps_{fd,g,t}- \frac{N_{g,t+1}}{N_{g,t}}\eps_{fd,g,t+1}\right)}.$$

\begin{thm}\label{thm:main_fd}
Suppose that Assumptions \ref{hyp:supp_gt}-\ref{hyp:common_trends} hold. Then,
\begin{align*}
\beta_{fd} & =E\left[\sum_{(g,t):D_{g,t}=1}\frac{N_{g,t}}{N_1}w_{fd,g,t}\Delta_{g,t}\right].
\end{align*}
\end{thm}

Theorem \ref{thm:main_fd} shows that under Assumption \ref{hyp:common_trends}, $\beta_{fd}$ is equal to a weighted sum of the ATEs in each treated $(g,t)$ cell with potentially some strictly negative weights, just as $\beta_{fe}$. We now characterize the $(g,t)$ cells whose ATEs are weighted negatively by $\beta_{fd}$. To do so, we focus on staggered adoption designs, as outside of this case it is more difficult to characterize those cells. Our characterization relies on the fact that for every $t\in \{2,...,T\}$, $\eps_{fd,g,t}=D_{g,t}-D_{g,t-1}-\left(D_{.,t}-D_{.,t-1}\right)$. $\eps_{fd,g,t}$ is the difference between the change of the treatment in group $g$ between $t-1$ and $t$, and the average change of the treatment across all groups.

\begin{prop}\label{prop:staggered_CT_fd}
Suppose that Assumptions \ref{hyp:supp_gt}-\ref{hyp:sharp} and \ref{hyp:staggered_adoption_design} hold
and for all $g$, $N_{g,t}$ does not depend on $t$. Then, for all $(g,t)$ such that $D_{g,t}=1$, $w_{fd,g,t}<0$ if and only if $D_{g,t-1}=1$ and $D_{.,t}-D_{.,t-1}>D_{.,t+1}-D_{.,t}$ (with the convention that $D_{.,T+1}=D_{.,T}$).
\end{prop}

Proposition \ref{prop:staggered_CT_fd} shows that for all $t\in \{2,...,T-1\}$ such that the increase in the proportion of treated units is larger from $t-1$ to $t$ than from $t$ to $t+1$, the period-$t$ ATE of groups already treated in $t-1$ receives a negative weight. Moreover, if, at period $T$, at least one group becomes treated, the ATE of groups already treated in $T-1$ also receives a negative weight. Therefore, the treatment effect arising at the date when a group starts receiving the treatment does not receive a negative weight, only long-run treatment effects do. Then, negative weights are a concern when instantaneous and long-run treatment effects may differ. Proposition \ref{prop:staggered_CT_fd} also shows that the prevalence of negative weights depends on how the number of groups that start receiving the treatment at date $t$ evolves with $t$. Assume for instance that this number decreases with $t$: many groups start receiving the treatment at date 1, a bit less start at date 2, etc., a case hereafter referred to as the ``more early adopters'' case. Then, if $N_{g,t}$ is constant across $(g,t)$, $D_{.,t}-D_{.,t-1}$ is decreasing in $t$, and all the long-run treatment effects receive negative weights, except maybe those of period $T$ if $D_{.,T}=D_{.,T-1}$. Conversely, assume that the number of groups that start receiving the treatment at date $t$ increases with $t$: few groups start receiving the treatment at date 1, a bit more start at date 2, etc., a case hereafter referred to as the ``more late adopters'' case. Then, if $N_{g,t}$ is constant across $(g,t)$, $D_{.,t}-D_{.,t-1}$ is increasing in $t$, and only the period-$T$ long-run treatment effects receive negative weights. Overall, negative weights are much more prevalent in the ``more early adopters'' than in the ``more late adopters'' case.

\medskip
We now come back to general sharp designs where the treatment may not follow a staggered adoption.
Let $\widetilde{\beta}_{fd}=E\left(\widehat{\beta}_{fd}\middle|\bm{D}\right)$ denote the expectation of $\widehat{\beta}_{fd}$ conditional on the vector of treatment assignments $\bm{D}$. Just as for $\widetilde{\beta}_{fe}$, one can show that the minimal value of $\sigma(\boldsymbol{\widetilde{\Delta}})$ compatible with $\widetilde{\beta}_{fd}$ and $\widetilde{\Delta}^{TR}=0$ is
$\underline{\sigma}_{fd}=|\widetilde{\beta}_{fd}|/\sigma(\boldsymbol{w_{fd}}),$
where
\begin{align*}
\sigma(\boldsymbol{w_{fd}}) =&\left(\sum_{(g,t):D_{g,t}=1}\frac{N_{g,t}}{N_1}\left(w_{fd,g,t}-1\right)^2\right)^{1/2}
\end{align*}
is the standard deviation of the $\boldsymbol{w_{fd}}$-weights. One can also show that $\underline{\underline{\sigma}}_{fd}$, the minimal value of $\sigma(\boldsymbol{\widetilde{\Delta}})$ compatible with $\widetilde{\beta}_{fd}$ and $\widetilde{\Delta}_{g,t}$ of a different sign than $\widetilde{\beta}_{fd}$ for all $(g,t)$, has the same expression as $\underline{\underline{\sigma}}_{fe}$, except that one needs to replace the weights $w_{g,t}$ by the weights $w_{fd,g,t}$ in its definition. Estimators of $\underline{\sigma}_{fe}$ and $\underline{\sigma}_{fd}$ (or $\underline{\underline{\sigma}}_{fe}$ and $\underline{\underline{\sigma}}_{fd}$) can then be used to determine which of $\widehat{\beta}_{fe}$ or $\widehat{\beta}_{fd}$ is more robust to heterogeneous treatment effects.

\medskip
Finally, and similarly to the result shown in Corollary \ref{cor:RH} for $\beta_{fe}$, $\beta_{fd}$ is equal to $\delta^{TR}$ under common trends and the following assumption:
\begin{hyp}\label{hyp:HTE_fd}
($\boldsymbol{w_{fd}}$ uncorrelated with $\boldsymbol{\widetilde{\Delta}}$) $E\left[\sum_{(g,t):D_{g,t}=1}\frac{N_{g,t}}{N_1}(w_{fd,g,t}-1)(\Delta_{g,t}-\Delta^{TR})\right]=0$.
\end{hyp}
Note that under the common trends assumption, one can jointly test Assumption \ref{hyp:HTE_fd} and Assumption \ref{hyp:HTE_fe}, the assumption that the weights attached to $\beta_{fe}$ are uncorrelated with the $\Delta_{g,t}$s: if $\widehat{\beta}_{fe}$ and $\widehat{\beta}_{fd}$ are significantly different, at least one of these two assumptions must fail. In the second application we revisit in Section \ref{sec:appli}, $\widehat{\beta}_{fe}$ and $\widehat{\beta}_{fd}$ are significantly different.

% section two_way_fixed_effects regressions (end)

\section{An alternative estimator}\label{sec:alternative_estimand} % (fold)

In this section, we show that it is possible to estimate a well-defined causal effect even if treatment effects are heterogeneous across groups or over time. Let
\begin{align*}
\delta^S&=E\left[\frac{1}{N_S}\sum_{(i,g,t):t\geq 2,D_{g,t}\ne D_{g,t-1}} \left[Y_{i,g,t}(1)-Y_{i,g,t}(0)\right]\right],
\end{align*}
with $N_S=\sum_{(g,t):t\geq 2,D_{g,t}\ne D_{g,t-1}}N_{g,t}$. $\delta^S$ is the ATE of all switching cells.
In staggered adoption designs, $\delta^S$ is the average of the treatment effect at the time when a group starts receiving the treatment, across all groups that become treated at some point.

\medskip
We now show that $\delta^S$ can be unbiasedly estimated by a weighted average of DID estimators. This result holds under the following supplementary assumptions.
\begin{hyp}\label{hyp:strong_exogeneity2}
	(Strong exogeneity for $Y(1)$) For all $(g,t)\in \{1,...,G\}\times\{2,...,T\}$, \\ $E(Y_{g,t}(1)-Y_{g,t-1}(1)|D_{g,1},...,D_{g,T})=E(Y_{g,t}(1)-Y_{g,t-1}(1))$.
\end{hyp}
Assumption \ref{hyp:strong_exogeneity2} is the equivalent of Assumption \ref{hyp:strong_exogeneity}, for the potential outcome with treatment. It requires that the shocks affecting a group's $Y_{g,t}(1)$ be mean independent of that group's treatment sequence.
\begin{hyp}\label{hyp:common_trends_bis_sharp}
(Common trends for $Y(1)$)
\medskip
For $t\geq 2$, $E(Y_{g,t}(1) - Y_{g,t-1}(1))$ does not vary across $g$.
\end{hyp}

Again, Assumption \ref{hyp:common_trends_bis_sharp} is the equivalent of Assumption \ref{hyp:common_trends}, for the potential outcome with treatment. It requires that between each pair of consecutive periods, the expectation of the outcome with treatment follow the same evolution over time in every group. Assumptions \ref{hyp:strong_exogeneity2} and \ref{hyp:common_trends_bis_sharp} ensure that one can reconstruct the potential outcome that groups leaving the treament between $t-1$ and $t$ would have experienced if they had remained treated. In staggered adoption designs, Assumption \ref{hyp:strong_exogeneity2} and \ref{hyp:common_trends_bis_sharp} are not necessary for identification, because no group leaves the treatment. Together, Assumptions \ref{hyp:common_trends} and \ref{hyp:common_trends_bis_sharp} imply that the ATE follows the same evolution over time in every group: $E\left(\Delta_{g,t}\right)=\eta_t+\theta_g$.\footnote{\label{foot:DE} It should be possible to weaken Assumptions \ref{hyp:strong_exogeneity2}-\ref{hyp:common_trends_bis_sharp}, in particular to account for dynamic effects where $\Delta_{g,t}$ may depend on $(D_{g,1},...,D_{g,t-1})$. This would however introduce complications that are beyond the scope of this paper.} This still allows for heterogeneous treatment effects across groups and over time.\footnote{Imposing Assumptions \ref{hyp:strong_exogeneity2} and \ref{hyp:common_trends_bis_sharp} does not change the decompositions obtained in Theorems \ref{thm:main_sharp} and \ref{thm:main_fd}. $Y_{g,t}(1)$ is observed for all the treated $(g,t)$ cells entering these decompositions, so those assumptions do not bring identifying information for those cells.}

\begin{hyp} (Existence of ``stable'' groups) For all $t\in\{2,...,T\}$:
\begin{enumerate}\label{hyp:existence_stable_sharp}
\item If there is at least one $g\in \{1,...,G\}$ such that $D_{g,t-1}=0$ and $D_{g,t}=1$, then there exists at least one $g'\ne g, g'\in \{1,...,G\}$ such that $D_{g',t-1}=D_{g',t}=0$.
\item If there is at least one $g\in \{1,...,G\}$ such that $D_{g,t-1}=1,D_{g,t}=0$, then there exists at least one $g'\ne g, g'\in \{1,...,G\}$ such that $D_{g',t-1}=D_{g',t}=1$.
    \end{enumerate}
\end{hyp}
The first point of the stable groups assumption requires that between each pair of consecutive time periods, if there is a ``joiner'' (i.e., a group switching from being untreated to treated), then there should be another group that is untreated at both dates. The second point requires that between each pair of consecutive time periods, if there is a ``leaver'' (i.e., a group switching from being treated to untreated), then there should be another group that is treated at both dates.

\medskip
Notice that under Assumption \ref{hyp:existence_stable_sharp}, groups' treatments are not independent, so Assumption \ref{hyp:independent_groups} cannot hold. Accordingly, we replace Assumption \ref{hyp:independent_groups} by Assumption \ref{hyp:independent_groups2} below. Assumption \ref{hyp:independent_groups2} requires that conditional on its own treatments, a group's outcomes be mean independent of the other groups' treatments. It is weaker than Assumption \ref{hyp:independent_groups}. Assumption \ref{hyp:existence_stable_sharp} is necessary to show that our estimator is unbiased, but it is not necessary to show that it is consistent. Accordingly, in Section \ref{sec:inference} of the Web Appendix, we show that our estimator is consistent under Assumption \ref{hyp:independent_groups}. For every $g\in \{1,...,G\}$, let $\bm{D}_g=(D_{1,g},...,D_{T,g})$.
\begin{hyp}\label{hyp:independent_groups2} (Mean independence between a group's outcome and other groups treatments) For all $g$ and $t$, $E(Y_{g,t}(0)|\bm{D})=E(Y_{g,t}(0)|\bm{D}_g)$ and $E(Y_{g,t}(1)|\bm{D})=E(Y_{g,t}(1)|\bm{D}_g)$.
\end{hyp}
We can now define our estimator. For all $t\in\{2,...,T\}$ and for all $(d,d')\in \{0,1\}^2$, let
\begin{equation}
N_{d,d',t}=\sum_{g:D_{g,t}=d,D_{g,t-1}=d'}N_{g,t}
	\label{eq:N_ddt}
\end{equation}
denote the number of observations
with treatment $d'$ at period $t-1$ and $d$ at period $t$. Let
\begin{align*}
\DID_{+,t}&=\sum_{g:D_{g,t}=1,D_{g,t-1}=0}\frac{N_{g,t}}{N_{1,0,t}}\left(Y_{g,t}-Y_{g,t-1}\right)-\sum_{g:D_{g,t}=D_{g,t-1}=0}\frac{N_{g,t}}{N_{0,0,t}}\left(Y_{g,t}-Y_{g,t-1}\right),\\
\DID_{-,t}&=\sum_{g:D_{g,t}=D_{g,t-1}=1}\frac{N_{g,t}}{N_{1,1,t}}\left(Y_{g,t}-Y_{g,t-1}\right)-\sum_{g:D_{g,t}=0,D_{g,t-1}=1}\frac{N_{g,t}}{N_{0,1,t}}\left(Y_{g,t}-Y_{g,t-1}\right).
\end{align*}
Note that $\DID_{+,t}$ is not defined when there is no group such that $D_{g,t}=1,D_{g,t-1}=0$, or no group such that $D_{g,t}=0,D_{g,t-1}=0$. In such instances, we let $\DID_{+,t}=0$. Similarly, let $\DID_{-,t}=0$ when there is no group such that $D_{g,t}=1,D_{g,t-1}=1$ or no group such that $D_{g,t}=0,D_{g,t-1}=1$. Finally, let
$$\DIDM = \sum_{t=2}^{T}\left(\frac{N_{1,0,t}}{N_S}\DID_{+,t}+\frac{N_{0,1,t}}{N_S}\DID_{-,t}\right).$$

\begin{thm}\label{thm:alternative}
If Assumptions \ref{hyp:supp_gt}, \ref{hyp:sharp}, \ref{hyp:strong_exogeneity}, \ref{hyp:common_trends}, and \ref{hyp:strong_exogeneity2}-\ref{hyp:independent_groups2} hold, $E\left[\DIDM\right]=\delta^S$.
\end{thm}
In Section \ref{sec:inference} of the Web Appendix, we also show that when $G$ goes to infinity, $\DIDM$ is a consistent and asymptotically normal estimator of $\delta^S$. The $\DIDM$ estimator is computed by the \st{fuzzydid} and \st{did\_multiplegt} Stata packages.

\medskip
Here is the intuition underlying Theorem \ref{thm:alternative}. $\DID_{+,t}$ compares the evolution of the mean outcome between $t-1$ and $t$ in two sets of groups: the joiners, and those remaining untreated. Under Assumptions \ref{hyp:strong_exogeneity} and \ref{hyp:common_trends}, $\DID_{+,t}$ estimates the joiners' treatment effect. Similarly, $\DID_{-,t}$ compares the evolution of the outcome between $t-1$ and $t$ in two sets of groups: those remaining treated, and the leavers. Under Assumptions \ref{hyp:strong_exogeneity2} and \ref{hyp:common_trends_bis_sharp}, it estimates the leavers' treatment effect. Finally, $\DIDM$ is a weighted average of those DIDs. Note that in staggered designs, there are no groups whose treatment decreases over time, so $\DIDM$ is only a weighted average of the $\DID_{+,t}$ estimators. Note also that one can separately estimate the joiners' and the leavers' treatment effect, by computing separately weighted averages of the $\DID_{+,t}$ and $\DID_{-,t}$ estimators. The former estimator only relies on Assumptions \ref{hyp:strong_exogeneity} and \ref{hyp:common_trends}, while the latter only relies on Assumptions \ref{hyp:strong_exogeneity2} and \ref{hyp:common_trends_bis_sharp}.

\medskip
$\DIDM$ is related to two other estimators. First, it is related to the Wald-TC estimator in point 2 of Theorem S1 in the Web Appendix of \cite{deChaisemartin15b}, but the weighting of $\DID_{+,t}$ and $\DID_{-,t}$ therein differs. As a result, $\DIDM$ estimates $\Delta^S$ under weaker assumptions. $\DIDM$ is also related to the multi-period DID estimator in \cite{imai2018}. However, the multi-period DID estimator is a weighted average of the $\DID_{+,t}$, so it does not estimate the leavers' treatment effect, and applies to a smaller population. Besides, \cite{imai2018} do not establish the properties of their estimator. Finally, they do not generalize it to non-binary treatments, something we do in Section \ref{sub:TC_non_binary} of the Web Appendix.

\medskip
There may be a bias-variance trade-off between $\DIDM$ and the two-way fixed effects regression estimators. For instance, assume that Regression \ref{reg1} is correctly specified:
\begin{eqnarray*}
&&Y_{g,t}(0)=\alpha_g+\lambda_t+\eps_{g,t},\nonumber\\
&&Y_{g,t}(1)-Y_{g,t}(0)=\delta,\nonumber\\
&&E(\eps_{g,t}|\bm{D})=0. \label{eq:linear_model_paper}
\end{eqnarray*}
Then, if the errors $\eps_{g,t}$ are homoskedastic and uncorrelated, it follows from the Gauss-Markov theorem that $\widehat{\beta}_{fe}$ is the linear estimator of $\delta$, the constant treatment effect parameter, with the lowest variance. As $\DIDM$ is also an unbiased linear estimator of $\delta$, the variance of  $\widehat{\beta}_{fe}$ must be lower than that of $\DIDM$. With heteroskedastic or correlated errors, one can construct examples where the variance of $\widehat{\beta}_{fe}$ is higher than that of $\DIDM$, but this still suggests that $\DIDM$ may often have a larger variance than that of $\widehat{\beta}_{fe}$, as we find in our applications in Section \ref{sec:appli}.

\medskip
$\DIDM$ uses groups whose treatment is stable to infer the trends that would have affected switchers if their treatment had not changed. This strategy could fail, if switchers experience different trends than groups whose treatment is stable. To assess if this is a serious concern, we propose to use the following placebo estimator, that essentially compares the outcome's evolution from $t-2$ to $t-1$, in groups that switch and do not switch treatment between $t-1$ and $t$. This placebo estimator is defined under a modified version of Assumption \ref{hyp:existence_stable_sharp}.
\begin{hyp} (Existence of ``stable'' groups for the placebo test) For all $t\in\{3,...,T\}$:
\begin{enumerate}
\item If there is at least one $g\in \{1,...,G\}$ such that $D_{g,t-2}=D_{g,t-1}=0$ and $D_{g,t}=1$, then there exists at least one $g'\ne g, g'\in \{1,...,G\}$ such that $D_{g',t-2}=D_{g',t-1}=D_{g',t}=0$.
\item If there is at least one $g\in \{1,...,G\}$ such that $D_{g,t-2}=D_{g,t-1}=1,D_{g,t}=0$, then there exists at least one $g'\ne g, g'\in \{1,...,G\}$ such that $D_{g',t-2}=D_{g',t-1}=D_{g',t}=1$.
    \end{enumerate}
\label{hyp:existence_stable_sharp_placebo}
\end{hyp}

For all $t\in\{2,...,T\}$ and for all $(d,d',d'')\in \{0,1\}^3$, let
\begin{align*}
N_{d,d',d'',t}&=\sum_{g:D_{g,t}=d,D_{g,t-1}=d',D_{g,t-2}=d''}N_{g,t}
\end{align*}
denote the number of observations with treatment status $d''$ at period $t-2$, $d'$ at period $t-1$, and $d$ at period $t$. Let
\footnotesize
\begin{align*}
N^{\pl}_{S}&=\sum_{(g,t):t\geq 3,D_{g,t}\ne D_{g,t-1}=D_{g,t-2}}N_{g,t}, \\
\DID^{\pl}_{+,t}&=\sum_{g:D_{g,t}=1,D_{g,t-1}=D_{g,t-2}=0}\frac{N_{g,t}}{N_{1,0,0,t}}\left(Y_{g,t-1}-Y_{g,t-2}\right)-\sum_{g:D_{g,t}=D_{g,t-1}=D_{g,t-2}=0}\frac{N_{g,t}}{N_{0,0,0,t}}\left(Y_{g,t-1}-Y_{g,t-2}\right), \\
\DID^{\pl}_{-,t}&=\sum_{g:D_{g,t}=D_{g,t-1}=D_{g,t-2}=1}\frac{N_{g,t}}{N_{1,1,1,t}}\left(Y_{g,t-1}-Y_{g,t-2}\right)-\sum_{g:D_{g,t}=0,D_{g,t-1}=D_{g,t-2}=1}\frac{N_{g,t}}{N_{0,1,1,t}}\left(Y_{g,t-1}-Y_{g,t-2}\right).
\end{align*}
\normalsize
When there is no group such that $D_{g,t}=1,D_{g,t-1}=D_{g,t-2}=0$ or no group such that $D_{g,t}=D_{g,t-1}=D_{g,t-2}=0$, we let $\DID^{\pl}_{+,t}=0$, and we adopt the same convention for $\DID^{\pl}_{-,t}=0$. Let
\begin{align*}
\DIDM^{\pl} & = \sum_{t=3}^{T}\left(\frac{N_{1,0,0,t}}{N^{\pl}_{S}}\DID^{\pl}_{+,t}+\frac{N_{0,1,1,t}}{N^{\pl}_{S}}\DID^{\pl}_{-,t}\right).
\end{align*}
\begin{thm}\label{thm:alternative_placebo}
If Assumptions \ref{hyp:supp_gt}, \ref{hyp:sharp}, \ref{hyp:strong_exogeneity}, \ref{hyp:common_trends}, \ref{hyp:strong_exogeneity2}, \ref{hyp:common_trends_bis_sharp}, \ref{hyp:independent_groups2}, and \ref{hyp:existence_stable_sharp_placebo} hold, then $E\left[\DIDM^{\pl}\right] =0$.
\end{thm}
$\DID^{\pl}_{+,t}$ compares the evolution of the mean outcome from $t-2$ to $t-1$ in two sets of groups: those  untreated at $t-2$ and $t-1$ but treated at $t$, and those untreated at $t-2$, $t-1$, and $t$. If Assumptions \ref{hyp:strong_exogeneity} and \ref{hyp:common_trends} hold, then $E\left[\DID^{\pl}_{+,t}\right]=0$. Similarly, if Assumptions \ref{hyp:strong_exogeneity2} and \ref{hyp:common_trends_bis_sharp} hold, $E\left[\DID^{\pl}_{-,t}\right]=0$. Then, $E\left[\DIDM^{\pl}\right]=0$ is a testable implication of Assumptions \ref{hyp:strong_exogeneity}, \ref{hyp:common_trends}, \ref{hyp:strong_exogeneity2}, and \ref{hyp:common_trends_bis_sharp}, so finding $\DIDM^{\pl}$ significantly different from 0 would imply that those assumptions are violated: groups that switch treatment experience different trends before that switch than the groups used to reconstruct their counterfactual trends when they switch.\footnote{See also \cite{callaway2018}, who propose another placebo test in staggered adoption designs.} Note that $\DIDM^{\pl}$ compares the trends of switching and stable groups one period before the switch. One can define other placebo estimators comparing those trends, say, two or three periods before the switch. $\DIDM^{\pl}$ and all those other placebo estimators are computed by the \st{did\_multiplegt} Stata package.

% section an_alternative_estimator (end)

\section{Extensions}\label{sec:extensions}

In this section, we briefly review some of the extensions in our Web Appendix. First, we show that the decomposition of $\beta_{fe}$ in Theorem \ref{thm:main_sharp} can be extended to fuzzy designs where the treatment varies within $(g,t)$ cells, to applications with a non binary treatment, and to two-way fixed effects regressions with control variables.\footnote{The decomposition of $\beta_{fd}$ in Theorem \ref{thm:main_fd} can also be extended to all of those cases.} In fuzzy designs or with a non-binary treatment, the weights in Theorem \ref{thm:main_sharp} remain essentially unchanged.

%\medskip
%We also consider the inclusion of covariates $X_{g,t}$ in the regression. Specifically, we study  the coefficient of $D_{g,t}$ in a regression of $Y_{i,g,t}$ on group and period fixed effects, $D_{g,t}$, and $X_{g,t}$. The assumptions have to be adapted to this case. In particular, the strong exogeneity and common trend conditions now write
%$$E\left(Y_{g,t}(0)\middle|\bm{D}_g, \bm{X}_g\right) - E\left(Y_{g,t-1}(0)\middle|\bm{D}_g, \bm{X}_g\right)=(X_{g,t}-X_{g,t-1})'\gamma+\lambda_t,$$
%for some vector $\gamma$ and constant $\lambda_t$, and where we let $\bm{X}_g=(X_{g,1},...,X_{g,T})$. Hence, including covariates allows for different trends across groups, but it requires that those differential evolutions are fully accounted for by a linear model in $X_{g,t}-X_{g,t-1}$, the change in a group's covariates. In such a setting, a similar reasoning as that used to derive Theorem \ref{thm:main_sharp} applies. We then obtain the same decomposition, except that in the weights, the residual $\eps_{g,t}$ is replaced by $\eps^X_{g,t}$, the residual of observations in cell $(g,t)$ in the regression of $D_{g,t}$ on group and period fixed effects and $X_{g,t}$.

\medskip
We also consider two-way fixed effects regressions with covariates. Specifically, we study  the coefficient of $D_{g,t}$ in a regression of $Y_{i,g,t}$ on group and period fixed effects, $D_{g,t}$, and a vector of covariates $X_{g,t}$. We show that a result very similar to Theorem \ref{thm:main_sharp} applies to that coefficient, up to two differences. First, including covariates allows for different trends across groups, provided those differential trends are fully accounted for by a linear model in $X_{g,t}-X_{g,t-1}$, the change in a group's covariates. Specifically, instead of Assumptions \ref{hyp:strong_exogeneity} and \ref{hyp:common_trends}, one needs to assume that
$$E\left(Y_{g,t}(0)\middle|\bm{D}_g, \bm{X}_g\right) - E\left(Y_{g,t-1}(0)\middle|\bm{D}_g, \bm{X}_g\right)=(X_{g,t}-X_{g,t-1})'\gamma+\lambda_t,$$
for some vector $\gamma$ and constant $\lambda_t$, and where $\bm{X}_g=(X_{g,1},...,X_{g,T})$. Importantly, when the covariates are group-specific linear trends, the equation above is equivalent to
$$E\left(Y_{g,t}(0)\middle|\bm{D}_g, \bm{X}_g\right) - E\left(Y_{g,t-1}(0)\middle|\bm{D}_g, \bm{X}_g\right)=\gamma_g+\lambda_t,$$
meaning that from $t-1$ to $t$, the evolution of $Y(0)$ in group $g$ should deviate from its group-specific linear trend $\gamma_g$ by an amount $\lambda_t$ common to all groups. Second, the residual $\eps_{g,t}$ in the weights in Theorem \ref{thm:main_sharp} has to be replaced by $\eps^X_{g,t}$, the residual of observations in cell $(g,t)$ in the regression of $D_{g,t}$ on group and period fixed effects and $X_{g,t}$. Some of the corresponding weights may still be negative, as in Theorem \ref{thm:main_sharp}. Overall, two-way fixed effects regressions with covariates may rely on a more plausible common trends assumptions than those without covariates, but they still require that the treatment effect be homogeneous, across time and between groups.

\medskip
Third, we show that under the common trends assumption and the assumption that the ATE of a $(g,t)$ cell does not change over time, $\beta_{fe}$ and $\beta_{fd}$ identify weighted sums of the ATEs of the $(g,t)$ cells whose treatment changes between $t-1$ and $t$. In sharp designs, the weights attached to $\beta_{fd}$ are all positive, while for $\beta_{fe}$, the same only holds in staggered adoption designs.

\medskip
Fourth, we show that our $\DIDM$ estimator can easily be extended to non-binary, discrete treatments. Then, we define it as a weighted average of DIDs comparing the evolution of the outcome in groups whose treatment went from $d$ to $d'$ between $t-1$ and $t$ and in groups with a treatment of $d$ at both dates, across all possible values of $d$, $d'$, and $t$.

\medskip
Finally, our \st{twowayfeweights}, \st{fuzzydid}, and \st{did\_multiplegt} Stata packages can handle all of those extensions.

\section{Applicability, and applications}\label{sec:appli}

\subsection{Applicability}\label{sub:litreview}

We conducted a review of all papers published in the American Economic Review (AER) between 2010 and 2012 to assess the importance of two-way fixed effects regressions in economics. Over these three years, the AER published 337 papers. Out of these 337 papers, 33 or 9.8\% of them estimate the FE or FD Regression, or other regressions resembling closely those regressions.
When one withdraws from the denominator theory papers and lab experiments, the proportion of papers using these regressions raises to 19.1\%.

\begin{table}[H]
\begin{center}
\caption{Papers using two-way fixed effects regressions published in the AER}
\begin{tabular}{l c c c c}
\hline \hline
 & 2010 & 2011 & 2012 & Total\\
Papers using two-way fixed effects regressions & 5 & 14 & 14 & 33  \\
\% of published papers  & 5.2\% & 	12.2\% & 11.2\% &	9.8\% \\
\% of empirical papers, excluding lab experiments & 12.8\% &	23.0\% &	19.2\% & 19.1\%\\
\hline \hline
\end{tabular}\label{table_lit_review}
\end{center}
\footnotesize{\textit{Notes}. This table reports the number of papers using two-way fixed effects regressions published in the AER from 2010 to 2012.}
\end{table}

Table \ref{table_lit_review2} shows descriptive statistics about the 33 2010-2012 AER papers estimating two-way fixed effects regressions. Panel A shows that 13 use the FE regression; six use the FD regression; six use regressions the FE or FD regression with several treatment variables; three use the FE or FD 2SLS regression discussed in Section \ref{sub:2sls} of the Web Appendix; five use other regressions that we deemed sufficiently close to the FE or FD regression to include them in our count.\footnote{For instance, two papers use regressions with three-way fixed-effects instead of two-way fixed effects.} Panel B shows that more than three fourths of those papers consider sharp designs, while less than one fourth consider fuzzy designs. Finally, Panel C assesses whether, in those applications, there are groups whose exposure to the treatment remains stable between each pair of consecutive time periods, the condition that has to be met to be able to compute the $\DIDM$ estimator. For about a half of the papers, reading the paper was not enough to assess this with certainty. We then assessed whether they presumably have stable groups or not. Overall, 12 papers have stable groups, 14 presumably have stable groups, five presumably do not have stable groups, and two do not have stable groups.

\medskip
In Section \ref{sec_appendix_litreview} of the Web Appendix, we review each of the 33 papers. We explain where two-way fixed effects regressions are used in the paper, and we detail our assessment of whether the design is a sharp or a fuzzy design, and of whether the stable groups assumption holds or not.

\begin{table}[H]
\begin{center}
\caption{Descriptive statistics on two-way fixed effects papers}
\begin{tabular}{l c }
\hline \hline
 & \# Papers \\
\textit{Panel A. Estimation method} & \\
Fixed-effects OLS regression & 13 \\
First-difference OLS regression  & 6 \\
Fixed-effects or first-difference OLS regression, with several treatment variables  & 6 \\
Fixed-effects or first-difference 2LS regression & 3 \\
Other regression & 5 \\
\\
\textit{Panel B. Research design} & \\
Sharp design & 26 \\
Fuzzy design  & 7 \\
\\
\textit{Panel C. Are there stable groups?} & \\
Yes & 12 \\
Presumably yes & 14 \\
Presumably no & 5 \\
No & 2 \\
\hline \hline
\end{tabular}\label{table_lit_review2}
\end{center}
\footnotesize{\textit{Notes}. This table reports the estimation method and the research design used in the 33 papers using two-way fixed effects regressions published in the AER from 2010 to 2012, and whether those papers have stable groups.}
\end{table}

%\section{Applications}\label{sec:applications}

\subsection{Application to \cite{gentzkow2011}}

\cite{gentzkow2011} study the effect of newspapers on voters' turnout in US presidential elections between 1868 and 1928. They regress the first-difference of the turnout rate in county $g$ between election years $t-1$ and $t$ on state-year fixed effects and on the first difference of the number of newspapers available in that county. This corresponds to Regression \ref{reg2}, with state-year fixed effects as controls. As reproduced in Table \ref{table_Gentzkow} below, \cite{gentzkow2011} find that $\widehat{\beta}_{fd}=0.0026$ (s.e.= $9\times 10^{-4}$). According to this regression, one more newspaper increased voters' turnout by 0.26 percentage points. On the other hand, $\widehat{\beta}_{fe}=-0.0011$ (s.e.= $0.0011$). $\widehat{\beta}_{fe}$ and $\widehat{\beta}_{fd}$ are significantly different (t-stat=2.86).

\medskip
We use the \st{twowayfeweights} Stata package, downloadable with its help file from the SSC repository, to estimate the weights attached to $\widehat{\beta}_{fe}$. 6,212  are strictly positive, 4,161 are strictly negative. The negative weights sum to -0.53. $\widehat{\underline{\sigma}}_{fe}=3\times 10^{-4}$, meaning that $\beta_{fe}$ and the ATT may be of opposite signs if the standard deviation of the ATEs across all the treated $(g,t)$ cells is equal to $0.0003$.\footnote{The number of newspapers is not binary, so strictly speaking, in this application the parameter of interest is the average causal response parameter introduced in Section \ref{sub:non_binary_ordered_treatment} of our Web Appendix, rather than the ATT.}  $\widehat{\underline{\underline{\sigma}}}_{fe}=7\times 10^{-4}$, meaning that $\beta_{fe}$ may be of a different sign than the ATEs of all the treated $(g,t)$ cells if the standard deviation of those ATEs is equal to $0.0007$. We also estimate the weights attached to $\widehat{\beta}_{fd}$. 5,472 are strictly positive, and 4,605  are strictly negative. The negative weights sum to -1.43. $\widehat{\underline{\sigma}}_{fd}=4\times 10^{-4}$, and $\widehat{\underline{\underline{\sigma}}}_{fd}=6\times 10^{-4}$.

\medskip
Therefore, $\beta_{fe}$ and $\beta_{fd}$ can only receive a causal interpretation if the weights attached to them are uncorrelated with the intensity of the treatment effect in each county$\times$election-year cell (Assumptions \ref{hyp:HTE_fe} and \ref{hyp:HTE_fd}, respectively). This is not warranted. First, as $\widehat{\beta}_{fe}$ and $\widehat{\beta}_{fd}$ significantly differ, Assumptions \ref{hyp:HTE_fe} and \ref{hyp:HTE_fd} cannot jointly hold. Moreover, the weights attached to $\widehat{\beta}_{fe}$ and $\widehat{\beta}_{fd}$ are correlated with variables that are likely to be themselves associated with the intensity of the treatment effect in each cell. For instance, the correlation between the weights attached to $\widehat{\beta}_{fd}$ and $t$, the year variable, is equal to $-0.06$ (t-stat=-3.28). The effect of newspapers may be different in the last than in the first years of the panel. For instance, new means of communication, like the radio, appear in the end of the period under consideration, and may diminish the effect of newspapers.\footnote{\label{foot:radio_Gentzkow} In fact, \cite{gentzkow2011} analyze the 1868 to 1928 period separately from later periods, because the growth of the radio may have changed newspapers' effects.} This would lead to a violation of Assumption \ref{hyp:HTE_fd}.

\medskip
The stable groups assumption holds: between each pair of consecutive elections, there are counties where the number of newspapers does not change.
We use the \st{fuzzydid} Stata package, downloadable with its help file from the SSC repository, to estimate a modified version of our $\DIDM$ estimator, that accounts for the fact that the number of newspapers is not binary (see section \ref{sub:non_binary_ordered_treatment} of our Web Appendix, where we define this modified estimator). We include state-year fixed effects as controls in our estimation. We find that $\DIDM=0.0043$, with a standard error of $0.0015$. $\DIDM$ is 66\% larger than $\widehat{\beta}_{fd}$, and the two estimators are significantly different at the 10\% level (t-stat=1.69). $\DIDM$ is also of a different sign than $\widehat{\beta}_{fe}$.

\medskip
Our $\DIDM$ estimator  only relies on a common trends assumption. To assess its plausibility, we compute $\DIDM^{\pl}$, the placebo estimator introduced in Section \ref{sec:alternative_estimand}.\footnote{Again, we need to slightly modify $\DIDM^{\pl}$ to account for the fact that the number of newspapers is not binary.} As shown in Table \ref{table_Gentzkow} below, our placebo estimator is small and not significantly different from 0, meaning that counties where the number of newspapers increased or decreased between $t-1$ and $t$ did not experience significantly different trends in turnout from $t-2$ to $t-1$ than counties where that number was stable. Our placebo estimator is estimated on a subset of the data: for each pair of consecutive time periods $t-1$ and $t$, we only keep counties where the number of newspapers did not change between $t-2$ and $t-1$. Still, almost 80\% of the county $\times$ election-year observations are used in the computation of the placebo estimator. Moreover, when reestimated on this subsample, the $\DIDM$ estimator is very close to the $\DIDM$ estimator in the full sample.
\begin{table}[H]
\begin{center}
\caption{Estimates of the effect of one additional newspaper on turnout}
\begin{tabular}{l c c c}
\hline \hline
 & Estimate & Standard error & N \\
$\widehat{\beta}_{fd}$ & 0.0026 & 0.0009 & 15,627 \\
$\widehat{\beta}_{fe}$  & -0.0011  & 0.0011 & 16,872 \\
$\DIDM$ & 0.0043 & 0.0015 & 16,872 \\
$\DIDM^{\pl}$ & -0.0009 & 0.0016 & 13,221 \\
$\DIDM$, on placebo subsample & 0.0045 & 0.0019 & 13,221 \\
\hline \hline
\end{tabular}\label{table_Gentzkow}
\end{center}
\footnotesize{\textit{Notes}. This table reports estimates of the effect of one additional newspaper on turnout, as well as a placebo estimate of the common trends assumption underlying $\DIDM$. Estimators are computed using the data of \cite{gentzkow2011}, with state-year fixed effects as controls. Standard errors are clustered by county. To compute the $\DIDM$ estimators, the number of newspapers is grouped into 4 categories: 0, 1, 2, and more than 3.}
\end{table}

\subsection{The effect of union membership on wages}

A number of articles have estimated the effect of union membership on wages using panel data and controlling for workers' fixed effects. For instance, \cite{jakubson1991} has found a 8.3\% union membership premium using that strategy, in a sample of American males from the PSID followed from 1976 to 1980. \cite{vella1998whose} estimate a similar regression and find similar results, in a sample of young American males from the NLSY followed from 1980 to 1987.\footnote{The fixed effects regression is not the main specification in \cite{vella1998whose}. The authors favor instead a dynamic selection model.}

\medskip
We use the data in \cite{vella1998whose} to compute various estimators of the union wage premium. As union status is often measured with error \citep[see, e.g.][]{freeman1984,card1996effect}, we discard changes in union status happening twice in three consecutive years. Specifically, for individuals with $D_{i,t-1}=0$, $D_{i,t}=1$, and $D_{i,t+1}=0$, we replace $D_{i,t}$ by 0. Similarly, for individuals with $D_{i,t-1}=1$, $D_{i,t}=0$, and $D_{i,t+1}=1$, we replace $D_{i,t}$ by 1. Doing so, we discard half of the union status changes in the initial data.\footnote{\label{foot:rob_union_appli} Keeping the original data does not change much the results presented below, except that the placebo estimator $\DIDM^{\pl,2}$ becomes significant.}

\medskip
We start by estimating a two-way fixed effects regression of wages on union membership with worker and year fixed effects. Table \ref{table_union} below shows that $\widehat{\beta}_{fe}=0.107$ (s.e.= $0.030$), a result close to that of the worker fixed effects regressions in \cite{jakubson1991} and \cite{vella1998whose}.

\medskip
Then, we estimate the weights attached to $\widehat{\beta}_{fe}$. 820 are strictly positive, 196 are strictly negative, but the negative weights only sum to -0.01. Still, $\widehat{\underline{\sigma}}_{fe}=0.097$, meaning that $\beta_{fe}$ and the ATT may be of opposite signs if the standard deviation of the treatment effect across the unionized worker $\times$ year observations is equal to $0.097$, a substantial but still possible amount of heterogeneity. The weights are negatively correlated with workers' years of schooling (correlation =$-0.12$, t-stat =$-1.88$). The union premium may be lower for more educated workers \citep[see][]{freeman1984unions}, as they may be less substitutable than less educated ones. Then, $\widehat{\beta}_{fe}$ may overestimate $\delta^{TR}$, the average union premium across all unionized worker $\times$ year observations. We also find that $\widehat{\beta}_{fd}=0.060$ (s.e.= $0.032$) and that $\widehat{\beta}_{fe}$ and $\widehat{\beta}_{fd}$ significantly differ (t-stat=1.91),\footnote{The standard error of $\widehat{\beta}_{fe}-\widehat{\beta}_{fd}$ is computed with a worker-level clustered bootstrap.} thus casting further doubt on Assumptions \ref{hyp:HTE_fe} and \ref{hyp:HTE_fd}.

\medskip
The stable groups assumption holds: between each pair of consecutive years, there are workers whose union membership status does not change. We therefore compute our $\DIDM$ estimator. Table \ref{table_union} shows that it is equal to $0.041$ (s.e.$=0.034$). $\DIDM$ is significantly different from $\widehat{\beta}_{fe}$ (t-stat=2.60) and $\widehat{\beta}_{fd}$ (t-stat=2.36).\footnote{The standard errors of $\widehat{\beta}_{fe}-\DIDM$ and $\widehat{\beta}_{fd}-\DIDM$ are computed with a worker-level clustered bootstrap.} As discussed in Section \ref{sec:alternative_estimand}, we can also estimate separately the union premium for workers joining and leaving a union, something that was previously done by \cite{freeman1984}. The joiners' effect estimate is equal to $0.059$ (s.e.$=0.053$), the leavers' effect is equal to $0.021$ (s.e.$=0.044$), and the two estimates do not significantly differ (t-stat$=0.55$).

\medskip
$\DIDM$ relies on a common trends assumption. To assess its plausibility, we compute $\DIDM^{\pl}$, the placebo estimator introduced in Section \ref{sec:alternative_estimand}. $\DIDM^{\pl}$ compares the wage growth of workers changing and not changing their union status one period before that change. We also compute $\DIDM^{\pl,2}$ and $\DIDM^{\pl,3}$, two other placebo estimators performing the same comparison two and three periods before the change. As shown in Table \ref{table_union} below, $\DIDM^{\pl}$ is large, positive, and significant (t-stat=2.49). On the other hand $\DIDM^{\pl,2}$ and $\DIDM^{\pl,3}$ are smaller and insignificant. Workers that become unionized start experiencing a differential positive pre-trend one year before becoming unionized. This differential pre-trend mostly comes from union joiners: for them, the placebo estimator is equal to $0.119$ (s.e.$=0.051$), while for union leavers the placebo is smaller ($0.061$) and insignificant (s.e.$=0.057$). Therefore, the placebos suggest that even the already small and insignificant $\DIDM$ estimator may overestimate the union premium, due to a positive pre-trend. In fact, the estimate of leavers' effect, for which there is no evidence of a pre-trend, is very close to 0. Overall, our results indicate that there may not be a significant union wage premium.
\begin{table}[H]
\begin{center}
\caption{Estimates of the union premium}
\begin{tabular}{l c c c}
\hline \hline
 & Estimate & Standard error & N \\
$\widehat{\beta}_{fe}$  & 0.107  & 0.030 & 4,360 \\
$\widehat{\beta}_{fd}$ & 0.060 & 0.032 & 3,815 \\
$\DIDM$ & 0.041 & 0.034 & 3,815 \\
$\DIDM^{\pl}$ & 0.094 & 0.038 & 3,101 \\
$\DIDM^{\pl,2}$ & -0.041 & 0.030 & 2,458 \\
$\DIDM^{\pl,3}$ & -0.004 & 0.033 & 1,881 \\
\hline \hline
\end{tabular}\label{table_union}
\end{center}
\footnotesize{\textit{Notes}. This table reports estimates of the effect of the union premium, as well as placebo estimators of the common trends assumption. Estimators are computed using the data of \cite{vella1998whose}. Standard errors are clustered at the worker level.}
\end{table}

\section{Conclusion}\label{sec:conclusion} % (fold)

Almost 20\% of empirical articles published in the AER between 2010 and 2012 use regressions with groups and period fixed effects to estimate treatment effects. In this paper, we show that under a common trends assumption, those regressions estimate weighted sums of the treatment effect in each group and period. The weights may be negative: in one application, we find that almost 50\% of the weights are negative. The negative weights are an issue when the treatment effect is heterogeneous, between groups or over time. Then, one could have that the treatment's coefficient in those regressions is negative while the treatment effect is positive in every group and time period. We therefore propose a new estimator to address this problem. This estimator estimates the treatment effect in the groups that switch treatment, at the time when they switch. It does not rely on any treatment effect homogeneity condition. It is computed by the \st{fuzzydid} and \st{did\_multiplegt} Stata packages. In the two applications we revisit, this estimator is significantly and economically different from the two-way fixed effects estimators.

% section conclusion (end)

\newpage
\bibliography{biblio}

\newpage
\appendix

\section{Proofs}

\subsection*{One useful lemma} % (fold)
\label{sub:additional_notation_and_two_useful_lemmas}

Our results rely on the following lemma.
\begin{lem}\label{lem:IV-DID}
If Assumptions \ref{hyp:supp_gt}-\ref{hyp:common_trends} hold, for all $(g,g',t,t')\in \{1,...,G\}^2\times \{1,...,T\}^2,$
\begin{align*}
&E\left(Y_{g,t}\middle|\bm{D}\right)-E\left(Y_{g,t'}\middle|\bm{D}\right)-\left(E\left(Y_{g',t}\middle|\bm{D}\right)-E\left(Y_{g',t'}\middle|\bm{D}\right)\right)\\
=&D_{g,t}E\left(\Delta_{g,t}\middle|\bm{D}\right)-D_{g,t'}E\left(\Delta_{g,t'}\middle|\bm{D}\right)-\left(D_{g',t}E\left(\Delta_{g',t}\middle|\bm{D}\right)-D_{g',t'}E\left(\Delta_{g',t'}\middle|\bm{D}\right)\right).
\end{align*}
\end{lem}

\subsubsection*{Proof of Lemma \ref{lem:IV-DID}}

For all $(g,t)\in \{1,...,G\}\times \{1,...,T\},$
\begin{align*}
E\left(Y_{g,t}\middle|\bm{D}\right)=&E\left(\frac{1}{N_{g,t}}\sum_{i=1}^{N_{g,t}}Y_{i,g,t}\middle|\bm{D}\right)\\
=&E\left(\frac{1}{N_{g,t}}\sum_{i=1}^{N_{g,t}}\left(Y_{i,g,t}(0)+D_{i,g,t}(Y_{i,g,t}(1)-Y_{i,g,t}(0))\right)\middle|\bm{D}\right)\\
=&E\left(Y_{g,t}(0)\middle|\bm{D}\right)+D_{g,t}E\left(\Delta_{g,t}\middle|\bm{D}\right)\\
=&E\left(Y_{g,t}(0)\middle|\bm{D}_g\right)+D_{g,t}E\left(\Delta_{g,t}\middle|\bm{D}\right),
\end{align*}
where the third equality follows from Assumption \ref{hyp:sharp}, and the fourth from Assumption \ref{hyp:independent_groups}.
Therefore,
\begin{align*}
&E\left(Y_{g,t}\middle|\bm{D}\right)-E\left(Y_{g,t'}\middle|\bm{D}\right)-\left(E\left(Y_{g',t}\middle|\bm{D}\right)-E\left(Y_{g',t'}\middle|\bm{D}\right)\right)\\
=& E\left(Y_{g,t}(0)-Y_{g,t'}(0)\middle|\bm{D}_g\right)-E\left(Y_{g',t}(0)-Y_{g',t'}(0)\middle|\bm{D}_{g'}\right)\\ +&D_{g,t}E\left(\Delta_{g,t}\middle|\bm{D}\right)-D_{g,t'}E\left(\Delta_{g,t'}\middle|\bm{D}\right)-\left(D_{g',t}E\left(\Delta_{g',t}\middle|\bm{D}\right)-D_{g',t'}E\left(\Delta_{g',t'}\middle|\bm{D}\right)\right)\\
=& E\left(Y_{g,t}(0)-Y_{g,t'}(0)\right)-E\left(Y_{g',t}(0)-Y_{g',t'}(0)\right)\\ +&D_{g,t}E\left(\Delta_{g,t}\middle|\bm{D}\right)-D_{g,t'}E\left(\Delta_{g,t'}\middle|\bm{D}\right)-\left(D_{g',t}E\left(\Delta_{g',t}\middle|\bm{D}\right)-D_{g',t'}E\left(\Delta_{g',t'}\middle|\bm{D}\right)\right)\\
=&  D_{g,t}E\left(\Delta_{g,t}\middle|\bm{D}\right)-D_{g,t'}E\left(\Delta_{g,t'}\middle|\bm{D}\right)-\left(D_{g',t}E\left(\Delta_{g',t}\middle|\bm{D}\right)-D_{g',t'}E\left(\Delta_{g',t'}\middle|\bm{D}\right)\right),
\end{align*}
where the first equality follows from Assumption \ref{hyp:independent_groups}, the second from the linearity of the conditional expectation and Assumption \ref{hyp:strong_exogeneity}, and the third from Assumption \ref{hyp:common_trends}.

\subsection*{Proof of Theorem \ref{thm:main_sharp}}

It follows from the Frisch-Waugh theorem and the definition of $\eps_{g,t}$ that
\begin{equation}\label{eq:eq1_beta1_sharp}
E\left(\widehat{\beta}_{fe}\middle|\bm{D}\right)=\frac{\sum_{g,t}N_{g,t}\eps_{g,t}E\left(Y_{g,t}\middle|\bm{D}\right)}{\sum_{g,t}N_{g,t}\eps_{g,t}D_{g,t}}.
\end{equation}
Now, by definition of $\eps_{g,t}$ again,
\begin{align}
&\sum_{t=1}^{T}N_{g,t}\eps_{g,t}=0\text{ for all }g\in \{1,...,G\}, \label{eq:residu_centreG}\\
&\sum_{g=1}^{G}N_{g,t}\eps_{g,t}=0\text{ for all }t\in \{1,...,T\}. \label{eq:residu_centreT}
\end{align}
Then,
\begin{align}
& \sum_{g,t}N_{g,t}\eps_{g,t}E\left(Y_{g,t}\middle|\bm{D}\right) \nonumber \\
=&  \sum_{g,t}N_{g,t}\eps_{g,t}\left(E\left(Y_{g,t}\middle|\bm{D}\right)-E\left(Y_{g,1}\middle|\bm{D}\right)-E\left(Y_{1,t}\middle|\bm{D}\right)+E\left(Y_{1,1}\middle|\bm{D}\right)\right) \label{eq:GoodmanBacon} \\
=&  \sum_{g,t}N_{g,t}\eps_{g,t}\left(D_{g,t}E\left(\Delta_{g,t}\middle|\bm{D}\right)-D_{g,1}E\left(\Delta_{g,1}\middle|\bm{D}\right)-D_{1,t}E\left(\Delta_{1,t}\middle|\bm{D}\right)+D_{1,1}E\left(\Delta_{1,1}\middle|\bm{D}\right)\right) \nonumber \\
=&  \sum_{g,t}N_{g,t}\eps_{g,t}D_{g,t}E\left(\Delta_{g,t}\middle|\bm{D}\right) \nonumber \\
=&  \sum_{(g,t):D_{g,t}=1}N_{g,t}\eps_{g,t}E\left(\Delta_{g,t}\middle|\bm{D}\right). \label{eq:num_beta1_sharp}
\end{align}
The first and third equalities follow from Equations \eqref{eq:residu_centreG} and \eqref{eq:residu_centreT}.
The second equality follows from Lemma \ref{lem:IV-DID}.
The fourth equality follows from Assumption \ref{hyp:sharp}.
Finally, Assumption \ref{hyp:sharp} implies that
\begin{equation}
\sum_{g,t}N_{g,t}\eps_{g,t}D_{g,t}=\sum_{(g,t):D_{g,t}=1}N_{g,t}\eps_{g,t}.	
	\label{eq:denom_beta1_sharp}
\end{equation}
Combining \eqref{eq:eq1_beta1_sharp}, \eqref{eq:num_beta1_sharp}, \eqref{eq:denom_beta1_sharp} yields
\begin{equation}\label{eq:thm1_cond}
E\left(\widehat{\beta}_{fe}\middle|\mathbf{D}\right) =  \sum_{(g,t):D_{g,t}=1}\frac{N_{g,t}}{N_1}w_{g,t} E\left(\Delta_{g,t}\middle|\bm{D}\right).
\end{equation}
Then, the result follows from the law of iterated expectations.

\subsection*{Proof of Proposition \ref{prop:staggered_CT_fe}}

If for all $t\geq 2$ $N_{g,t}/N_{g,t-1}$ does not depend on $t$, then it follows from the first order conditions attached to Regression \ref{reg1} and a few lines of algebra that $\eps_{g,t}=D_{g,t}-D_{g,.}-D_{.,t}+D_{.,.}$. Therefore, $w_{g,t}$ is proportional to $D_{g,t}-D_{g,.}-D_{.,t}+D_{.,.}$. Then, for all $(g, t,t')$ such that $D_{g,t}=D_{g,t'}=1$, $D_{.,t}>D_{.,t'}$ implies $w_{g,t}<w_{g,t'}$. Similarly, for all $(g, g',t)$ such that $D_{g,t}=D_{g',t}=1$, $D_{g,.}>D_{g',.}$ implies $w_{g,t}<w_{g',t}$.

\subsection*{Proof of Corollary \ref{cor:sensib}}

\subsubsection*{Proof of the first point}

We start by proving the first point. If the assumptions of the corollary hold and $\widetilde{\Delta}^{TR}=0$, then
$$\left\{\begin{array}{rcl}
	\widetilde{\beta}_{fe}& =&\sum_{(g,t):D_{g,t}=1}\frac{N_{g,t}}{N_1}w_{g,t}\widetilde{\Delta}_{g,t}, \\[2mm]
	0 & = & \sum_{(g,t):D_{g,t}=1}\frac{N_{g,t}}{N_1}\widetilde{\Delta}_{g,t},
\end{array}\right.$$
where the first equality follows from \eqref{eq:thm1_cond}.
These two conditions and the Cauchy-Schwarz inequality imply
$$|\widetilde{\beta}_{fe}|  = \left|\sum_{(g,t):D_{g,t}=1}\frac{N_{g,t}}{N_1}(w_{g,t}-1)(\widetilde{\Delta}_{g,t}-\widetilde{\Delta}^{TR})\right| \leq \sigma(\boldsymbol{w})\sigma(\boldsymbol{\widetilde{\Delta}}).$$
Hence, $\sigma(\boldsymbol{\widetilde{\Delta}})\geq \underline{\sigma}_{fe}$.

\medskip
Now, we prove that we can rationalize this lower bound. Let us define
$$\widetilde{\Delta}^{TR}_{g,t} = \frac{\widetilde{\beta}_{fe}\left(w_{g,t} - 1 \right)}{\sigma^2(\boldsymbol{w})}.$$
Then,
$$\widetilde{\Delta}^{TR}=\sum_{(g,t):D_{g,t}=1}\frac{N_{g,t}}{N_1}\frac{\widetilde{\beta}_{fe}\left(w_{g,t} - 1 \right)}{\sigma^2(\boldsymbol{w})} = \frac{\widetilde{\beta}_{fe}}{\sigma^2(\boldsymbol{w})}\left(\sum_{(g,t):D_{g,t}=1}\frac{N_{g,t}}{N_1}w_{g,t} - 1 \right)= 0,$$
as it follows from the definition of $w_{g,t}$ that $\sum_{(g,t):D_{g,t}=1}\frac{N_{g,t}}{N_1}w_{g,t}=1$.

\medskip
Similarly,
\begin{align*}
\sum_{(g,t):D_{g,t}=1}\frac{N_{g,t}}{N_1}w_{g,t}\frac{\widetilde{\beta}_{fe}\left(w_{g,t} - 1 \right)}{\sigma^2(\boldsymbol{w})}& =\frac{\widetilde{\beta}_{fe}}{\sigma^2(\boldsymbol{w})} \sum_{(g,t):D_{g,t}=1}\frac{N_{g,t}}{N_1}w_{g,t}\left(w_{g,t} - 1\right)\\
 & =\frac{\widetilde{\beta}_{fe}}{\sigma^2(\boldsymbol{w})} \sum_{(g,t):D_{g,t}=1}\frac{N_{g,t}}{N_1}\left(w_{g,t} - 1\right)^2\\
& = \widetilde{\beta}_{fe},
\end{align*}
where the second equality follows again from the fact that $\sum_{(g,t):D_{g,t}=1}\frac{N_{g,t}}{N_1}w_{g,t}=1$.

\subsubsection*{Proof of the second point}

We first suppose that $\widetilde{\beta}_{fe}>0$. We seek to solve:
\begin{align*}
\min_{\widetilde{\Delta}_{(1)},...,\widetilde{\Delta}_{(n)}} \sum_{i=1}^n\frac{N_{(i)}}{N_1}\left(\widetilde{\Delta}_{(i)}-\widetilde{\Delta}^{TR}\right)^2 \quad \text{ s.t. } \widetilde{\beta}_{fe}& =\sum_{i=1}^n\frac{N_{(i)}}{N_1}w_{(i)}\widetilde{\Delta}_{(i)},\\
\widetilde{\Delta}_{(i)}&\leq 0\text{ for all }i\in\{1,...,n\}.
\end{align*}
This is a quadratic programming problem, with a matrix that is symmetric positive but not definite. Hence, by \cite{frank1956} and the fact that the linear term in the quadratic problem is 0, the solution exists if and only if the set of constraints is not empty. If $w_{(n)}\geq 0$, the set of constraints is empty because $\sum_{i=1}^n \frac{N_{(i)}}{N_1} w_{(i)} \widetilde{\Delta}_{(i)} \leq 0 < \widetilde{\beta}_{fe}$. On the other hand, if $w_{(n)}<0$, this set is non-empty since it includes $(0,...,0, \widetilde{\beta}_{fe}/(P_{(n)}w_{(n)}))$.

\medskip
We now derive the corresponding bound. For that purpose, remark that
$$\sum_{i=1}^n \frac{N_{(i)}}{N_1}\left(\widetilde{\Delta}_{(i)} - \sum_{i=1}^n \frac{N_{(i)}}{N_1}\widetilde{\Delta}_{(i)}\right)^2 = \sum_{i=1}^n \frac{N_{(i)}}{N_1}\widetilde{\Delta}_{(i)}^2 - \left(\sum_{i=1}^n \frac{N_{(i)}}{N_1}\widetilde{\Delta}_{(i)}\right)^2.$$
The Karush–Kuhn–Tucker necessary conditions for optimality are that for all $i$:
\begin{align*}
&\widetilde{\Delta}_{(i)} = \widetilde{\Delta}^{TR} + \lambda w_{(i)} - \gamma_{(i)}, \nonumber  \\
&\sum_{i=1}^n \frac{N_{(i)}}{N_1} w_{(i)} \widetilde{\Delta}_{(i)} = \widetilde{\beta}_{fe}, \nonumber \\
&\gamma_{(i)} \geq 0, \nonumber \\
&\gamma_{(i)} \widetilde{\Delta}_{(i)} = 0,
\end{align*}
where $\widetilde{\Delta}^{TR}=\sum_{i=1}^n \frac{N_{(i)}}{N_1} \widetilde{\Delta}_{(i)}$, $2\lambda$ is the Lagrange multiplier of the constraint $\sum_{i=1}^n \frac{N_{(i)}}{N_1} w_{(i)} \widetilde{\Delta}_{(i)}=\widetilde{\beta}_{fe}$ and $2\frac{N_{(i)}}{N_1} \gamma_{(i)}$ is the Lagrange multiplier of the constraint $\widetilde{\Delta}_{(i)} \leq 0$.

\medskip
These constraints imply that $\widetilde{\Delta}_{(i)} = 0$ if and only if $\widetilde{\Delta}^{TR} + \lambda w_{(i)} \geq 0$. Therefore, if $\widetilde{\Delta}^{TR} + \lambda w_{(i)} < 0$, $\widetilde{\Delta}_{(i)} \ne 0$ so $\gamma_{(i)}=0$, and $\widetilde{\Delta}_{(i)} = \widetilde{\Delta}^{TR} + \lambda w_{(i)}$. Therefore,
\begin{align}\label{eq:circular_Deltai}
\widetilde{\Delta}_{(i)}=&\min(\widetilde{\Delta}^{TR} + \lambda w_{(i)},0).
\end{align}
This equation implies that $\widetilde{\Delta}_{(i)} \leq \widetilde{\Delta}^{TR} + \lambda w_{(i)}$, which in turn implies that $\widetilde{\Delta}^{TR} \leq \widetilde{\Delta}^{TR} + \lambda$, so $\lambda \geq 0$.

\medskip
As a result, $\widetilde{\Delta}^{TR} + \lambda w_{(i)}$ is decreasing in $i$, and because $x\mapsto \min(x,0)$ is increasing, $\widetilde{\Delta}_{(i)}$ is also decreasing in $i$. Then $\widetilde{\Delta}_{(n)}<0$: otherwise one would have $\widetilde{\Delta}_{(i)}=0$ for all $i$ which would imply $\widetilde{\beta}_{fe}=0$, a contradiction. Let $s=\min\{i\in\{1,...,n\}:\widetilde{\Delta}_{(i)}<0\}$. Using again \eqref{eq:circular_Deltai}, we get
$$\widetilde{\Delta}^{TR} = \sum_{i\geq s}\frac{N_{(i)}}{N_1}\widetilde{\Delta}_{(i)} = P_s \widetilde{\Delta}^{TR} + \lambda S_s.$$
Therefore,
\begin{align}\label{eq:Delta}
\widetilde{\Delta}^{TR} = \frac{\lambda S_s}{1-P_s}.
\end{align}
Hence, plugging $\widetilde{\Delta}$ in \eqref{eq:circular_Deltai}, we obtain that for all $i\geq s$,
$$\widetilde{\Delta}_{(i)} = \lambda\left\{\frac{S_s}{1-P_s} + w_{(i)}\right\}.$$
Finally, using again \eqref{eq:circular_Deltai}, we obtain
$$\widetilde{\beta}_{fe} =  \sum_{i\geq s}\frac{N_{(i)}}{N_1} w_{(i)} \widetilde{\Delta}_{(i)} =  \lambda\left\{\frac{S^2_s}{1-P_s} + T_s\right\}.$$
Thus,
$$\lambda = \frac{\widetilde{\beta}_{fe}}{T_s + S^2_s/(1-P_s)}.$$
Then,  using what precedes,
\begin{align*}
\underline{\underline{\sigma}}_{fe}^2 = & \sum_{i\geq s} \frac{N_{(i)}}{N_1} \left(\lambda w_{(i)}\right)^2 + \sum_{i<s} \frac{N_{(i)}}{N_1} \left(\widetilde{\Delta}^{TR}\right)^2 \nonumber\\
= & \lambda^2 T_s + (1-P_s) \left( \frac{\lambda S_s}{1-P_s}\right)^2 \nonumber\\
= & \lambda^2\left[T_s + \frac{S^2_s}{1-P_s}\right] \nonumber\\
= & \frac{\widetilde{\beta}_{fe}^2}{T_s + S^2_s/(1-P_s)}.
\end{align*}
The result follows, once noted that Equations \eqref{eq:circular_Deltai} and \eqref{eq:Delta} imply that $s=\min\{i\in\{1,...,n\}:w_{(i)}<-S_{(i)}/(1-P_{(i)})\}$.

\medskip
Finally, consider the case $\widetilde{\beta}_{fe}<0$. By letting $\widetilde{\Delta}_{(i)}'=-\widetilde{\Delta}_{(i)}$ and $\widetilde{\beta}_{fe}'=-\widetilde{\beta}_{fe}$, we have
$$\underline{\underline{\sigma}}_{fe} = \min_{\widetilde{\Delta}'_{(1)}\leq 0,...,\widetilde{\Delta}'_{(n)}\leq 0} \sum_{i=1}^n \frac{N_{(i)}}{N_1} \widetilde{\Delta}'_{(i)}{}^2 - \left(\sum_{i=1}^n \frac{N_{(i)}}{N_1}\widetilde{\Delta}'_{(i)}\right)^2 \; \text{s.t. } \sum_{i=1}^n \frac{N_{(i)}}{N_1} w_{(i)} \widetilde{\Delta}'_{(i)} = \widetilde{\beta}_{fe}'.$$
This is the same program as before, with $\widetilde{\beta}_{fe}'$ instead of $\widetilde{\beta}_{fe}$. Therefore, by the same reasoning as before, we obtain
\begin{align*}
\underline{\underline{\sigma}}_{fe}^2 = \frac{(\widetilde{\beta}_{fe}')^2}{T_s + S^2_s/(1-P_s)}=\frac{\widetilde{\beta}_{fe}^2}{T_s + S^2_s/(1-P_s)}.
\end{align*}

\subsection*{Proof of Corollary \ref{cor:RH}}
We have
\begin{align*}
\beta_{fe}& =E\left(\sum_{(g,t):D_{g,t}=1}\frac{N_{g,t}}{N_1}w_{g,t}\widetilde{\Delta}_{g,t}\right)\\
& =E\left(\left(\sum_{(g,t):D_{g,t}=1}\frac{N_{g,t}}{N_1}w_{g,t}\right)\widetilde{\Delta}^{TR}\right)\\
&=E\left(\widetilde{\Delta}^{TR}\right)\\
& =\delta^{TR}.
\end{align*}
The first equality follows from the law of iterated expectations and \eqref{eq:thm1_cond}. The second equality follows from Assumption \ref{hyp:HTE_fe}. By the definition of $w_{g,t}$, $\sum_{(g,t):D_{g,t}=1}\frac{N_{g,t}}{N_1}w_{g,t}=1$, hence the third equality. The fourth equality follows from the law of iterated expectations.

\subsection*{Proof of Theorem \ref{thm:main_fd}}

It follows from the Frisch-Waugh theorem and the definition of $\eps_{fd,g,t}$ that
\begin{equation}\label{eq:eq1_beta2_sharp}
E\left(\widehat{\beta}_{fd}\middle|\bm{D}\right)=\frac{\sum_{(g,t):t\geq 2}N_{g,t}\eps_{fd,g,t}\left(E\left(Y_{g,t}\middle|\bm{D}\right)-E\left(Y_{g,t-1}\middle|\bm{D}\right)\right)}{\sum_{(g,t):t\geq 2}N_{g,t}\eps_{fd,g,t}\left(D_{g,t}-D_{g,t-1}\right)}.
\end{equation}
Now, by definition of $\eps_{fd,g,t}$ again,
\begin{equation}
\sum_{g=1}^{G}N_{g,t}\eps_{fd,g,t}=0\text{ for all }t\in \{2,...,T\}.
\label{eq:residu_centreT_fd}	
\end{equation}
Then,
\begin{align}\label{eq:eq2_beta2_sharp}
& \sum_{(g,t):t\geq 2}N_{g,t}\eps_{fd,g,t}\left(E\left(Y_{g,t}\middle|\bm{D}\right)-E\left(Y_{g,t-1}\middle|\bm{D}\right)\right) \nonumber\\
=&\sum_{(g,t):t\geq 2}N_{g,t}\eps_{fd,g,t}\left(E\left(Y_{g,t}\middle|\bm{D}\right)-E\left(Y_{g,t-1}\middle|\bm{D}\right)-E\left(Y_{1,t}\middle|\bm{D}\right)-E\left(Y_{1,t-1}\middle|\bm{D}\right)\right) \nonumber\\
=&\sum_{(g,t):t\geq 2}N_{g,t}\eps_{fd,g,t}\left(D_{g,t}\widetilde{\Delta}_{g,t} - D_{g,t-1}\widetilde{\Delta}_{g,t-1} - D_{1,t}\widetilde{\Delta}_{1,t} + D_{1,t-1}\widetilde{\Delta}_{1,t-1}\right) \nonumber\\
=&\sum_{(g,t):t\geq 2}N_{g,t}\eps_{fd,g,t}\left(D_{g,t}\widetilde{\Delta}_{g,t} - D_{g,t-1}\widetilde{\Delta}_{g,t-1}\right) \nonumber\\
=&\sum_{g,t}\left(N_{g,t}\eps_{fd,g,t}-N_{g,t+1}\eps_{fd,g,t+1}\right)D_{g,t}\widetilde{\Delta}_{g,t}\nonumber\\
=&\sum_{(g,t):D_{g,t}=1}N_{g,t}\left(\eps_{fd,g,t}-\frac{N_{g,t+1}}{N_{g,t}}\eps_{fd,g,t+1}\right)\widetilde{\Delta}_{g,t}.
\end{align}
The first and third equalities follow from \eqref{eq:residu_centreT_fd}. The second equality follows
from Lemma \ref{lem:IV-DID}. The fourth equality follows from a summation by part, and from the fact $\eps_{fd,g,1}=\eps_{fd,g,T+1}=0$. The fifth equality follows from Assumption \ref{hyp:sharp}.

\medskip
A similar reasoning yields
\begin{align}\label{eq:eq3_beta2_sharp}
&\sum_{(g,t):t\geq 2}N_{g,t}\eps_{fd,g,t}\left(D_{g,t}-D_{g,t-1}\right)=\sum_{(g,t):D_{g,t}=1}N_{g,t}\left(\eps_{fd,g,t}-\frac{N_{g,t+1}}{N_{g,t}}\eps_{fd,g,t+1}\right).
\end{align}
Combining \eqref{eq:eq1_beta2_sharp}, \eqref{eq:eq2_beta2_sharp}, \eqref{eq:eq3_beta2_sharp}, and the law of iterated expectations yields the result.

\subsection*{Proof of Proposition \ref{prop:staggered_CT_fd}}

It follows from the first order conditions attached to Regression \ref{reg2} and a few lines of algebra that $\eps_{fd,g,t}=D_{g,t}-D_{g,t-1}-D_{.,t}+D_{.,t-1}$. Therefore, under Assumption \ref{hyp:staggered_adoption_design} and if $N_{g,t}$ does not vary across $t$, one has that for all $(g,t)$ such that $D_{g,t}=1, 1\leq t\leq T-1$, $w_{fd,g,t}$ is proportional to $1-D_{g,t-1}-\left(2D_{.,t}-D_{.,t-1}-D_{.,t+1}\right)$. $D_{.,t}-D_{.,t-1}\leq 1$, and under Assumption \ref{hyp:staggered_adoption_design} $D_{.,t}-D_{.,t+1}\leq 0$, so $1-D_{g,t-1}-\left(2D_{.,t}-D_{.,t-1}-D_{.,t+1}\right)$ can only be strictly negative if $D_{g,t-1}=1$. Then, for all $(g,t)$ such that $D_{g,t}=1, 1\leq t\leq T-1$, $w_{fd,g,t}$ is strictly negative if and only if $D_{g,t-1}=1$ and $2D_{.,t}-D_{.,t-1}-D_{.,t+1}>0$.

\medskip
Similarly, when $t=T$, under the same assumptions as above, one has that for all $g$ such that $D_{g,T}=1$, $w_{fd,g,T}$ is proportional to $1-D_{g,T-1}-(D_{.,T}-D_{.,T-1}).$ $D_{.,T}-D_{.,T-1}\leq 1$, so $1-D_{g,T-1}-(D_{.,T}-D_{.,T-1})$ can only be strictly negative if $D_{g,T-1}=1$. Then, $w_{fd,g,T}$ is strictly negative if and only if $D_{g,T-1}=1$ and $D_{.,T}-D_{.,T-1}>0$.

\medskip
Finally, when $t=1$, one has that for all $g$ such that $D_{g,1}=1$, $D_{g,2}=1$ under Assumption \ref{hyp:staggered_adoption_design}, so $w_{fd,g,1}$ is proportional to $D_{.,2}-D_{.,1},$ which is greater than 0 under Assumption \ref{hyp:staggered_adoption_design}.

\subsection*{Proof of Theorem \ref{thm:alternative}} % (fold)
\label{sub:proof_of_theorem_ref_thm_alternative}

First, by definition of $\DIDM$,
\begin{align}\label{eq:Wald_TC_weak_exo}
E\left(\DIDM\right) =&\sum_{t=2}^{T}E\left(\left(\frac{N_{1,0,t}}{N_S}E\left(\DID_{+,t}\middle|\bm{D}\right)+\frac{N_{0,1,t}}{N_S}E\left(\DID_{-,t}\middle|\bm{D}\right)\right)\right).
\end{align}
Let $t$ be greater than 2, and let us focus for now on the case where there is at least one $g_1$ such that $D_{g_1,t-1}=0$ and $D_{g_1,t}=1$. Then Assumption \ref{hyp:existence_stable_sharp} ensures that there is a least another group $g_2$ such that $D_{g_2,t-1}=D_{g_2,t}=0$. For every $g$ such that $D_{g,t-1}=0$ and $D_{g,t}=1$, we have
\begin{align}
	E\left(Y_{g,t}-Y_{g,t-1}\middle|\bm{D}\right) = & E\left(\Delta_{g,t}\middle|\bm{D}\right) + E\left(Y_{g,t}(0)-Y_{g,t-1}(0)\middle|\bm{D}\right). \label{eq:for_TC1}
\end{align}
Under Assumptions \ref{hyp:independent_groups2}, \ref{hyp:strong_exogeneity}, and \ref{hyp:common_trends}, for all $t\geq 2$, there exists a real number $\psi_{0,t}$ such that for all $g$
\begin{align}\label{eq:phi}
E\left(Y_{g,t}(0)-Y_{g,t-1}(0)\middle|\bm{D}\right)=E\left(Y_{g,t}(0)-Y_{g,t-1}(0)\middle|\bm{D}_g\right)=E\left(Y_{g,t}(0)-Y_{g,t-1}(0)\right)=\psi_{0,t}.
\end{align}
Then,
\begin{align}\label{eq:thmWaldTCsharp}
	&N_{1,0,t}E\left(\DID_{+,t}\middle|\bm{D}\right)\nonumber\\
=& \sum_{g:D_{g,t}=1,D_{g,t-1}=0} N_{g,t}E\left(\Delta_{g,t}\middle|\bm{D}\right) + \sum_{g:D_{g,t}=1,D_{g,t-1}=0} N_{g,t}E\left(Y_{g,t}(0)-Y_{g,t-1}(0)\middle|\bm{D}\right) \nonumber \\
	& - \frac{N_{1,0,t}}{N_{0,0,t}}\sum_{g:D_{g,t}=D_{g,t-1}=0} N_{g,t} E\left(Y_{g,t}(0)-Y_{g,t-1}(0)\middle|\bm{D}\right) \nonumber\\
	= & \sum_{g:D_{g,t}=1,D_{g,t-1}=0} N_{g,t}E\left(\Delta_{g,t}\middle|\bm{D}\right) +
	\psi_{0,t}\left(\sum_{g:D_{g,t}=1,D_{g,t-1}=0} N_{g,t}
	- \frac{N_{1,0,t}}{N_{0,0,t}}\sum_{g:D_{g,t}=D_{g,t-1}=0} N_{g,t}\right)\nonumber\\
	= & \sum_{g:D_{g,t}=1,D_{g,t-1}=0} N_{g,t}E\left(\Delta_{g,t}\middle|\bm{D}\right).
\end{align}
The first equality follows by \eqref{eq:for_TC1}, the second by \eqref{eq:phi}, and the third after some algebra. If there is no $g$ such that $D_{g,t-1}=0$ and $D_{g,t}=1$, \eqref{eq:thmWaldTCsharp} still holds, as $\DID_{+,t}=0$ in this case.

\medskip
A similar reasoning yields
\begin{align}\label{eq:DIDminus}
N_{0,1,t}E\left(\DID_{-,t}\middle|\bm{D}\right) = \sum_{g:D_{g,t}=0,D_{g,t-1}=1} N_{g,t} E\left(\Delta_{g,t}\middle|\bm{D}\right)
\end{align}
Plugging \eqref{eq:thmWaldTCsharp} and \eqref{eq:DIDminus} into \eqref{eq:Wald_TC_weak_exo} yields
\begin{align*}
	E(\DIDM) = & \sum_{t=2}^{T}E\left(E\left(\frac{1}{N_S}\left(\sum_{g:D_{g,t}=1,D_{g,t-1}=0} N_{g,t} \Delta_{g,t}+\sum_{g:D_{g,t}=0,D_{g,t-1}=1} N_{g,t} \Delta_{g,t}\right)\middle|\bm{D}\right)\right) \\
	= & \delta^S.
\end{align*}

\subsection*{Proof of Theorem \ref{thm:alternative_placebo}} % (fold)
\label{sub:proof_of_theorem_ref_thm_alternative_placebo}

First, as with $\DIDM$, we have
\small
\begin{align}\label{eq:Wald_TC_placebo_weak_exo}
E\left(\DIDM^{\pl}\right)
=&\sum_{t=3}^{T}E\left(\left(\frac{N_{1,0,0,t}}{N^{\pl}_S}E\left(\DID^{\pl}_{+,t}\middle|\bm{D}\right)+\frac{N_{0,1,1,t}}{N^{\pl}_S}E\left(\DID^{\pl}_{-,t}\middle|\bm{D}\right)\right)\right).
\end{align}
\normalsize
Let $t$ be greater than 3, and let us for now focus on the case where there exists at least one $g_1$ such that $D_{g_1,t-2}=D_{g_1,t-1}=0$ and $D_{g_1,t}=1$. Then Assumption \ref{hyp:existence_stable_sharp_placebo} ensures that there is a least another group $g_2$ such that $D_{g_2,t-2}=D_{g_2,t-1}=D_{g_2,t}=0$. Then,
\begin{align}\label{eq:thmWaldTCsharp_placebo}
&N_{1,0,0,t}E\left(\DID^{\pl}_{+,t}\middle|\bm{D}\right)\nonumber\\
=& \sum_{g:D_{g,t}=1,D_{g,t-1}=D_{g,t-2}=0} N_{g,t}E\left(Y_{g,t-1}(0) - Y_{g,t-2}(0)\middle|\bm{D}\right) \nonumber\\
& - \frac{N_{1,0,0,t}}{N_{0,0,0,t}}\sum_{g:D_{g,t}=D_{g,t-1}=D_{g,t-2}=0} N_{g,t} E\left(Y_{g,t-1}(0) - Y_{g,t-2}(0)\middle|\bm{D}\right) \nonumber\\
= & \psi_{0,t-1}\left(\sum_{g:D_{g,t}=1,D_{g,t-1}=D_{g,t-2}=0} N_{g,t}- \frac{N_{1,0,0,t}}{N_{0,0,0,t}}\sum_{g:D_{g,t}=D_{g,t-1}=D_{g,t-2}=0} N_{g,t}\right)\nonumber\\
	= & 0.
\end{align}
The second equality follows by \eqref{eq:phi}, and the third follows after some algebra. If there exists no $g$ such that $D_{g,t-2}=D_{g,t-1}=0$ and $D_{g,t}=1$, \eqref{eq:thmWaldTCsharp_placebo} still holds, as $\DID^{\pl}_{+,t}=0$ in this case.

\medskip
A similar reasoning yields
\begin{align}\label{eq:thmWaldTCsharp_placebo2}
&N_{0,1,1,t}E\left(\DID^{\pl}_{-,t}\middle|\bm{D}\right) = 0
\end{align}
The result follows after plugging \eqref{eq:thmWaldTCsharp_placebo} and \eqref{eq:thmWaldTCsharp_placebo2} into \eqref{eq:Wald_TC_placebo_weak_exo}.

\newpage


\section*{Supplementary material}

# Web appendix of "Two-way fixed effects estimators with heterogeneous treatment effects"

Clément de Chaisemartin[*]    Xavier D'Haultfœuille[†]

February 13, 2020

**Abstract**

In this web appendix, we first discuss whether common trends necessarily implies homogenous treatment effect. Second, we show that the decomposition in Theorem 1 in the paper extends to fuzzy designs, to regressions with covariates, to regressions with a non-binary treatment, and we derive another decomposition under the supplementary assumption that the treatment effect does not change over time. Third, we extend the $DID_M$ estimator to non-binary treatments. Fourth, we discuss inference. Fifth, we review all the papers included in our survey of papers published in the AER between 2010 and 2012 (see Section 6 of the paper). Finally, the last section gathers the proofs of all the additional results in this Web Appendix.

## 1 Can common trends hold with heterogeneous treatment effects?

Throughout the paper, we assume that groups experience common trends, but that the effect of the treatment may be heterogeneous between groups and / or over time. We now discuss two examples where this may happen. We then argue that the mechanisms behind these examples are fairly general. Thus treatment effects are often likely to be heterogeneous, even when common trends are plausible.

First, assume one wants to learn the effect of the minimum wage on the employment levels of some US counties. For simplicity, let us assume that the minimum wage can only take two values, a low and a high value. Also, let us assume that there are only two periods, the 90s and the 2000s. Between these two periods, the amount of competition from China for the US industry increased substantially. Thus, for the common trends assumption to hold for counties A and B, the effect of that increase in competition should be the same on averge in those two

[*]University of California at Santa Barbara, clementdechaisemartin@ucsb.edu
[†]CREST, xavier.dhaultfoeuille@ensae.fr

counties, in the counterfactual state of the world where A and B have a low minimum wage at both dates. For that to be true, the economy of those two counties should be pretty similar. For instance, if A has a very service-oriented economy, while B has a very industry-oriented economy, it is unlikely that their employment levels will react similarly to Chinese competition.

Now, if the economies of A and B are similar, they should also have similar effects of the minimum wage on employment, thus implying that the treatment effect is homogenous between groups. On the other hand, the treatment effect may vary over time. For instance, the drop in the employment levels of A and B due to Chinese competition will probably be higher if their minimum wage is high than if their minimum wage is low. This is equivalent to saying that the effect of the minimum wage on employment diminishes from the first to the second period: due to Chinese competition in the second period, the minimum wage may have a more negative effect on employment then.[1]

Second, assume one wants to learn the effect of a job training program implemented in some US counties on participants' wages. Let us suppose that individuals self-select into the training according to a Roy model:

$$D_{i,g,t} = 1\{Y_{i,g,t}(1) - Y_{i,g,t}(0) > c_{g,t}\}, \tag{23}$$

where $c_{g,t}$ represents the cost of the training for individuals in county $g$ and period $t$. We consider fuzzy designs such as this one in the next section. Here, the common trends condition requires that average wages without the training follow the same evolution in all counties. As above, for this to hold counties used in the analysis should have similar economies, so let us assume that those counties are actually identical copies of each other: at each period, their distribution of wages without and with the training is the same. Therefore, $(g,t) \mapsto E(Y_{g,t}(1) - Y_{g,t}(0))$ is constant. However, $c_{g,t}$ may vary across counties and over time: some counties may subsidize the training more than others, and some counties may change their subsidies over time. Then, $(g,t) \mapsto E(Y_{i,g,t}(1) - Y_{i,g,t}(0)|D_{i,g,t} = 1)$ will not be constant, despite the fact that all counties in the sample have similar economies and experience similar trends on their wages.

Overall, when the treatment is assigned at the group × period level as in the minimum wage example, the economic restrictions underlying the common trends assumption may also imply homogeneous treatment effect between groups. However, those restrictions usually do not imply that the treatment effect is constant over time. Moreover, when the treatment is assigned at the individual level, as in the job training example, the economic restrictions underlying the common trends assumption neither imply homogeneous treatment effects between groups, nor homogeneous treatment effects over time.

---

[1] To simplify our discussion, in this example we consider only two counties. But in order to estimate consistently average treatment effects in the presence of county-specific shocks, the number of groups should tend to infinity, as in Section 5 below.

## 2 Results in fuzzy designs

In this section, the research design may be fuzzy: the treatment may vary within $(g,t)$ cells. For instance, Enikolopov et al. (2011) study the effect of having access to an independent TV channel in Russia, and in each Russian region some people have access to that channel while other people do not.

### 2.1 Generalizing the decomposition of $\beta_{fe}$ to fuzzy designs

For any $(g,t) \in \{1,...,G\} \times \{1,...,T\}$, let

$$\Delta_{g,t}^{TR} = \frac{1}{N_{g,t}D_{g,t}} \sum_{i:D_{i,g,t}=1} [Y_{i,g,t}(1) - Y_{i,g,t}(0)]$$

denote the average treatment effect across the treated units of cell cell $(g,t)$. One has

$$\delta^{TR} = E\left(\sum_{g,t} \frac{N_{g,t}D_{g,t}}{N_1} \Delta_{g,t}^{TR}\right),$$

which generalizes (2) to fuzzy designs.[2] Theorem 1 shows that $\beta_{fe}$ is also equal to the expectation of a weighted sum of the $\Delta_{g,t}^{TR}$s. Let

$$w_{g,t}^{TR} = \frac{\varepsilon_{g,t}}{\sum_{g,t} \frac{N_{g,t}D_{g,t}}{N_1}\varepsilon_{g,t}}.$$

**Theorem S1** *Suppose that Assumptions 1 and 3-5 hold. Then,*

$$\beta_{fe} = E\left(\sum_{g,t} \frac{N_{g,t}D_{g,t}}{N_1} w_{g,t}^{TR} \Delta_{g,t}^{TR}\right).$$

Theorem S1 shows that in fuzzy designs, $\beta_{fe}$ is equal to the expectation of a weighted sum of the ATTs in each $(g,t)$ cell. Again, some of the weights may be strictly negative. Note that under Assumption 2, Theorem S1 reduces to Theorem 1 in the paper.

The weights have a simple expression in the following special case.

**Assumption S1** *(Heterogenous adoption)* $T = 2$ and for all $g \in \{1,...,G\}$, $D_{g,2} > D_{g,1} = 0$.

Assumption S1 is satisfied in applications with two time periods, and where all groups are fully untreated at $t = 1$ and partly treated at $t = 2$. This type of design often arises in practice, for instance when an innovation is heterogeneously adopted by various groups.

---
[2]Any equation with a numbering lower than (23) refers to an equation in the paper.

**Proposition S1** *If Assumptions 1 and S1 hold and $N_{g,2}/N_{g,1}$ does not vary across $g$, then*[3]

$$w_{g,2}^{TR} = \frac{D_{g,2} - D_{.,2}}{\sum_{g=1}^{G} \frac{N_{g,2}}{N_1}(D_{g,2} - D_{.,2})^2}.$$

Proposition S1 shows that in the heterogeneous adoption design, $\beta_{fe}$ assigns negative weights to the period-two ATT of groups with a mean treatment lower than the mean treatment in the full population. The reason why negative weights arise is intuitive. With two periods, the FE regression is equivalent to a regression of the first difference of the outcome on the period-two treatment in each group. This regression compares the evolution of the outcome in more- and less-treated groups. Doing so, it subtracts the treatment effect of the less-treated groups, hence the negative weights. Negative weights are a concern if the ATTs of the less- and more-treated groups systematically differ. This could be the case if treatment is determined by a Roy selection model. Then, the groups with the highest proportion of treated units could also be those where the ATT is the highest. On the other hand, if the proportion of treated units is randomly assigned to each group, negative weights are not a concern.[4]

The DID$_M$ estimator can also be generalized to fuzzy designs, see point 2 of Theorem S1 in the Web Appendix of de Chaisemartin and D'Haultfœuille (2018) for further details.

## 2.2 Application to Enikolopov et al. (2011)

Enikolopov et al. (2011) study the effect of NTV, an independent TV channel introduced in 1996 in Russia, on the share of the electorate voting for opposition parties. NTV's coverage rate was heterogeneous across subregions: while a large fraction of the population received NTV in urbanized subregions, a smaller fraction received it in more rural subregions. The authors estimate the FE regression: they regress the share of votes for opposition parties in the 1995 and 1999 elections in Russian subregions on subregion fixed effects, an indicator for the 1999 election, and on the share of the population having access to NTV in each subregion at the time of the election. In 1995, the share of the population having access to NTV was equal to 0 in all subregions, while in 1999 it was strictly greater than 0 everywhere. Therefore, the authors' research design corresponds exactly to the heterogenous adoption design discussed above. Enikolopov et al. (2011) find that $\widehat{\beta}_{fe} = 6.65$ (s.e.= 1.40). According to this regression, increasing the share of the population having access to NTV from 0 to 100% increases the share of votes for the opposition parties by 6.65 percentage points. Because there are only two time periods in the data and the regression is not weighted by subregions' populations, $\widehat{\beta}_{fe} = \widehat{\beta}_{fd}$.

We use the `twowayfeweights` Stata package, downloadable with its help file from the SSC repository, to compute the weights attached to $\widehat{\beta}_{fe}$. In 1995, all the weights are equal to zero

---
[3] Under Assumption S1, $D_{g,1} = 0$, so $w_{g,1}^{TR}$ does not enter in the decomposition in Theorem 1.
[4] Corollaries 1 and 2 extend directly to fuzzy settings.

because NTV does not exist yet. In 1999, 918 weights (47.4%) are strictly positive, while 1,020 (52.6%) are strictly negative. The negative weights sum to -2.26. $\widehat{\sigma}_{fe} = 0.91$: $\beta_{fe}$ and $\delta^{TR}$ may be of opposite signs if the standard deviation of the effect of NTV across subregions is above 0.91 percentage point. $\widehat{\underline{\sigma}}_{fe} = 1.23$: $\beta_{fe}$ may be of a different sign than the treatment effect in every subregion if the standard deviation of the effect of NTV across subregions is above 1.23 percentage point, a plausible amount of treatment effect heterogeneity.

Therefore, $\beta_{fe}$ can only receive a causal interpretation if the effect of NTV is constant across subregions, or if the weights attached to it are uncorrelated with the intensity of that effect in each subregion (Assumption 7). These assumptions are not warranted. First, we estimate $\widehat{\beta}_{fe}$ again, weighting the regression by subregions' populations. We obtain $\widehat{\beta}_{fe} = 14.89$, more than twice its value in the unweighted regression, and the difference between the coefficients is significant (t-stat=2.46). Therefore, we can reject the null that the treatment effect is constant: if the treatment effect was constant across subregions, the weighting would not matter so both the unweighted and the weighted regressions would estimate the same parameter. Second, the weights attached to $\widehat{\beta}_{fe}$ are correlated with variables that are likely to be themselves associated with the intensity of the effect in each subregion. For instance, the correlation between the weights and subregions' populations is equal to 0.35 (t-stat=14.01). The effect of NTV may be higher in less populated subregions, as those regions are more rural and fewer other sources of information may be available there. This would lead to a violation of Assumption 7.

## 3 Extensions of the decomposition results

We consider hereafter several extensions of our decompositions of $\beta_{fe}$ and $\beta_{fd}$ in the paper. First, we consider decompositions under the common trends assumption, and under the assumption that the treatment effect is stable over time. Second, we extend our decompositions to ordered treatments. Third, we investigate the effect of including covariates in the regression. Fourth, we study two-way fixed effects 2SLS regressions. Throughout, we focus on sharp designs to ease the exposition. Nevertheless, all the results generalize to fuzzy designs.

### 3.1 $\beta_{fe}$ and $\beta_{fd}$ as weighted sums of ATEs of switching groups

We first show that under an additional condition, $\beta_{fe}$ and $\beta_{fd}$ can be written as a weighted sum of the ATEs of switching groups.

**Assumption S2** *(Stable treatment effect) For all $g$ and $t \geq 2$,*

$$E\left(\Delta_{g,t}|\boldsymbol{D}\right) D_{g,t-1} = E\left(\Delta_{g,t-1}|\boldsymbol{D}\right) D_{g,t-1}.$$

5

The stable treatment effect assumption requires that the ATE of every group treated in $t-1$ does not change from $t-1$ to $t$. By iteration, the ATE of a group treated for instance from period $t_0$ to $T$ is unrestricted before $t_0$ but should be constant from $t_0$ to $T$. Assumption S2 rules out the possibility that the treatment effect changes over time. Therefore, it may be implausible and should be carefully discussed.

We now show that under the common trend and stable treatment effects assumptions, $\beta_{fe}$ and $\beta_{fd}$ may identify weighted averages of ATEs. Let $N_S = \sum_{(g,t):D_{g,t}\neq D_{g,t-1}} N_{g,t}$ and, for all $g$ and $t \geq 2$,

$$w_{g,t}^S = \frac{(D_{g,t}-D_{g,t-1})\sum_{t'\geq t}\frac{N_{g,t'}}{N_{g,t}}\varepsilon_{g,t'}}{\sum_{(g,,t):t\geq 2}\left[\frac{N_{g,t}}{N_S}(D_{g,t}-D_{g,t-1})\sum_{t'\geq t}\frac{N_{g,t'}}{N_{g,t}}\varepsilon_{g,t'}\right]},$$

$$w_{fd,g,t}^S = \frac{(D_{g,t}-D_{g,t-1})\varepsilon_{fd,g,t}}{\sum_{(g,,t):t\geq 2}\frac{N_{g,t}}{N_S}(D_{g,t}-D_{g,t-1})\varepsilon_{fd,g,t}}.$$

**Theorem S2** *Suppose that Assumptions 1-5 and S2 hold. Then,*

$$\beta_{fe} = E\left(\sum_{(g,t):D_{g,t}\neq D_{g,t-1},t\geq 2}\frac{N_{g,t}}{N_S}w_{g,t}^S\Delta_{g,t}\right),$$

$$\beta_{fd} = E\left(\sum_{(g,t):D_{g,t}\neq D_{g,t-1},t\geq 2}\frac{N_{g,t}}{N_S}w_{fd,g,t}^S\Delta_{g,t}\right).$$

*Moreover, $w_{fd,g,t}^S \geq 0$ for all $g$ and $t \geq 2$. If Assumption 6 holds and $N_{g,t}/N_{g,t-1}$ does not vary across $g$ for all $t \geq 2$, $w_{g,t}^S \geq 0$ for all $g$ and $t \geq 2$.*

Theorem S2 shows that in sharp designs, under the common trends and stable treatment effect assumptions, $\beta_{fe}$ and $\beta_{fd}$ identify weighted sums of ATEs of switching cells. The weights differ from those in Theorems 1 and 2. Now the weights attached to $\beta_{fe}$ are all positive in staggered adoption designs, while the weights attached to $\beta_{fd}$ are all positive in all sharp designs. Therefore, in staggered adoption (resp. sharp) designs, $\beta_{fe}$ (resp. $\beta_{fd}$) relies on the assumption that the treatment effect is stable over time, but it does not require that treatment effects be homogeneous between groups.

## 3.2 Non-binary, ordered treatment

We now consider the case where the treatment takes a finite number of ordered values, $D_{i,g,t} \in \{0,1,...,\overline{d}\}$, and show that Theorem 1 can easily be extended to this case.[5] We need to define

---

[5]Theorem 2 can also be extended to the case of a non-binary treatment.

potential outcomes for all the possible treatment values. For instance $Y_{i,g,t}(\bar{d})$ is the counterfactual outcome of observation $i$ in cell $(g,t)$ if she receives treatment value $\bar{d}$. We also need to modify the treatment effect parameters we consider. In lieu of $\delta^{TR}$, we consider the average causal response (ACR) on the treated,

$$\delta^{ACR} = E\left(\frac{1}{N_1}\sum_{i,g,t} Y_{i,g,t}(D_{g,t}) - Y_{i,g,t}(0)\right).$$

Similarly, for all $(g,t)$ such that $D_{g,t} \neq 0$, we consider, instead of $\Delta_{g,t}$,

$$\Delta^{ACR}_{g,t} = \frac{1}{N_{g,t} D_{g,t}} \sum_{i=1}^{N_{g,t}} \left[Y_{i,g,t}(D_{g,t}) - Y_{i,g,t}(0)\right].$$

Then, similarly to (2), the following decomposition holds:

$$\delta^{ACR} = E\left(\sum_{(g,t):D_{g,t}\neq 0} \frac{N_{g,t} D_{g,t}}{N_1} \Delta^{ACR}_{g,t}\right).$$

Let $w^O_{g,t} = \frac{\varepsilon_{g,t}}{\sum_{g,t} \frac{N_{g,t} D_{g,t}}{N_1} \varepsilon_{g,t}}$. Note that if the treatment is binary, $w^O_{g,t} = w_{g,t}$.

**Theorem S3** *Suppose that Assumptions 1-5 hold and $D_{i,g,t} \in \{0,...,\bar{d}\}$. Then,*

$$\beta_{fe} = E\left(\sum_{(g,t):D_{g,t}\neq 0} \frac{N_{g,t} D_{g,t}}{N_1} w^O_{g,t} \Delta^{ACR}_{g,t}\right).$$

Theorem S3 shows that under Assumption 5, when the treatment is not binary $\beta_{fe}$ identifies a weighted sum of the ACRs in all the $(g,t)$ cells that are not untreated. Then, since the proof of Corollary 1 does not rely on the nature of the treatment, Corollary 1 directly applies to ordered treatments as well, by just replacing $w_{g,t}$ and $N_{g,t}$ by $w^O_{g,t}$ and $N_{g,t} D_{g,t}$, respectively. Corollary 2 extends as well to this set-up, by simply modifying the no-correlation condition appropriately.

### 3.3 Including covariates

Often times, researchers also include a vector of covariates $X_{g,t}$ as control variables in their regression. In this section, we show that our Theorem 1 can be extended to this case.[6] We start by redefining Regression 1 in this context.

---

[6]Theorem 2 can also be extended to regressions with covariates.

**Regression 1X** *(Fixed-effects regression with covariates)*

Let $\widehat{\beta}_{fe}^{X}$ denote the coefficient of $D_{g,t}$ in an OLS regression of $Y_{i,g,t}$ on group and period fixed effects, $D_{g,t}$, and $X_{g,t}$. Let $\beta_{fe}^{X} = E\left(\widehat{\beta}_{fe}^{X}\right)$.

Then, we need to modify Assumptions 3-5. Hereafter, we let $\boldsymbol{X}_g = (X_{g,1}, ..., X_{g,t})$.

**Assumption S3** *(Independent groups with covariates)* The vectors $(Y_{g,t}(0), Y_{g,t}(1), D_{g,t}, X_{g,t})_{1 \leq t \leq T}$ are mutually independent.

**Assumption S4** *(Strong exogeneity and common trends with covariates)* There is a vector $\gamma$ of same dimension as $X_{g,t}$ such that

$$E\left(Y_{g,t}(0) - Y_{g,t-1}(0) - (X_{g,t} - X_{g,t-1})'\gamma | \boldsymbol{D}_g, \boldsymbol{X}_g\right) = E\left(Y_{g,t}(0) - Y_{g,t-1}(0) - (X_{g,t} - X_{g,t-1})'\gamma\right)$$

and $E\left(Y_{g,t}(0) - Y_{g,t-1}(0) - (X_{g,t} - X_{g,t-1})'\gamma\right)$ does not vary across $g$.

Rearranging, Assumption S4 requires that

$$E\left(Y_{g,t}(0) | \boldsymbol{D}_g, \boldsymbol{X}_g\right) - E\left(Y_{g,t-1}(0) | \boldsymbol{D}_g, \boldsymbol{X}_g\right) = (X_{g,t} - X_{g,t-1})'\gamma + \lambda_t,$$

for some constant $\lambda_t$. Then, Assumption S4 allows for the possibility that groups experience different evolutions of their $Y_{g,t}(0)$ over time, but it requires that those differential evolutions are fully accounted for by a linear model in $X_{g,t} - X_{g,t-1}$, the change in a group's covariates. Assumption S4 is implied by the linear model that is often invoked to justify the use of the FE regression with covariates. For instance, the use of Regression 1X is often justified by the following model:

$$Y_{g,t}(0) = \gamma_g + \lambda_t + X'_{g,t}\gamma + \eta_{g,t}, \quad E(\eta_{g,t} | \boldsymbol{D}_g, \boldsymbol{X}_g) = 0. \tag{24}$$

Equation (24) implies Assumption S4, but it does not imply Assumption 5.

An interesting special case is when the control variables are group-specific linear trends. Then, Assumption s4 requires that for all $t \geq 2$,

$$E\left(Y_{g,t}(0) | \boldsymbol{D}_g, \boldsymbol{X}_g\right) - E\left(Y_{g,t-1}(0) | \boldsymbol{D}_g, \boldsymbol{X}_g\right) = \gamma_g + \lambda_t,$$

for some constants $\gamma_g$ and $\lambda_t$. From $t-1$ to $t$, the evolution of $Y(0)$ in group $g$ should deviate from its group-specific linear trend $\gamma_g$ by an amount $\lambda_t$ common to all groups. Then, Assumption S4 is a "common deviation from linear trends" assumption, which may be more plausible than the standard common trends assumption.

Let $\varepsilon_{g,t}^{X}$ denote the residual of observations in cell $(g,t)$ in the regression of $D_{g,t}$ on group and period fixed effects and $X_{g,t}$. One can show that if the regressors in Regression 1 are not collinear, the average value of $\varepsilon_{g,t}^{X}$ across all treated $(g,t)$ cells differs from 0: $\sum_{(g,t):D_{g,t}=1}(N_{g,t}/N_1)\varepsilon_{g,t}^{X} \neq 0$. Then, let

$$w_{g,t}^{X} = \frac{\varepsilon_{g,t}^{X}}{\sum_{(g,t):D_{g,t}=1} \frac{N_{g,t}}{N_1} \varepsilon_{g,t}^{X}}.$$

**Theorem S4** *Suppose that Assumptions 1-2 and S3-S4 hold. Then,*

$$\beta_{fe}^X = E\left[\sum_{(g,t):D_{g,t}=1} \frac{N_{g,t}}{N_1} w_{g,t}^X \Delta_{g,t}\right].$$

Theorem S4 shows that under a modified version of the common trends assumption accounting for the covariates, $\beta_{fe}^X$ identifies a weighted sum of the $\Delta_{g,t}^{TR}$s, as $\beta_{fe}$ in Theorem 1, with different but still potentially negative weights.[7] Assumption S4 may be more plausible than Assumption 5, but adding covariates may increase the prevalence of negative weights, or the correlation between the weights and the $\Delta_{g,t}$s, thus making $\beta_{fe}^X$ less robust to heterogeneous effects than $\beta_{fe}$.

## 3.4 2SLS regressions

Researchers have sometimes estimated 2SLS versions of Regressions 1 and 2. Our main conclusions also apply to those regressions. Let $\widehat{\beta}_{fe}^{2SLS}$ denote the coefficient of $D_{i,g,t}$ in a 2SLS regression of $Y_{g,t}$ on group and period fixed effects and $D_{i,g,t}$, using a variable $Z_{g,t}$ constant within each group × period as the instrument for $D_{i,g,t}$. $Z_{g,t}$ typically represents an incentive for treatment allocated at the group × period level. For instance, Duflo (2001) studies the effect of years of schooling on wages in Indonesia, using a primary school construction program as an instrument. Specifically, she estimates a 2SLS regression of wages on cohort and district of birth fixed effects and years of schooling, using the interaction of belonging to a cohort entering primary school after the program was completed and the number of schools constructed in one's district of birth as the instrument for years of schooling.

Remark that $\widehat{\beta}_{fe}^{2SLS} = \widehat{\beta}_{fe}^Y / \widehat{\beta}_{fe}^D$, where $\widehat{\beta}_{fe}^Y$ (resp. $\widehat{\beta}_{fe}^D$) is the coefficient of $Z_{g,t}$ in the reduced-form regression of $Y_{g,t}$ (resp. the first-stage regression of $D_{g,t}$) on group and period fixed effects and $Z_{g,t}$. Then let $\beta_{fe}^{2SLS} = E[\widehat{\beta}_{fe}^Y]/E[\widehat{\beta}_{fe}^D]$.[8] Following Imbens and Angrist (1994), for any $z \in \text{Supp}(Z)$ let $D_{i,g,t}(z)$ denote the potential treatment of unit $i$ in $(g,t)$ if $Z_{i,g,t} = z$. It follows from Theorem 1 that under a common trends assumption on $D_{i,g,t}(0)$, $E[\widehat{\beta}_{fe}^D]$ is equal to a weighted sum of the average effects of the instrument on the treatment in each group and time period, with potentially many negative weights. Similarly, under a common trends assumption

---

[7]In a previous version of this paper, we had shown that under a different, and arguably less natural, common trends assumption, $\beta_{fe}^X$ identifies a weighted sum of the $\Delta_{g,t}^{TR}$, with the same weights as in Theorem 1. We thank an anonymous referee for pointing out issues with the common trends assumption we had previously proposed.

[8]We do not consider here $E[\widehat{\beta}_{fe}^{2SLS}]$, as the 2SLS estimator may not have an expectation. Moreover, under conditions similar to those imposed in Section 5 of the paper, $\beta_{fe}^{2SLS}$ is the probability limit of $\widehat{\beta}_{fe}^{2SLS}$, which makes $\beta_{fe}^{2SLS}$ the proper estimand here.

on $Y_{i,g,t}(D_{i,g,t}(0))$ instead of $Y_{i,g,t}(0)$, $E[\widehat{\beta}_{fe}^Y]$ is equal to a weighted sum of the average effects of the instrument on the outcome, again with potentially many negative weights. For instance, in Duflo (2001), under a common trends assumption on $D_{i,g,t}(0)$, the number of years of schooling individuals would complete if zero new schools were constructed in their district, the first stage coefficient identifies a weighted sum of the effect of one new school on years of schooling in every district, with many negative weights.[9]

Hence, it is only if the average effects of $Z_{g,t}$ on $Y_{i,g,t}$ and $D_{i,g,t}$ are constant across groups and periods, or if the weights are uncorrelated to treatment effects as in Assumption 7, that the reduced-form and first-stage coefficients respectively identify the average effect of $Z_{i,g,t}$ on $Y_{i,g,t}$ and $D_{i,g,t}$. Then, this implies that $\beta_{fe}^{2SLS}$ identifies, under the conditions in Imbens and Angrist (1994), the LATE of $D_{i,g,t}$ on $Y_{i,g,t}$ among units that comply with the instrument.[10]

## 4 Extending the DID$_\text{M}$ estimator

Theorem 3 can be extended to the case where $D_{i,g,t}$ is not binary but takes values in $\mathcal{D} = \{0, ..., \overline{d}\}$. The causal effect we consider is the switchers' causal response

$$\delta^{SCR} = E\left(\frac{1}{N_{D,S}} \sum_{(i,g,t): t \geq 2, D_{g,t} \neq D_{g,t-1}} [Y_{i,g,t}(\max(D_{g,t}, D_{g,t-1})) - Y_{i,g,t}(\min(D_{g,t}, D_{g,t-1}))]\right),$$

where $N_{D,S} = \sum_{(g,t): t \geq 2} N_{g,t} |D_{g,t} - D_{g,t-1}|$. Note that $\delta^{SCR} = \delta^S$ when $D_{i,g,t}$ is binary.

We identify $\delta^{SCR}$ under the following two conditions, which generalize Assumptions 4-5 and 9-12 to non-binary treatments.

**Assumption S5** *(Mean independence between a group's outcome and other groups treatments))* *For all $(d, g, t) \in \mathcal{D} \times \{1, ..., G\} \times \{1, ..., T\}$, $E(Y_{g,t}(d)|\boldsymbol{D}) = E(Y_{g,t}(d)|\boldsymbol{D}_g)$.*

**Assumption S6** *(Strong exogeneity) For all $(d, g, t) \in \mathcal{D} \times \{1, ..., G\} \times \{2, ..., T\}$, $E(Y_{g,t}(d) - Y_{g,t-1}(d)|\boldsymbol{D}_g) = E(Y_{g,t}(d) - Y_{g,t-1}(d))$.*

**Assumption S7** *(Common trends) For every d, for all $t \geq 2$ and g, $E(Y_{g,t}(d) - Y_{g,t-1}(d))$ does not vary across g.*

---

[9]New schools were constructed in every district, so this application falls into the heterogeneous adoption case.

[10]In the special case with two groups and two periods, a binary incentive for treatment, and where only group 1 in period 1 receives the incentive, de Chaisemartin (2010) and Hudson et al. (2015) show that in a 2SLS regression of $Y_{i,g,t}$ on $1\{g = 2\}$, $1\{t = 2\}$ and $D_{i,g,t}$, using $Z_{g,t} = 1\{g = 2\}1\{t = 2\}$ as the instrument, the coefficient of $D_{i,g,t}$ identifies a LATE under common trends assumptions on $Y_{i,g,t}(D_{i,g,t}(0))$ and $D_{i,g,t}(0)$. However, the discussion above shows that this result does not generalize to applications with multiple groups and periods or a non-binary instrument, as in Duflo (2001) where the number of new schools constructed varies across districts.

**Assumption S8** *(Existence of "stable" groups)* For all $t \in \{2, ..., T\}$, for all $(d, d') \in \mathcal{D}^2$, $d \neq d'$, if there is at least one $g \in \{1, ..., G\}$ such that $D_{g,t-1} = d$ and $D_{g,t} = d'$, then there exists at least one $g' \neq g$, $g' \in \{1, ..., G\}$ such that $D_{g',t-1} = D_{g',t} = d$.

When the treatment takes a large number of values, Assumption S8 may be violated. A solution, then, is to consider a modified treatment variable $\widetilde{D}_{g,t} = h(D_{g,t})$ that groups together several values of $D_{g,t}$, to ensure that Assumption S8 holds for $\widetilde{D}_{g,t}$. For instance, if the treatment can be equal to 0, 1, 2, or 3, and there is a group whose treatment switches from 2 to 3 between periods 1 and 2, but no group whose treatment remains equal to 2 between those two dates, one may define $\widetilde{D}_{g,t} = \min(D_{g,t}, 2)$ if there is a group whose treatment is equal to 3 at periods 1 and 2. Then, Theorem S5 below still holds, after replacing $D_{g,t}$ by $\widetilde{D}_{g,t}$ in the $\text{DID}_{d,d',t}$ estimators defined below, and if Assumption S7 is replaced by the requirement that $E(Y_{g,t}(d) - Y_{g,t-1}(d))$ only depends on $t$ and $h(d)$.

In order to define $\text{DID}_M$ in this context, let us introduce, for all $(d, d', t) \in \mathcal{D}^2 \times \{2, ..., T\}$,

$$\text{DID}_{d,d',t} = [1\{d < d'\} - 1\{d' < d\}] \left[ \sum_{(g,t): D_{g,t}=d', D_{g,t-1}=d, t\geq 2} \frac{N_{g,t}}{N_{d,d',t}} [Y_{g,t} - Y_{g,t-1}] \right.$$
$$\left. - \sum_{(g,t): D_{g,t}=D_{g,t-1}=d, t\geq 2} \frac{N_{g,t}}{N_{d,d,t}} [Y_{g,t} - Y_{g,t-1}] \right],$$

where $N_{d,d',t}$ is defined as in (3) for any $(d, d') \in \mathcal{D}^2$. Then

$$\text{DID}_M = \sum_{t=2}^{T} \sum_{(d,d') \in \mathcal{D}^2, d \neq d'} \frac{N_{d,d',t}}{N_{D,S}} \text{DID}_{d,d',t}.$$

If the treatment is binary, the $\text{DID}_M$ estimator defined above is equal to that defined in Section 4 of the paper.

**Theorem S5** *Suppose that $D_{i,g,t} \in \mathcal{D}$ and Assumptions 1-2 and S5-S8 hold. Then $E[\text{DID}_M] = \delta^{SCR}$.*

Theorem S5 generalizes Theorem 3 to non-binary treatments. We can also extend Theorem 4 in the same way to construct placebo tests of Assumption S7.

Finally, Theorem 3 can also be extended to the case with covariates. Under versions of Assumptions 10 and 11 written conditional on $X$, a conditional version of the $\text{DID}_M$ estimator is consistent for $\delta^S$ under the common support condition $\text{Supp}(X_{d,g,t}) = \text{Supp}(X)$. We refer to de Chaisemartin and D'Haultfœuille (2018) for further details.

# 5 Statistical properties of $\text{DID}_M$ and inference on $\delta^S$

In this section, we establish the asymptotic properties of $\text{DID}_M$ and construct confidence intervals on $\delta^S$ based on $\text{DID}_M$. We consider an asymptotic framework where the number of groups $G$ tends to infinity. To define the confidence intervals, let $P_{d,d',t} = N_{d,d',t}/G$ and

$$Q_{d,d',t} = \frac{1}{G} \sum_g N_{g,t} \mathbb{1}\{D_{g,t} = d, D_{g,t-1} = d'\} (Y_{g,t} - Y_{g,t-1}).$$

Then, let $\widehat{\sigma}^2 = \sum_g \widehat{\psi}_g^2 / G$, with

$$\widehat{\psi}_g = \frac{G}{N_S} \sum_{t>1} N_{g,t} \left[ \mathbb{1}\{D_{g,t} \neq D_{g,t-1}\}(Y_{g,t} - Y_{g,t-1} - \text{DID}_M) - \widehat{\psi}_{g,t}^B \right],$$

$$\widehat{\psi}_{g,t}^B = \frac{1}{P_{0,0,t}} \left[ \mathbb{1}\{D_{g,t} > D_{g,t-1}\}Q_{0,0,t} + P_{1,0,t}\mathbb{1}\{D_{g,t} = D_{g,t-1} = 0\} \left(Y_{g,t} - Y_{g,t-1} - \frac{Q_{0,0,t}}{P_{0,0,t}}\right) \right]$$

$$+ \frac{1}{P_{1,1,t}} \left[ \mathbb{1}\{D_{g,t} < D_{g,t-1}\}Q_{1,1,t} + P_{0,1,t}\mathbb{1}\{D_{g,t} = D_{g,t-1} = 1\} \left(Y_{g,t} - Y_{g,t-1} - \frac{Q_{1,1,t}}{P_{1,1,t}}\right) \right].$$

We consider confidence intervals of the form

$$\text{CI}_{1-\alpha}(\delta^S) = \left[ \text{DID}_M - z_{1-\alpha/2} \frac{\widehat{\sigma}}{\sqrt{G}}, \text{DID}_M + z_{1-\alpha/2} \frac{\widehat{\sigma}}{\sqrt{G}} \right],$$

where $z_{1-\alpha/2}$ denotes the quantile of order $1 - \alpha/2$ of a standard normal variable.

We now establish the asymptotic properties of $\text{DID}_M$ and $\text{CI}_{1-\alpha}(\delta^S)$ under the following assumptions. Hereafter, we denote $\boldsymbol{U} = (P_{0,0,1}, Q_{0,0,1}, ..., P_{1,1,T}, Q_{1,1,T})$.

**Assumption S9** *(Existence of moments and limits)* $\sup_{g,t} N_{g,t} < +\infty$ and $\sup_{(d,g,t)} E(Y_{g,t}^4(d)) < +\infty$. $\lim_G E[\boldsymbol{U}]$ and $\lim_G G \times V(\boldsymbol{U})$ exist.

**Assumption S10** *(Positive probability of "stable" groups and existence of switchers)* For all $(d, g, t) \in \{0, 1\} \times (\mathbb{N}\backslash\{0\}) \times \{2, ..., T\}$, $\Pr(D_{g,t} = 1-d, D_{g,t-1} = d) > 0$ implies $\lim_G E[P_{1-d,d,t}] > 0$ and $\lim_G E[P_{d,d,t}] > 0$. Moreover, $\lim_G E[P_{0,1,t} + P_{1,0,t}] > 0$ for at least one $t$.

Assumption S9 imposes the (uniform) existence of moments of order 4 of $Y_{g,t}(d)$, and that some non-random averages converge as $G$ tends to infinity. These assumptions ensure that we can apply law of large numbers and central limit theorems in our set-up where groups are independent but not necessarily identically distributed. Assumption S10 imposes that when at least one group switches from $d$ to $1 - d$ with a positive probability on a given period, then on average over all groups, the limit probabilities of switching from $d$ to $1 - d$ will be positive as $G \to \infty$. The limit probability of remaining at $d$ will also be positive. This latter condition may be seen as a weaker version of Assumption 11, as it imposes the existence of "stable" groups only

with probability tending to one as $G \to \infty$. The last condition in Assumption S10 simply states that asymptotically, the proportion of switchers is strictly postive.

The following result shows that under these conditions, $\text{DID}_M$ is asymptotically normal, and that $\text{CI}_{1-\alpha}(\delta^S)$ is asymptotically conservative.

**Theorem S6** *Suppose that Assumptions 1-5, 9-10 and S9-S10 hold. Then, as $G \to \infty$,*

$$\sqrt{G}\left(\text{DID}_M - \delta^S\right) \xrightarrow{d} \mathcal{N}(0, \sigma^2),$$

*with $\sigma^2$ defined in (38) below. Moreover,*

$$\limsup_{G \to \infty} \Pr\left(\delta^S \in \text{CI}_{1-\alpha}(\delta^S)\right) \geq 1 - \alpha.$$

Theorem S6 shows that $\text{DID}_M$ is an asymptotically normal estimator of $\delta^S$ when the number of groups tends to infinity, provided the outcomes and treatments are independent across groups. As is usually the case for estimators constructed using independent but not identically distributed random variables (see e.g. Liu and Singh, 1995), the asymptotic variance of $\text{DID}_M$ can only be conservatively estimated. As a result, the confidence interval we propose is asymptotically conservative.

## 6  Detailed literature review

We now review the 33 papers that use two-way fixed effects or closely related regressions that we found in our literature review. For each paper, we use the following presentation:
**Authors (year), Title.** *Where the two-way fixed effects estimator is used in the paper.*
Description of the two-way fixed effects estimator used in the paper, and how it relates to Regression 1 or 2. Assessment of whether the stable groups assumption holds in this paper. Assessment of whether the research design is sharp or fuzzy.

1. **Chandra et al. (2010), Patient Cost-Sharing and Hospitalization Offsets in the Elderly.** *First line of Tables 2 and 3.*
   In the regressions in the first line of Tables 2 and 3, the outcomes (e.g. a measure of utilization for plan p in month t) are regressed on plan fixed effects, month fixed effects, and an indicator of whether plan p had increased copayments in month t (see regression equation at the bottom of page 198). This regression corresponds to Regression 1. The period analyzed runs from January 2000 to September 2003. The stable groups assumption is satisfied until January 2002, when the HMO plans also become treated. This is a sharp design.

13

2. **Duggan and Morton (2010), The Effect of Medicare Part D on Pharmaceutical Prices and Utilization.** *Tables 2 and 3.*

   In regression Equation (1), the dependent variable is the change in the price of drug j between 2003 and 2006, the explanatory variables are the Medicare market share for drug j in 2003, and some control variables. This regression corresponds to Regression 2, with some control variables. The stable groups assumption is presumably not satisfied: it seems unlikely that there are drugs whose Medicare market share in 2003 is equal to 0. This is a sharp design.

3. **Aizer (2010), The Gender Wage Gap and Domestic Violence.** *Table 2.*

   In regression Equation (2), the dependent variable is the log of female assaults among females of race r in county c and year t, and the explanatory variables are race, year, county, race × year, race × county, and county × year fixed effects, as well as the gender wage gap in county c, year t, and race r, and some control variables. This regression is a "three-way fixed effects" version of Regression 1, with some control variables. The stable groups assumption is presumably satisfied: it seems likely that between each pair of consecutive years, there are counties where the gender wage gap does not change. This is a fuzzy design: the treatment of interest is the gender wage gap in a couple (see the bargaining model in Appendix 1), which varies within (year,county) cells.

4. **Algan and Cahuc (2010), Inherited Trust and Growth.** *Figure 4.*

   Figure 4 presents a regression of changes in income per capita from 1935 to 2000 on changes in inherited trust over the same period and a constant. This regression corresponds to Regression 2. The stable groups assumption is satisfied: there are countries where inherited trust does not change from 1935 to 2000. This is a sharp design.

5. **Ellul et al. (2010), Inheritance Law and Investment in Family Firms.** *Table 7.*

   In the regressions presented in Table 7, the dependent variable is the capital expenditure of firm j in year t, and the explanatory variables are firm fixed effects, an indicator for whether year t is a succession period for firm j, some controls, and three treatment variables: the interaction of the succession indicator with the level of investor protection in the country where firm j is located, the interaction of the succession indicator with the level of inheritance laws permissiveness in the country where firm j is located, and the interaction of the succession indicator with the level of inheritance laws permissiveness and the level of investor protection in the country where firm j is located. This regression is similar to Regression 1 with controls, except that it has three treatment variables. The stable groups assumption is presumably not satisfied: for instance, it seems unlikely that there are countries with no investor protection at all. This is a sharp design.

6. **Bustos (2011), Trade Liberalization, Exports, and Technology Upgrading: Ev-**

**idence on the Impact of MERCOSUR on Argentinean Firms.** *Tables 3 to 12.*

In regression Equation (11), the dependent variable is the change in exporting status of firm i in sector j between 1992 and 1996, and the explanatory variables are the change in trade tariffs in Brasil for products in sector j over the same period, and some control variables. This regression corresponds to Regression 2, with some controls. The stable groups assumption is presumably satisfied: it seems likely that there are sectors where trade tariffs in Brasil did not change between 1992 and 1996. This is a sharp design.

7. **Anderson and Sallee (2011), Using Loopholes to Reveal the Marginal Cost of Regulation: The Case of Fuel-Economy Standards.** *Table 5 Column 2.*

   In the regression in Table 5 Column (2), the dependent variable is an indicator for whether a car sold is a flexible fuel vehicle, and the explanatory variables are state and month fixed effects, the percent of gas stations that have ethanol fuel in each month × state, and some controls. This regression corresponds to Regression 1. The stable groups assumption is presumably satisfied: it seems likely that between each pair of consecutive months, there are states where the percent ethanol availability does not change. This a fuzzy design: the treatment of interest is whether a car buyer has access to ethanol fuel, which varies within (month,state) cells.

8. **Bagwell and Staiger (2011), What Do Trade Negotiators Negotiate About? Empirical Evidence from the World Trade Organization.** *Table 3, OLS columns.*

   In regression equations (15a) and (15b), the dependent variable is the ad valorem tariff level bound by country c on product g, while the explanatory variables are country and product fixed effects, and two treatment variables which vary at the country × product level. These regressions are similar to Regression 1, except that they have two treatment variables. The stable groups assumption is not applicable here, as none of the two sets of fixed effects included in the regression correspond to an ordered variable. This is a sharp design.

9. **Zhang and Zhu (2011), Group Size and Incentives to Contribute: A Natural Experiment at Chinese Wikipedia.** *Tables 3 and 4, Columns 4-6.*

   In the regression in, say, Table 3 Column (4), the dependent variable is the total number of contributions to Wikipedia by individual i at period t, regressed on individual fixed effects, an indicator for whether period t is after the Wikipedia block, the interaction of this indicator and a measure of social participation by individual i, and some controls. This regression corresponds to Regression 1 with some controls. The stable groups assumption is satisfied: there are individuals with a social participation measure equal to 0. This is a sharp design.

10. **Hotz and Xiao (2011), The Impact of Regulations on the Supply and Quality**

**of Care in Child Care Markets.** *Table 7, Columns 4 and 5.*

In Regression Equation (1), the dependent variable is the outcome for market m in state s and year t, and the explanatory variables are state and year fixed effects, various measures of regulations in state s in year t, and some controls. This regression corresponds to Regression 1 with several treatment variables and with some controls. The stable groups assumption is presumably satisfied: between each pair of consecutive years, it is likely that there are states whose regulations do not change. This is a sharp design.

11. **Mian and Sufi (2011), House Prices, Home Equity-Based Borrowing, and the US Household Leverage Crisis.** *Tables 2 and 3.*

    In Regression Equation (1), the dependent variable is the change in homeowner leverage from 2002 to 2006 for individual i living in zip code z in MSA m, and the dependent variable is the change in the house price for that individual, instrumented by MSA-level housing supply elasticity. This regression is the 2SLS version of Regression 2, with some controls. The stable groups assumption is presumably not satisfied: it is unlikely that some MSAs have an housing supply elasticity equal to 0. This is a sharp design.

12. **Wang (2011), State Misallocation and Housing Prices: Theory and Evidence from China.** *Table 5, Panel A.*

    In regression Equation (15), the dependent variable is the quantity of housing services in household i's residence in year t, while the explanatory variables are an indicator for period t being after the reform, a measure of mismatch in household i, the interaction of the measure of mismatch and the time indicator, and some controls. This regression is similar to Regression 1 with some controls, except that it has a measure of mismatch in household i instead of household fixed effects. The stable groups assumption is presumably satisfied: it is likely that some households have a mismatch equal to 0. This is a sharp design.

13. **Duranton and Turner (2011), The Fundamental Law of Road Congestion: Evidence from US Cities.** *Table 5.*

    In the regressions presented in, say, the first column of Table 5, the dependent variable is the change in vehicle kilometers traveled in MSA s between decades t and t-1, and the explanatory variables are the change in kilometers of roads in MSA s between decades t and t-1, and decade effects. This regression corresponds to Regression 2. The stable groups assumption is presumably satisfied: it is likely that between each pair of consecutive decades, there are some MSAs where the kilometers of roads do not change. This is a sharp design.

14. **Acemoglu et al. (2011), The Consequences of Radical Reform: The French Revolution.** *Table 3.*

In regression Equation (1), the dependent variable is urbanization in polity j at time t, while the explanatory variables are time and polity fixed effects, and the number of years of French presence in polity j interacted with the time effects. This regression corresponds to Regression 1. The stable groups assumption is satisfied as there are several polities that did not experience any year of French presence. This is a sharp design.

15. **Baum-Snow and Lutz (2011), School Desegregation, School Choice, and Changes in Residential Location Patterns by Race.** *Tables 2 to 6.*
In regression Equation (1), the dependent variable is, say, whites public school enrolment in MSA j in year t, while the explanatory variables are MSA and region × time fixed effects, and an indicator for whether MSA j is desegregated. This regression corresponds to Regression 1 with controls. The stable groups assumption is satisfied: between each pair of consecutive years, there are MSAs whose desegregation status does not change. This is a sharp design.

16. **Dinkelman (2011), The Effects of Rural Electrification on Employment: New Evidence from South Africa.** *Tables 4 and 5 Columns 5-8, Table 8 Columns 3-4, Table 9 Column 2, and Table 10 Columns 2, 4, and 6.*
In regression Equation (3), the dependent variable is, say, the first difference of the female employment rate for community j between periods 1 and 2, and the explanatory variables are district fixed effects, the change of electrification status of community j between periods 1 and 2, and some statistical controls. The land gradient in community j is used as an instrument for the change in electrification. This regression corresponds to the 2SLS version of Regression 2 with some controls. The stable groups assumption is presumably satisfied: it is likely that there are communities whose land gradient is 0. This is a sharp design.

17. **Enikolopov et al. (2011), Media and Political Persuasion: Evidence from Russia.** *Table 3.*
In regression Equation (5), the dependent variable is the share of votes for party j in election-year t and subregion s, and the explanatory variables are subregion and election fixed effects, and the share of people having access to NTV in subregion s in election-year t. This regression corresponds to Regression 1. The stable groups assumption is not satisfied: the share of people having access to NTV strictly increases in all regions between 1995 and 1999, the two elections used in the analysis. This a fuzzy design: the treatment of interest is whether a person has access to NTV, which varies within (subregion,year) cells.

18. **Fang and Gavazza (2011), Dynamic Inefficiencies in an Employment-Based Health Insurance System: Theory and Evidence.** *Tables 2, 3, 5, and 6, Column 3.*
In regression Equation (7), the dependent variable is the health expenditures of individual j working in industry i in period t and region r, and the explanatory variables are individual

17

effects, region specific time effects, and the job tenure of individual j. The death rate of establishments in industry i in period t and region r is used as an instrument for the job tenure of individual j. This regression is the 2SLS version of Regression 2 with controls. The stable groups assumption is presumably satisfied: between each pair of consecutive years, it is likely that there are some industry × region pairs where the death rate of establishments does not change. This a fuzzy design: the instrument of interest is whether a person's former employee closed down over the current year, which varies within (industry,year) cells.

19. **Gentzkow et al. (2011), The Effect of Newspaper Entry and Exit on Electoral Politics.** *Tables 2 and 3.*

    In regression Equation (2), the dependent variable is the change in voter turnout in county c between elections year t and t-1, and the explanatory variables are state × year effects, and the change in the number of newspapers in county c between t and t-1. This regression corresponds to Regression 2 with controls. The stable groups assumption is satisfied: between each pair of consecutive years, there are some counties where the number of newspapers does not change. This is a sharp design.

20. **Bloom et al. (2012), Americans Do IT Better: US Multinationals and the Productivity Miracle.** *Table 2, Columns 6-8.*

    In the regression in, say, Column 6 of Table 2, the dependent variable is the log of output per worker in firm i in period t, while the explanatory variables are firms and time fixed effects, the log of the amount of IT capital per employee $\ln(C/L)$, the interaction of $\ln(C/L)$ and an indicator for whether the firm is owned by a US multinational, the interaction of $\ln(C/L)$ and an indicator for whether the firm is owned by a non-US multinational, and some controls. This regression is similar to Regression 1 with some controls, except that it has three treatment variables. The stable groups assumption is presumably satisfied: between each pair of consecutive years, it is likely that there are some firms where the amount of IT capital per employee $\ln(C/L)$ does not change. This is a sharp design.

21. **Simcoe (2012), Standard Setting Committees: Consensus Governance for Shared Technology Platforms.** *Table 4, Columns 1-3.*

    In regression Equation (5), the dependent variable is a measure of time to consensus for project i submitted to committee j, while the explanatory variables are an indicator for projects submitted to the standards track, a measure of distributional conflict, the interaction of the standards track and distributional conflict, and some controls variables. This regression is similar to Regression 1 with some controls, except that it has a measure of distributional conflict instead of committee fixed effects. The stable groups assumption is presumably not satisfied: it is unlikely that there is any committee where the measure of

distributional conflict is equal to 0. This is a sharp design.

22. **Moser and Voena (2012), Compulsory Licensing: Evidence from the Trading with the Enemy Act.** *Table 2.*
In the regression equation in the beginning of Section III, the dependent variable is the number of patents by US inventors in patent class c at period t, and the explanatory variables are patent class and time fixed effects, the interaction of period t being after the trading with the enemy act and the number of licensed patents in class c, and some control variables. This regression corresponds to Regression 1 with some controls. The stable groups assumption is satisfied: there are patent classes where no patent was licensed. This is a sharp design.

23. **Forman et al. (2012), The Internet and Local Wages: A Puzzle.** *Tables 2 and 4.*
In regression Equation (1), the dependent variable is the difference between log wages in 2000 and 1995 in county i, and the explanatory variables are the proportion of businesses using Internet in county i in 2000, and control variables. This regression corresponds to Regression 2 with some controls. The stable groups assumption is satisfied: there are counties with no Internet investment in 2000. This a fuzzy design: the treatment of interest is whether a business uses Internet, which varies within (county,year) cells.

24. **Besley and Mueller (2012), Estimating the Peace Dividend: The Impact of Violence on House Prices in Northern Ireland.** *Table 1, Columns 3 and 5-7.*
In regression Equation (1), the dependent variable is the price of houses in region r at time t, while the explanatory variables are region and time fixed effects, and the number of people killed because of the civil war in region r at time t-1. This regression corresponds to Regression 1. The stable groups assumption is presumably satisfied: between each pair of consecutive years, it is likely that there are some regions where the number of people killed because of the civil war does not change. This is a sharp design.

25. **Dafny et al. (2012), Paying a Premium on Your Premium? Consolidation in the US Health Insurance Industry.** *Table 3.*
In regression Equation (3), the dependent variable is the the concentration of the hospital industry in market m and year t, and explanatory variables are time fixed effects, market fixed effects, and the change in concentration in market m induced by a merger interacted with an indicator for t being after the merger. This regression corresponds to Regression 1. The stable groups assumption is satisfied: there are many markets where the merger did not change concentration. This is a sharp design.

26. **Hornbeck (2012), The Enduring Impact of the American Dust Bowl: Short- and Long-Run Adjustments to Environmental Catastrophe.** *Table 2.* In regression

19

Equation (1), the dependent variable is, say, the change in log land value in county c between period t and 1930, and the explanatory variables are state × year fixed effects, the share of county c in high erosion regions, the share of county c in medium erosion regions, and some control variables. This regression is similar to Regression 1 with controls, except that it has two treatment variables. The stable groups assumption is satisfied: many counties have 0% of their land situated in medium or high erosion regions. This a fuzzy design: the treatments of interest are whether a piece of land is in high or in medium erosion regions, which varies within (county,year) cells.

27. **Bajari et al. (2012), A Rational Expectations Approach to Hedonic Price Regressions with Time-Varying Unobserved Product Attributes: The Price of Pollution.** *Table 5.*

    In, say, the first regression equation in the bottom of page 1915, the dependent variable is the change in the price of house j between sales 2 and 3, and the explanatory variables are the change in various pollutants in the area around house j between sales 2 and 3, and some controls. This regression is similar to Regression 2 with controls, except that it has several treatment variables. The stable groups assumption is presumably satisfied: it is likely that for each pair of consecutive sales, there are houses where the level of each pollutant does not change. This is a sharp design.

28. **Dahl and Lochner (2012), The Impact of Family Income on Child Achievement: Evidence from the Earned Income Tax Credit.** *Table 3.*

    In regression Equation (4), the dependent variable is the change in test scores for child i between years a and a-1, while the explanatory variables are the change in the EITC income of her family and some controls, and the change in the expected EITC income of her family based on her family income in year a-1 is used to instrument for the actual change of her family's EITC income. This regression is a 2SLS version of Regression 2 with controls, except that it does not have years fixed effects. The stable groups assumption is presumably satisfied: it is likely that for each pair of consecutive years, there are children whose family's expected EITC income does not change. This is a sharp design.

29. **Imberman et al. (2012), Katrina's Children: Evidence on the Structure of Peer Effects from Hurricane Evacuees.** *Tables 3-6.*

    In regression Equation (1), the dependent variable is the test score of student i in school j in grade g and year t, and the explanatory variables are school and grade × year fixed effects, the fraction of Katrina evacuee students received by school j in grade g and year t, and some controls. This regression is a three-way fixed effects version of Regression 1. The stable groups assumption is satisfied: there are schools that did not receive any Katrina evacuee. This a fuzzy design: the treatment of interest is the proportion of evacuees in

one's class, which varies within (school,grade,year) cells.

30. **Chaney et al. (2012), The Collateral Channel: How Real Estate Shocks Affect Corporate Investment.** *Table 5.*
In regression Equation (1), the dependent variable is the value of investment in firm i and year t divided by the lagged book value of properties, plants, and equipments (PPE), and the explanatory variables are firm and time fixed effects and the market value of firm i in year t divided by its lagged PPE, and some controls. This regression corresponds to Regression 1, with some controls. The stable groups assumption is presumably satisfied: it is likely that between each pair of consecutive years, there are firms whose market value divided by their lagged PPE does not change. This is a sharp design.

31. **Aaronson et al. (2012), The Spending and Debt Response to Minimum Wage Hikes.** *Tables 1, 2, and 5.*
In regression Equation (1), the outcome variable is, say, income of household i at period t, and the explanatory variables are household and time fixed effects, and the minimum wage in the state where household i lives in period t. This regression corresponds to Regression 1. The stable groups assumption is satisfied: between each pair of consecutive periods, there are states where the minimum wage does not change. This is a sharp design.

32. **Brambilla et al. (2012), Exports, Export Destinations, and Skills.** *Table 5.*
In the regression in, say, the first column of Table 2, the dependent variable is a measure of skills in the labor force employed by firm i in industry j at period t, and the explanatory variables are firm and industry × period fixed effects, the ratio of exports to sales in firm i at period t, and some controls. This regression corresponds to Regression 1, with some controls. The stable groups assumption is presumably satisfied: it is likely that between each pair of consecutive periods, there are firms whose ratio of exports to sales does not change. This is a sharp design.

33. **Faye and Niehaus (2012), Political Aid Cycles.** *Table 3, Columns 4 and 5, and Tables 4 and 5.*
In regression Equation (2), the dependent variable is the amount of donations received by receiver r from donor d in year t, and the explanatory variables are donor × receiver fixed effects, an indicator for whether there is an election in country r in year t, a measure of alignment between the ruling political parties in countries r and d at $t$, and the interaction of the election indicator and the measure of alignment. This regression corresponds to Regression 1. The stable groups assumption is presumably not satisfied: it is unlikely that there are donor-receiver pairs that are perfectly unaligned. This is a sharp design.

# 7   Proofs

Theorem S1 relies on the following lemma.

**Lemma S1** *If Assumptions 1 and 3-5 hold, for all $(g, g', t, t') \in \{1, ..., G\}^2 \times \{1, ..., T\}^2$,*

$$E(Y_{g,t}|\bm{D}) - E(Y_{g,t'}|\bm{D}) - (E(Y_{g',t}|\bm{D}) - E(Y_{g',t'}|\bm{D}))$$
$$=D_{g,t}E\left(\Delta_{g,t}^{TR}|\bm{D}\right) - D_{g,t'}E\left(\Delta_{g,t'}^{TR}|\bm{D}\right) - \left(D_{g',t}E\left(\Delta_{g',t}^{TR}|\bm{D}\right) - D_{g',t'}E\left(\Delta_{g',t'}^{TR}|\bm{D}\right)\right).$$

*Proof of Lemma S1*

For all $(g, t) \in \{1, ..., G\} \times \{1, ..., T\}$,

$$E(Y_{g,t}|\bm{D}) = E\left(\frac{1}{N_{g,t}} \sum_{i=1}^{N_{g,t}} (Y_{i,g,t}(0) + D_{i,g,t}(Y_{i,g,t}(1) - Y_{i,g,t}(0))) \bigg| \bm{D}\right)$$
$$= E(Y_{g,t}(0)|\bm{D}) + D_{g,t}E\left(\Delta_{g,t}^{TR}|\bm{D}\right).$$

The end of the proof is the same as that of Lemma 1.

**Proof of Theorem S1**

The proof of that result is very similar to the proof of Theorem 1.

$$\beta_{fe} = E\left(\frac{\sum_{g,t} N_{g,t}\varepsilon_{g,t}Y_{g,t}}{\sum_{g,t} N_{g,t}\varepsilon_{g,t}D_{g,t}}\right)$$
$$= E\left(\frac{\sum_{g,t} N_{g,t}\varepsilon_{g,t}E(Y_{g,t}|\bm{D})}{\sum_{g,t} N_{g,t}\varepsilon_{g,t}D_{g,t}}\right)$$
$$= E\left(\frac{\sum_{g,t} N_{g,t}\varepsilon_{g,t}\left(E(Y_{g,t}|\bm{D}) - E(Y_{g,1}|\bm{D}) - E(Y_{1,t}|\bm{D}) + E(Y_{1,1}|\bm{D})\right)}{\sum_{g,t} N_{g,t}\varepsilon_{g,t}D_{g,t}}\right)$$
$$= E\left(\frac{\sum_{g,t} N_{g,t}\varepsilon_{g,t}\left(D_{g,t}E\left(\Delta_{g,t}^{TR}|\bm{D}\right) - D_{g,1}E\left(\Delta_{g,1}^{TR}|\bm{D}\right) - D_{1,t}E\left(\Delta_{1,t}^{TR}|\bm{D}\right) + D_{1,1}E\left(\Delta_{1,1}^{TR}|\bm{D}\right)\right)}{\sum_{g,t} N_{g,t}\varepsilon_{g,t}D_{g,t}}\right)$$
$$= E\left(\frac{\sum_{g,t} N_{g,t}\varepsilon_{g,t}D_{g,t}E\left(\Delta_{g,t}^{TR}|\bm{D}\right)}{\sum_{g,t} N_{g,t}\varepsilon_{g,t}D_{g,t}}\right)$$
$$= E\left(\frac{\sum_{g,t} N_{g,t}\varepsilon_{g,t}D_{g,t}\Delta_{g,t}^{TR}}{\sum_{g,t} N_{g,t}\varepsilon_{g,t}D_{g,t}}\right).$$

The second equality follows from the law of iterated expectations. The third and fifth equalities follow from Equations (5) and (6). The fourth equality follows from Lemma S1. The last equality follows from the law of iterated expectations.

**Proof of Proposition S1**

Assuming that $N_{g,2}/N_{g,1}$ does not vary across $g$ ensures that there exists a strictly positive real number $\phi$ such that $N_{g,2}/N_{g,1} = \phi$. Then,

$$
\begin{aligned}
\varepsilon_{g,2} &= D_{g,2} - D_{g,.} - D_{.,2} + D_{.,.} \\
&= D_{g,2} - \left(\frac{N_{g,1}}{N_{g,.}}D_{g,1} + \frac{N_{g,2}}{N_{g,.}}D_{g,2}\right) - D_{.,2} + \left(\frac{N_{.,1}}{N}D_{.,1} + \frac{N_{.,2}}{N}D_{.,2}\right) \\
&= D_{g,2} - \left(\frac{1}{1+\phi}D_{g,1} + \frac{\phi}{1+\phi}D_{g,2}\right) - D_{.,2} + \left(\frac{1}{1+\phi}D_{.,1} + \frac{\phi}{1+\phi}D_{.,2}\right) \\
&= \frac{1}{1+\phi}\left(D_{g,2} - D_{g,1} - D_{.,2} + D_{.,1}\right),
\end{aligned}
\qquad (25)
$$

where the first and third equalities follow from the fact $N_{g,2}/N_{g,1}$ does not vary across $g$. Then, the definition of $w_{g,2}^{TR}$, Equation (25) and Assumption 1 imply that

$$
\begin{aligned}
w_{g,2}^{TR} &= \frac{(D_{g,2} - D_{.,2})}{\sum_{g=1}^{G}\frac{N_{g,2}}{N_1}(D_{g,2} - D_{.,2})D_{g,2}} \\
&= \frac{(D_{g,2} - D_{.,2})}{\sum_{g=1}^{G}\frac{N_{g,2}}{N_1}(D_{g,2} - D_{.,2})^2}.
\end{aligned}
$$

**Proof of Theorem S2**

The proof relies on the lemma below, which we start by proving, before proving the theorem.

**Lemma S2** *If Assumptions 1-5 and S2 hold,*

$$
\begin{aligned}
&E(Y_{g,t}|\boldsymbol{D}) - E(Y_{g,t-1}|\boldsymbol{D}) - (E(Y_{g',t}|\boldsymbol{D}) - E(Y_{g',t-1}|\boldsymbol{D})) \\
&= (D_{g,t} - D_{g,t-1})E(\Delta_{g,t}|\boldsymbol{D}) - (D_{g',t} - D_{g',t-1})E(\Delta_{g',t}|\boldsymbol{D}).
\end{aligned}
$$

*Proof of Lemma S2*

By Lemma 1 and Assumption S2,

$$
\begin{aligned}
&E(Y_{g,t}|\boldsymbol{D}) - E(Y_{g,t-1}|\boldsymbol{D}) - (E(Y_{g',t}|\boldsymbol{D}) - E(Y_{g',t-1}|\boldsymbol{D})) \\
&= D_{g,t}E(\Delta_{g,t}|\boldsymbol{D}) - D_{g,t-1}E(\Delta_{g,t-1}|\boldsymbol{D}) - D_{g',t}E(\Delta_{g',t}|\boldsymbol{D}) + D_{g',t-1}E(\Delta_{g',t-1}|\boldsymbol{D}) \\
&= (D_{g,t} - D_{g,t-1})E(\Delta_{g,t}|\boldsymbol{D}) - (D_{g',t} - D_{g',t-1})E(\Delta_{g',t}|\boldsymbol{D}).
\end{aligned}
$$

*Proof of the decomposition for the fixed-effect regression*

First, we have

$$\sum_{g,t} N_{g,t}\varepsilon_{g,t}E\left(Y_{g,t}|\boldsymbol{D}\right)$$

$$=\sum_{g,t} N_{g,t}\varepsilon_{g,t}[E\left(Y_{g,t}|\boldsymbol{D}\right) - E\left(Y_{1,t}|\boldsymbol{D}\right)]$$

$$=\sum_{g}\sum_{t=2}^{T}\left[\sum_{t'\geq t} N_{g,t'}\varepsilon_{g,t'}\right][E\left(Y_{g,t}|\boldsymbol{D}\right) - E\left(Y_{g,t-1}|\boldsymbol{D}\right) - (E\left(Y_{1,t}|\boldsymbol{D}\right) - E\left(Y_{1,t-1}|\boldsymbol{D}\right))]$$

$$=\sum_{g}\sum_{t=2}^{T}\left[\sum_{t'\geq t} N_{g,t'}\varepsilon_{g,t'}\right][(D_{g,t} - D_{g,t-1})E\left(\Delta_{g,t}|\boldsymbol{D}\right) - (D_{1,t} - D_{1,t-1})E\left(\Delta_{1,t}|\boldsymbol{D}\right)]$$

$$=\sum_{(g,t):D_{g,t}\neq D_{g,t-1},t\geq 2}\left[N_{g,t}(D_{g,t} - D_{g,t-1})\sum_{t'\geq t}\frac{N_{g,t'}}{N_{g,t}}\varepsilon_{g,t'}\right]E\left(\Delta_{g,t}|\boldsymbol{D}\right). \qquad (26)$$

The first equality follows by (6). The second equality follows from summation by part and (6). The third equality follows from Lemma S2. The fourth equality stems from the fact that by (6), the terms with $g=1$ vanish.

Similarly,

$$\sum_{g,t} N_{g,t}\varepsilon_{g,t}D_{g,t} = \sum_{g}\sum_{t=2}^{T}\left[\sum_{t'\geq t} N_{g,t'}\varepsilon_{g,t'}\right][D_{g,t} - D_{g,t-1}]$$

$$= \sum_{(g,t):D_{g,t}\neq D_{g,t-1},t\geq 2} N_{g,t}(D_{g,t} - D_{g,t-1})\sum_{t'\geq t}\frac{N_{g,t'}}{N_{g,t}}\varepsilon_{g,t'}. \qquad (27)$$

The result follows by combining (4), (26), (27), and the law of iterated expectations.

*Proof of the decomposition for the first-difference regression*

First, we have

$$\sum_{(g,t):t\geq 2} N_{g,t}\varepsilon_{fd,g,t}\left(E\left(Y_{g,t}|\boldsymbol{D}\right) - E\left(Y_{g,t-1}|\boldsymbol{D}\right)\right)$$

$$=\sum_{(g,t):t\geq 2} N_{g,t}\varepsilon_{fd,g,t}\left(E\left(Y_{g,t}|\boldsymbol{D}\right) - E\left(Y_{g,t-1}|\boldsymbol{D}\right) - (E\left(Y_{1,t}|\boldsymbol{D}\right) - E\left(Y_{1,t-1}|\boldsymbol{D}\right))\right)$$

$$=\sum_{(g,t):t\geq 2} N_{g,t}\varepsilon_{fd,g,t}[(D_{g,t} - D_{g,t-1})E\left(\Delta_{g,t}|\boldsymbol{D}\right) - (D_{1,t} - D_{1,t-1})E\left(\Delta_{1,t}|\boldsymbol{D}\right)]$$

$$=\sum_{(g,t):t\geq 2} N_{g,t}\varepsilon_{fd,g,t}(D_{g,t} - D_{g,t-1})E\left(\Delta_{g,t}|\boldsymbol{D}\right).$$

The first equality follows from (14). The second equality follows from Lemma S2. The third equality follows from (14) again. The result follows by combining (13) with the last display, and using the law of iterated expectations.

*Proof that $w^S_{g,t} \geq 0$ under Assumption 6 and if $N_{g,t}/N_{g,t-1}$ does not depend on $g$*

Under Assumption 6, one has that $D_{g,t} = 1\{t \geq a_g\}$, with $a_g \in \{1, ..., T+1\}$. Therefore, given the form of $w^S_{g,t}$, we just have to prove that for all $g$,

$$\sum_{t \geq a_g} N_{g,t}\varepsilon_{g,t} \geq 0. \tag{28}$$

Because $N_{g,t}/N_{g,t-1}$ does not vary across $g$ for all $t \geq 2$, we have $N_{g,t} = N_{g,0}\gamma_t$ for some $\gamma_t \geq 0$. Moreover, $\varepsilon_{g,t} = D_{g,t} - D_{g,.} - D_{.,t} + D_{.,.}$. Let $\widetilde{\gamma}_t = \gamma_t/\sum_{t \geq 0}\gamma_t$, then $D_{g,.} = \sum_{t \geq a_g}\widetilde{\gamma}_t$, and $D_{.,.} = \sum_{t \geq 0}\widetilde{\gamma}_t D_{.,t}$. Hence,

$$\frac{1}{N_{g,0}\sum_t \gamma_t} \sum_{t \geq a_g} N_{g,t}\varepsilon_{g,t} = D_{g,.}(1 - D_{g,.} + D_{.,.}) - \sum_{t \geq a_g}\widetilde{\gamma}_t D_{.,t}$$

$$= D_{g,.}\left(1 - D_{g,.} + \sum_{t < a_g}\widetilde{\gamma}_t D_{.,t}\right) - \left(\sum_{t \geq a_g}\widetilde{\gamma}_t D_{.,t}\right)(1 - D_{g,.}). \tag{29}$$

Now, because $D_{.,t} \leq 1$,

$$\sum_{t \geq a_g}\widetilde{\gamma}_t D_{.,t} \leq \sum_{t \geq a_g}\widetilde{\gamma}_t = D_{g,.}.$$

Hence, in view of (29),

$$\frac{1}{N_{g,0}\sum_t \gamma_t}\sum_{t \geq a_g} N_{g,t}\varepsilon_{g,t} \geq D_{g,.}\sum_{t < a_g}\widetilde{\gamma}_t D_{.,t} \geq 0.$$

Therefore, (28) and the result follows.

*Proof that $w^S_{fd,g,t} \geq 0$*

We just have to focus on the cases where $D_{g,t} \neq D_{g,t-1}$. Note that $\varepsilon_{fd,g,t} = D_{g,t} - D_{g,t-1} - (D_{.,t} - D_{.,t-1})$. Then, if $D_{g,t} - D_{g,t-1} = 1$, the numerator of $w^S_{fd,g,t}$ has the same sign as $1 - (D_{.,t} - D_{.,t-1})$, which is positive. If $D_{g,t} - D_{g,t-1} = -1$, the numerator of $w^S_{fd,g,t}$ has the same sign as $1 + (D_{.,t} - D_{.,t-1})$, which is also positive. Because the denominator sums terms that are always positive, it is positive as well. The result follows.

**Proof of Theorem S3**

The reasoning is exactly the same as in Theorem 1, except that we rely on Lemma S3 below, instead of Lemma 1. We thus only prove Lemma S3.

**Lemma S3** *If Assumptions 1-5 hold and $D_{i,g,t} \in \{0,,...,\overline{d}\}$.*

$$E(Y_{g,t}|\boldsymbol{D}) - E(Y_{g,t'}|\boldsymbol{D}) - (E(Y_{g',t}|\boldsymbol{D}) - E(Y_{g',t'}|\boldsymbol{D}))$$
$$= D_{g,t}E\left(\Delta^{ACR}_{g,t}|\boldsymbol{D}\right) - D_{g,t'}E\left(\Delta^{ACR}_{g,t'}|\boldsymbol{D}\right) - \left(D_{g',t}E\left(\Delta^{ACR}_{g',t}|\boldsymbol{D}\right) - D_{g',t'}E\left(\Delta^{ACR}_{g',t'}|\boldsymbol{D}\right)\right).$$

*Proof of Lemma S3*

Under Assumption 2, we have $E(Y_{g,t}|\boldsymbol{D}) = E(Y_{g,t}(0)|\boldsymbol{D}) + E(Y_{g,t}(D_{g,t}) - Y_{g,t}(0)|\boldsymbol{D})$. The result follows by decomposing similarly the three other terms $E(Y_{g,t'}|\boldsymbol{D})$, $E(Y_{g',t}|\boldsymbol{D})$, and $E(Y_{g',t'}|\boldsymbol{D})$, using Assumptions 3-5, and finally using the definition of $\Delta_{g,t}^{ACR}$.

**Proof of Theorem S4**

The proof relies on the following lemma, that resembles Lemma 1 and that we do not prove.

**Lemma S4** *If Assumptions 1, 2, and S3-S4 hold, for all $(g, g', t, t') \in \{1, ..., G\}^2 \times \{1, ..., T\}^2$,*

$$E\left(Y_{g,t} - X'_{g,t}\gamma\big|\boldsymbol{X}, \boldsymbol{D}\right) - E\left(Y_{g,t'} - X'_{g,t'}\gamma\big|\boldsymbol{X}, \boldsymbol{D}\right)$$
$$- \left(E\left(Y_{g',t} - X'_{g',t}\gamma\big|\boldsymbol{X}, \boldsymbol{D}\right) - E\left(Y_{g',t'} - X'_{g',t'}\gamma\big|\boldsymbol{X}, \boldsymbol{D}\right)\right)$$
$$= D_{g,t}E(\Delta_{g,t}|\boldsymbol{X}, \boldsymbol{D}) - D_{g,t'}E(\Delta_{g,t'}|\boldsymbol{X}, \boldsymbol{D}) - \left(D_{g',t}E(\Delta_{g',t}|\boldsymbol{X}, \boldsymbol{D}) - D_{g',t'}E(\Delta_{g',t'}|\boldsymbol{X}, \boldsymbol{D})\right).$$

It follows from the Frisch-Waugh theorem and the definition of $\varepsilon_{g,t}^X$ that

$$E\left(\widehat{\beta}_{fe}\Big|\boldsymbol{X}, \boldsymbol{D}\right) = \frac{\sum_{g,t} N_{g,t}\varepsilon_{g,t}^X E(Y_{g,t}|\boldsymbol{X}, \boldsymbol{D})}{\sum_{g,t} N_{g,t}\varepsilon_{g,t}^X D_{g,t}}. \tag{30}$$

Now, by definition of $\varepsilon_{g,t}^X$ again,

$$\sum_{t=1}^T N_{g,t}\varepsilon_{g,t}^X = 0 \text{ for all } g \in \{1, ..., G\}, \tag{31}$$

$$\sum_{g=1}^G N_{g,t}\varepsilon_{g,t}^X = 0 \text{ for all } t \in \{1, ..., T\}, \tag{32}$$

$$\sum_{g,t} N_{g,t}\varepsilon_{g,t}^X X_{g,t} = 0. \tag{33}$$

Then,

$$\sum_{g,t} N_{g,t}\varepsilon_{g,t}^X E(Y_{g,t}|\boldsymbol{X}, \boldsymbol{D}))$$
$$= \sum_{g,t} N_{g,t}\varepsilon_{g,t}^X E\left(Y_{g,t} - X'_{g,t}\lambda\big|\boldsymbol{X}, \boldsymbol{D}\right))$$
$$= \sum_{g,t} N_{g,t}\varepsilon_{g,t}^X \left(E\left(Y_{g,t} - X'_{g,t}\lambda\big|\boldsymbol{X}, \boldsymbol{D}\right) - E\left(Y_{g,1} - X'_{g,1}\lambda\big|\boldsymbol{X}, \boldsymbol{D}\right)\right.$$
$$\left. - E\left(Y_{1,t} - X'_{1,t}\lambda\big|\boldsymbol{X}, \boldsymbol{D}\right) + E\left(Y_{1,1} - X'_{1,1}\lambda\big|\boldsymbol{X}, \boldsymbol{D}\right)\right)$$

26

The first equality follows from (33). The second follows from Equations (31) and (32). Hence,

$$\sum_{g,t} N_{g,t}\varepsilon^X_{g,t}E(Y_{g,t}|\boldsymbol{X},\boldsymbol{D})) = \sum_{g,t} N_{g,t}\varepsilon^X_{g,t}(D_{g,t}E(\Delta_{g,t}|\boldsymbol{X},\boldsymbol{D}) - D_{g,1}E(\Delta_{g,1}|\boldsymbol{X},\boldsymbol{D})$$
$$-D_{1,t}E(\Delta_{1,t}|\boldsymbol{X},\boldsymbol{D}) + D_{1,1}E(\Delta_{1,1}|\boldsymbol{X},\boldsymbol{D}))$$
$$= \sum_{g,t} N_{g,t}\varepsilon^X_{g,t}D_{g,t}E(\Delta_{g,t}|\boldsymbol{X},\boldsymbol{D})$$
$$= \sum_{(g,t):D_{g,t}=1} N_{g,t}\varepsilon^X_{g,t}E(\Delta_{g,t}|\boldsymbol{X},\boldsymbol{D}). \tag{34}$$

The first equality follows from Lemma 4. The second follows from Equations (31) and (32). The third follows from Assumption 2. Finally, Assumption 2 implies that

$$\sum_{g,t} N_{g,t}\varepsilon^X_{g,t}D_{g,t} = \sum_{(g,t):D_{g,t}=1} N_{g,t}\varepsilon^X_{g,t}. \tag{35}$$

Combining (30), (34), (35), and the law of iterated expectations yields the result.

**Proof of Theorem S5**

Reasoning as in the proof of Theorem 3, we get that for all $t \geq 2$,

$$N_{d,d',t}E(\text{DID}_{d,d',t}|\boldsymbol{D}) = \sum_{g:D_{g,t}=d',D_{g,t-1}=d} N_{g,t}E(Y_{g,t}(\max(d,d')) - Y_{g,t}(\min(d,d'))|\boldsymbol{D})$$
$$= \sum_{(i,g):D_{g,t}=d',D_{g,t-1}=d} E(Y_{i,g,t}(\max(D_{g,t},D_{g,t-1})) - Y_{i,g,t}(\min(D_{g,t},D_{g,t-1}))|\boldsymbol{D}).$$

For all $(g,t)$, there exists one $(d,d') \in \mathcal{D}^2$ such that $D_{g,t} = d'$ and $D_{g,t-1} = d$. Hence,

$$\sum_{t=2}^{T}\sum_{(d,d')\in\mathcal{D}^2:d\neq d'} N_{d,d',t}E(\text{DID}_{d,d',t}|\boldsymbol{D})$$
$$= \sum_{t=2}^{T}\sum_{(d,d')\in\mathcal{D}^2} N_{d,d',t}E(\text{DID}_{d,d',t}|\boldsymbol{D})$$
$$= \sum_{(i,g,t):t\geq 2} E(Y_{i,g,t}(\max(D_{g,t},D_{g,t-1})) - Y_{i,g,t}(\min(D_{g,t},D_{g,t-1}))|\boldsymbol{D}).$$

The result follows by definition of $\text{DID}_M$ and $\delta^{SCR}$, and the law of iterated expectations.

**Proof of Theorem S6**

*1. Asymptotic normality*

Let us define $P^S = N_S/G$ and

$$T^S = \frac{1}{G} \sum_{(g,t):t\geq 2} N_{g,t}\mathbb{1}\{D_{g,t} \neq D_{g,t-1}\} [Y_{g,t}(1) - Y_{g,t}(0)].$$

We prove the result in two steps. First, we prove that $\sqrt{G}\left(\text{DID}_M - E(T^S)/E(P^S)\right)$ is asymptotically normal. Second, we show that the difference between $E(T^S)/E(P^S)$ and $\delta^S$ is asymptotically negligible.

**Convergence of $\sqrt{G}\left(\text{DID}_M - E(T^S)/E(P^S)\right)$**

By Assumption S9,

$$\sup_{g,t} N_{g,t}^4 E[\mathbb{1}\{D_{g,t} = d, D_{g,t-1} = d'\}(Y_{g,t} - Y_{g,t-1})^4] < +\infty.$$

Thus, Lyapunov's condition for the central limit theorem holds, and because $\Sigma = \lim_g G \times V(\boldsymbol{U})$ exists,

$$\sqrt{G}(\boldsymbol{U} - E[\boldsymbol{U}]) \xrightarrow{d} \mathcal{N}(0, \Sigma),$$

By Assumption S9, $P_{d,d',t}^\infty = \lim_{G\to\infty} E(P_{d,d',t})$ exists. Then for $d \in \{0,1\}$, define $\mathcal{T}_d = \{t : P_{1-d,d,t}^\infty > 0\}$ and the event $\mathcal{D} = \{N_S > 0\} \cap \mathcal{D}_0 \cap \mathcal{D}_1$, with

$$\mathcal{D}_d = \{t \in \mathcal{T}_d \iff \min(N_{d,d,t}, N_{1-d,d,t}) > 0\}.$$

By the law of large numbers and Assumption S10, $N_S > 0$ with probability approaching one. Next, fix $d \in \{0,1\}$. If $t \in \mathcal{T}_d$, then, by the law of large numbers, $N_{1-d,d,t} > 0$ with probability approaching one. Moreover, for such a $t$, there exists $g$ such that $\Pr(D_{g,t} = 1-d, D_{g,t-1} = d) > 0$ and thus, by Assumption S10, $P_{d,d,t}^\infty > 0$. Then, by the law of large numbers again, $N_{d,d,t} > 0$ with probability approaching one. Conversely, if $\min(N_{d,d,t}, N_{1-d,d,t}) > 0$, then there exists $g$ such that $\Pr(D_{g,t} = 1-d, D_{g,t-1} = d) > 0$. Hence, by Assumption S10 again, $P_{1-d,d,t}^\infty > 0$. This shows that $\mathcal{D}_d$, and thus $\mathcal{D}$, holds with probability approaching one.

Now, by definition of $\text{DID}_M$, under $\mathcal{D}$ we have $\text{DID}_M = f(\boldsymbol{U})$, with

$$f(p_{0,0,1}, q_{0,0,1}, ..., p_{1,1,T}, q_{1,1,T}) = \frac{\sum_{d=0}^1 (-1)^d \sum_{t \in \mathcal{T}_d} q_{1-d,d,t} - (p_{1-d,d,t}/p_{d,d,t})q_{d,d,t}}{\sum_{t=1}^T p_{1,0,t} + p_{0,1,t}}, \quad (36)$$

for all $(p_{d,d',t})_{(d,d',t)}$ such that all denominators are strictly positive. By Assumption S9 again, $E[\boldsymbol{U}]$ converges to $\boldsymbol{U}^\infty$. Furthermore, $f$ is continuously differentiable in a neighborhood of $\boldsymbol{U}^\infty$. Thus, by the uniform delta method (see, e.g. van der Vaart, 2000, Theorem 3.8),

$$\sqrt{G}(\text{DID}_M - f(E[\boldsymbol{U}])) = J_f(\boldsymbol{U}^\infty) \times \sqrt{G}(\boldsymbol{U} - E[\boldsymbol{U}]) + o_P(1). \quad (37)$$

Finally, $f(E(\boldsymbol{U})) = f_N/f_D$, with

$$f_N = \frac{1}{G} \sum_{d=0}^{1} (-1)^d \sum_{t \in \mathcal{T}_d} \sum_g N_{g,t} \Bigg[ E\left(\mathbb{1}\{D_{g,t} = 1-d, D_{g,t-1} = d\}(Y_{g,t} - Y_{g,t-1})\right) \\ - \frac{E(P_{1-d,d,t})}{E(P_{d,d,t})} E\left(\mathbb{1}\{D_{g,t} = D_{g,t-1} = d\}(Y_{g,t} - Y_{g,t-1})\right) \Bigg]$$

$$f_D = \sum_{t \geq 2} E[P_{1,0,t} + P_{0,1,t}].$$

Reasoning as in the proof of Theorem 3 (see in particular Equations (20)-(21)) and noting that if $t \notin \mathcal{T}_d$, then, by Assumption S10 $\Pr(D_{g,t} = 1-d, D_{g,t-1} = d) = 0$, we get $f_N = E(T^S)$. Moreover, $f_D = E(P^S)$. Hence,

$$\sqrt{G}\left(\text{DID}_M - \frac{E(T^S)}{E(P^S)}\right) \xrightarrow{d} \mathcal{N}(0, \sigma^2), \text{ with } \sigma^2 = J_f(\boldsymbol{U}^\infty)\Sigma J_f(\boldsymbol{U}^\infty)'. \quad (38)$$

**Convergence to 0 of $\sqrt{G}\left(E(T^S)/E(P^S) - \delta^S\right)$**

Let us define $\boldsymbol{D}_G = (D_{g,t})_{(g,t):g \leq G, t=1\ldots T}$, $\widetilde{T} = E[T^S|\boldsymbol{D}_G]$ and $I = \mathbb{1}\{|P^S - E(P^S)| < \varepsilon_G\}$, where $\varepsilon_G > 0$ will be specified below. By Assumption S10, $\lim_{G \to \infty} E(P^S) > 0$. Thus, it suffices to prove that $\sqrt{G}\left(E(P^S)\delta^S - E(T^S)\right) \to 0$. Because $\delta^S = E[T^S/P^S]$, we have

$$E(P^S)\delta^S - E(T^S) = E\left[\widetilde{T}(E(P^S) - P^S)/P^S\right]$$
$$= E\left[\widetilde{T}I(E(P^S) - P^S)/P^S\right] + E\left[\widetilde{T}(1-I)(E(P^S) - P^S)/P^S\right]. \quad (39)$$

First, consider the second term on the right-hand side. By applying twice the Cauchy-Schwarz inequality, we get

$$\sqrt{G}|E[\widetilde{T}(1-I)(E(P^S) - P^S)/P^S]| \leq \sqrt{G}V(P^S)^{1/2}\left[E[(\widetilde{T}/P^S)^4]\Pr(I=0)\right]^{1/4}$$

By Assumption S9, $\sqrt{G}V(P^S)^{1/2}$ converges towards a finite limit. Thus, it suffices to show that the term into brackets tends to 0. To this end, note first that

$$E[T^S|\boldsymbol{D}_G] = \frac{1}{G} \sum_{(g,t):t \geq 2} N_{g,t}\mathbb{1}\{D_{g,t} \neq D_{g,t-1}\}E[Y_{g,t}(1) - Y_{g,t}(0)|\boldsymbol{D}_G].$$

Now, let $A_{g,t} = N_{g,t}\mathbb{1}\{D_{g,t} \neq D_{g,t-1}\}$ and $B_{g,t} = E[Y_{g,t}(1) - Y_{g,t}(0)|\boldsymbol{D}_G]$. By Assumption S9 and Jensen's inequality, $\sup_{(g,G,t):g \leq G} E[|B_{g,t}|^4] < +\infty$. Then

$$\left(\frac{\widetilde{T}}{P^S}\right)^4 \leq \left(\frac{\sum_{g,t} A_{g,t}|B_{g,t}|}{\sum_{g,t} A_{g,t}}\right)^4$$
$$\leq \left(\max_{g,t}|B_{g,t}|^4\right)$$
$$\leq \sum_{g,t}|B_{g,t}|^4.$$

29

Hence, $E[(\widetilde{T}/P^S)^4]\Pr(I=0) \leq K_1 G \Pr(I=0)$ for some constant $K_1 > 0$. Moreover, by Hoeffding's inequality,
$$\Pr(I=0) \leq 2\exp\left(\frac{-2G^2\varepsilon_G^2}{\sum_g N_{g,\cdot}^2}\right).$$

By Assumption S9, there exists $c > 0$ such that for all $G$, $1/G\sum_{g=1}^{G} N_{g,\cdot}^2 < 2c$. Then,
$$\frac{-2G^2\varepsilon_G^2}{\sum_g N_{g,\cdot}^2} \leq -\frac{G\varepsilon_G^2}{c}.$$

Let $\varepsilon_G = (C\ln(G)/G)^{1/2}$, for some $C > c$. Then, by what precedes, $G\Pr(I=0) \to 0$ and the second term of the right-hand side of (39) tends to zero.

Now, let us move to the first term of the right-hand side of (39). We have
$$\begin{aligned}
&|E[\widetilde{T}I(E(P^S)-P^S)/P^S]| \\
&= \left|E[I(\widetilde{T}-E(\widetilde{T}))(E(P^S)-P^S)/P^S] + E(\widetilde{T})\left(E(P^S)E(I/P^S)-E(I)\right)\right| \\
&\leq E(P^S)V(\widetilde{T})^{1/2}\left(E[I(1/P^S-1/E(P^S))^2]\right)^{1/2} \\
&\quad + |E(\widetilde{T})|E(P^S)|E(I/P^S)-E(I)/E(P^S)|.
\end{aligned} \tag{40}$$

We now prove that both terms on the right-hand side tend to zero. First, by Taylor expansions of $x \mapsto 1/x$ around $E(P^S)$, there exist $(P_1^S, P_2^S)$ in the interval between $P^S$ and $E(P^S)$ such that
$$\frac{1}{P^S} = \frac{1}{E(P^S)} - \frac{P^S-E(P^S)}{P_1^{S2}}, \tag{41}$$
$$\frac{1}{P^S} = \frac{1}{E(P^S)} - \frac{P^S-E(P^S)}{E(P^S)^2} + \frac{(P^S-E(P^S))^2}{P_2^{S3}}. \tag{42}$$

When $I=1$, $|P_1^S - E(P^S)| < \varepsilon_G$ and $|P_2^S - E(P^S)| < \varepsilon_G$. Recall also that $\lim_G E(P^S) > 0$ and $\varepsilon_G \to 0$. Then, in view of (41) and by Assumption S9,
$$E[I(1/P^S-1/E(P^S))^2] \leq \frac{V(P^S)}{(E(P^S)-\varepsilon_G)^4} \to 0.$$

Moreover, by definition of $\widetilde{T}$, $GV(\widetilde{T}) \leq GV(T^S)$, and the latter is bounded by Assumption S9. Therefore, the first term of the right-hand side of (40) tends to 0.

Now, multiplying (42) by $I$ and taking the expectation on both sides, we obtain:
$$\begin{aligned}
\sqrt{G}|E(I/P^S)-E(I)/E(P^S)| &\leq \sqrt{G}\left|\frac{E[I(P^S-E(P^S))]}{E(P^S)^2}\right| + \frac{\sqrt{G}V(P^S)}{(E(P^S)-\varepsilon_G)^3} \\
&\leq \left|\frac{[G\Pr(I=0)V(P^S)]^{1/2}}{E(P^S)^2}\right| + \frac{\sqrt{G}V(P^S)}{(E(P^S)-\varepsilon_G)^3} \\
&\to 0.
\end{aligned}$$

30

## 2. Validity of the confidence intervals

**There exists $\overline{\sigma}_G^2$ such that $\sum_g \widehat{\psi}_g^2/G - \overline{\sigma}_G^2 \xrightarrow{\mathbb{P}} 0$, with $\liminf_G \overline{\sigma}_G^2 \geq \sigma^2$.**

Let $Q_{d,d',t}^\infty = \lim_{G\to\infty} E[Q_{d,d',t}]$ and $\boldsymbol{\lambda}_g$ be the column vector such that $\boldsymbol{U} = \sum_{g=1}^G \boldsymbol{\lambda}_g/G$. Some tedious algebra show that $J_f(\boldsymbol{U}^\infty) \times \boldsymbol{\lambda}_g = \psi_g$, with

$$\psi_g = \frac{1}{P_S^\infty} \sum_{t>1} N_{g,t} \left[\mathbb{1}\{D_{g,t} \neq D_{g,t-1}\}(Y_{g,t} - Y_{g,t-1} - f(\boldsymbol{U}^\infty)) - \psi_{g,t}^B\right]$$

and

$$\psi_{g,t}^B = \frac{1}{P_{0,0,t}^\infty}\left[\mathbb{1}\{D_{g,t} > D_{g,t-1}\}Q_{0,0,t}^\infty + P_{1,0,t}^\infty \mathbb{1}\{D_{g,t} = D_{g,t-1} = 0\}\left(Y_{g,t} - Y_{g,t-1} - \frac{Q_{0,0,t}^\infty}{P_{0,0,t}^\infty}\right)\right]$$
$$+ \frac{1}{P_{1,1,t}^\infty}\left[\mathbb{1}\{D_{g,t} < D_{g,t-1}\}Q_{1,1,t}^\infty + P_{0,1,t}^\infty \mathbb{1}\{D_{g,t} = D_{g,t-1} = 1\}\left(Y_{g,t} - Y_{g,t-1} - \frac{Q_{1,1,t}^\infty}{P_{1,1,t}^\infty}\right)\right].$$

Hence, in view of (37)-(38), $\sigma^2 = \lim_G \sum_g V(\psi_g)/G$. Next, we show that $\sum_g \widehat{\psi}_g^2/G$ is asymptotically larger than $\sigma^2$. For that purpose, let $\overline{\sigma}_G^2 = \sum_g E(\psi_g^2)/G$ and remark that $\widehat{\psi}_g = J_f(\boldsymbol{U}) \times \boldsymbol{\lambda}_g$. Then

$$\frac{1}{G}\sum_g \widehat{\psi}_g^2 - \overline{\sigma}_G^2 = \frac{1}{G}\sum_g \left[\widehat{\psi}_g^2 - \psi_g^2\right] + \left[\frac{1}{G}\sum_g \psi_g^2 - \overline{\sigma}_G^2\right]. \tag{43}$$

Let $\lambda_{k,g}$ denote the $k$th coordinate of $\boldsymbol{\lambda}_g$. Assumption S9 ensures that

$$\sup_{k,g} E\left[\lambda_{k,g}^4\right] < +\infty. \tag{44}$$

Thus, we also have $\sup_g E(\psi_g^2) < +\infty$. Therefore, by the weak law of large numbers, the second term on the right-hand side of (43) converges to 0. Next,

$$\frac{1}{G}\sum_g \left[\widehat{\psi}_g^2 - \psi_g^2\right] = (J_f(\boldsymbol{U}) - J_f(\boldsymbol{U}^\infty)) \left[\frac{1}{G}\sum_g \boldsymbol{\lambda}_g \boldsymbol{\lambda}_g'\right](J_f(\boldsymbol{U}) + J_f(\boldsymbol{U}^\infty))'. \tag{45}$$

By (44) again and the weak law of large numbers, $\boldsymbol{U} \xrightarrow{\mathbb{P}} \boldsymbol{U}^\infty$. Moreover, $f$ is continuously differentiable in a vicinity of $\boldsymbol{U}^\infty$, Thus, by the continuous mapping theorem, $J_f(\boldsymbol{U}) - J_f(\boldsymbol{U}^\infty) \xrightarrow{\mathbb{P}} 0$. By (44) once again and the weak law of large numbers,

$$\frac{1}{G}\sum_g \left(\boldsymbol{\lambda}_g \boldsymbol{\lambda}_g' - E[\boldsymbol{\lambda}_g \boldsymbol{\lambda}_g']\right) \xrightarrow{\mathbb{P}} 0.$$

Moreover, by the Cauchy-Schwarz inequality and (44), we have, for all $k, \ell$,

$$\left|\frac{1}{G}\sum_g E[\lambda_{k,g}\lambda_{\ell,g}]\right| \leq \sup_{k,\ell,g} E[|\lambda_{k,g}\lambda_{\ell,g}|] \leq \sup_{k,\ell,g} E\left[|\lambda_{k,g}|^2\right] < +\infty.$$

As a result, $\sum_g \boldsymbol{\lambda}_g \boldsymbol{\lambda}_g'/G = O_P(1)$. Finally, because $J_f(\boldsymbol{U})$ converges in probability, $J_f(\boldsymbol{U}) + J_f(\boldsymbol{U}^\infty) = O_P(1)$. Thus, in view of (45), the first term on the right-hand side of (43) converges in probability to 0. Hence, we have proven that $\sum_g \widehat{\psi}_g^2/G - \overline{\sigma}_G^2 \xrightarrow{\mathbb{P}} 0$. Finally, $E(\psi_g^2) \geq V(\psi_g)$ and thus $\overline{\sigma}_G^2 - \sum_g V(\psi_g)/G \geq 0$. Therefore, $\liminf_G \overline{\sigma}_G^2 \geq \sigma^2$.

31

**$\mathrm{CI}_{1-\alpha}(\delta^S)$ is asymptotically conservative.**

By (38), the convergence to 0 of $\sqrt{G}\left(E(T^S)/E(P^S) - \delta^S\right)$, (37), and since $\psi_g = J_f(\boldsymbol{U}^\infty) \times \boldsymbol{\lambda}_g$,

$$\sqrt{G}\frac{\mathrm{DID}_{\mathrm{M}} - \delta^S}{\widehat{\sigma}} = \frac{\sigma}{\overline{\sigma}_G}\left[\frac{\overline{\sigma}_G}{\widehat{\sigma}}\frac{\frac{1}{\sqrt{G}}\sum_g \psi_g}{\sigma} + o_P(1)\right].$$

Let $Z_G$ denote the term into brackets. By the first step above and Slutsky's lemma $Z_G \xrightarrow{d} \mathcal{N}(0,1)$. Fix $\eta > 0$. Because $\liminf_G \overline{\sigma}_G^2 \geq \sigma^2$, there exists $G_0$ such that for every $G \geq G_0$,

$$\sqrt{G}\left|\frac{\mathrm{DID}_{\mathrm{M}} - \delta^S}{\widehat{\sigma}}\right| \leq (1+\eta)|Z_G|.$$

Then, letting $\Phi$ denote the cdf of the standard normal distribution, we get

$$\limsup_G \Pr\left(\delta^S \in \mathrm{CI}_{1-\alpha}\right) \geq \Phi\left(\frac{z_{1-\alpha/2}}{1+\eta}\right) - \Phi\left(\frac{z_{\alpha/2}}{1+\eta}\right).$$

The result follows by letting $\eta$ tend to zero.

# References

Aaronson, D., Agarwal, S. and French, E. (2012), 'The spending and debt response to minimum wage hikes', *American Economic Review* **102**(7), 3111–3139.

Acemoglu, D., Cantoni, D., Johnson, S. and Robinson, J. A. (2011), 'The consequences of radical reform: The french revolution', *American Economic Review* **101**(7), 3286–3307.

Aizer, A. (2010), 'The gender wage gap and domestic violence', *American Economic Review* **100**(4), 1847–1859.

Algan, Y. and Cahuc, P. (2010), 'Inherited trust and growth', *American Economic Review* **100**(5), 2060–2092.

Anderson, S. T. and Sallee, J. M. (2011), 'Using loopholes to reveal the marginal cost of regulation: The case of fuel-economy standards', *American Economic Review* **101**(4), 1375–1409.

Bagwell, K. and Staiger, R. W. (2011), 'What do trade negotiators negotiate about? empirical evidence from the world trade organization', *American Economic Review* **101**(4), 1238–1273.

Bajari, P., Fruehwirth, J. C., Kim, K. I. and Timmins, C. (2012), 'A rational expectations approach to hedonic price regressions with time-varying unobserved product attributes: The price of pollution', *American Economic Review* **102**(5), 1898–1926.

Baum-Snow, N. and Lutz, B. F. (2011), 'School desegregation, school choice, and changes in residential location patterns by race', *American Economic Review* **101**(7), 3019–3046.

Besley, T. and Mueller, H. (2012), 'Estimating the peace dividend: The impact of violence on house prices in northern ireland', *American Economic Review* **102**(2), 810–833.

Bloom, N., Sadun, R. and Van Reenen, J. (2012), 'Americans do it better: Us multinationals and the productivity miracle', *American Economic Review* **102**(1), 167–201.

Brambilla, I., Lederman, D. and Porto, G. (2012), 'Exports, export destinations, and skills', *American Economic Review* **102**(7), 3406–3438.

Bustos, P. (2011), 'Trade liberalization, exports, and technology upgrading: Evidence on the impact of mercosur on argentinian firms', *American Economic Review* **101**(1), 304–340.

Chandra, A., Gruber, J. and McKnight, R. (2010), 'Patient cost-sharing and hospitalization offsets in the elderly', *American Economic Review* **100**(1), 193–213.

Chaney, T., Sraer, D. and Thesmar, D. (2012), 'The collateral channel: How real estate shocks affect corporate investment', *American Economic Review* **102**(6), 2381–2409.

Dafny, L., Duggan, M. and Ramanarayanan, S. (2012), 'Paying a premium on your premium? consolidation in the us health insurance industry', *American Economic Review* **102**(2), 1161–1185.

Dahl, G. B. and Lochner, L. (2012), 'The impact of family income on child achievement: Evidence from the earned income tax credit', *American Economic Review* **102**(5), 1927–1956.

de Chaisemartin, C. (2010), A note on instrumented difference in differences. Working Paper.

de Chaisemartin, C. and D'Haultfœuille, X. (2018), 'Fuzzy differences-in-differences', *The Review of Economic Studies* **85**(2), 999–1028.

Dinkelman, T. (2011), 'The effects of rural electrification on employment: New evidence from south africa', *American Economic Review* **101**(7), 3078–3108.

Duflo, E. (2001), 'Schooling and labor market consequences of school construction in indonesia: Evidence from an unusual policy experiment', *American Economic Review* **91**(4), 795–813.

Duggan, M. and Morton, F. S. (2010), 'The effect of medicare part d on pharmaceutical prices and utilization', *American Economic Review* **100**(1), 590–607.

Duranton, G. and Turner, M. A. (2011), 'The fundamental law of road congestion: Evidence from us cities', *American Economic Review* **101**(6), 2616–2652.

Ellul, A., Pagano, M. and Panunzi, F. (2010), 'Inheritance law and investment in family firms', *American Economic Review* **100**(5), 2414–2450.

Enikolopov, R., Petrova, M. and Zhuravskaya, E. (2011), 'Media and political persuasion: Evidence from russia', *American Economic Review* **101**(7), 3253–3285.

Fang, H. and Gavazza, A. (2011), 'Dynamic inefficiencies in an employment-based health insurance system: Theory and evidence', *American Economic Review* **101**(7), 3047–77.

Faye, M. and Niehaus, P. (2012), 'Political aid cycles', *American Economic Review* **102**(7), 3516–3530.

Forman, C., Goldfarb, A. and Greenstein, S. (2012), 'The internet and local wages: A puzzle', *American Economic Review* **102**(1), 556–575.

Gentzkow, M., Shapiro, J. M. and Sinkinson, M. (2011), 'The effect of newspaper entry and exit on electoral politics', *American Economic Review* **101**(7), 2980–3018.

Hornbeck, R. (2012), 'The enduring impact of the american dust bowl: Short-and long-run adjustments to environmental catastrophe', *American Economic Review* **102**(4), 1477–1507.

Hotz, V. J. and Xiao, M. (2011), 'The impact of regulations on the supply and quality of care in child care markets', *American Economic Review* **101**(5), 1775–1805.

Hudson, S., Hull, P. and Liebersohn, C. (2015), Interpreting instrumented difference-in-differences, Technical report, Working Paper (available upon request).

Imbens, G. W. and Angrist, J. D. (1994), 'Identification and estimation of local average treatment effects', *Econometrica* **62**(2), 467–475.

Imberman, S. A., Kugler, A. D. and Sacerdote, B. I. (2012), 'Katrina's children: Evidence on the structure of peer effects from hurricane evacuees', *American Economic Review* **102**(5), 2048–2082.

Liu, R. Y. and Singh, K. (1995), 'Using iid bootstrap inference for general non-iid models', *Journal of statistical planning and inference* **43**(1-2), 67–75.

Mian, A. and Sufi, A. (2011), 'House prices, home equity-based borrowing, and the us household leverage crisis', *American Economic Review* **101**(5), 2132–2156.

Moser, P. and Voena, A. (2012), 'Compulsory licensing: Evidence from the trading with the enemy act', *American Economic Review* **102**(1), 396–427.

Simcoe, T. (2012), 'Standard setting committees: Consensus governance for shared technology platforms', *American Economic Review* **102**(1), 305–336.

van der Vaart, A. (2000), *Asymptotics Statistics*, Cambridge University Press.

Wang, S.-Y. (2011), 'State misallocation and housing prices: Theory and evidence from china', *American Economic Review* **101**(5), 2081–2107.

Zhang, X. M. and Zhu, F. (2011), 'Group size and incentives to contribute: A natural experiment at chinese wikipedia', *American Economic Review* **101**(4), 1601–1605.



\end{document}